\RequirePackage{rotating}
\documentclass[fleqn,usenatbib]{mnras}

\usepackage{newtxtext,newtxmath}
   
\usepackage{float}

\usepackage[T1]{fontenc}

\usepackage{graphicx}	
\usepackage{amsmath}	
\usepackage{amssymb}	
\usepackage{subcaption}
\usepackage{xcolor}
\usepackage{placeins}
\usepackage{pdflscape}
\setcitestyle{citesep={,}}
\usepackage{hyperref}

\newcommand{\degree}{$^{\circ}$}	

\title[{Source Counts and Sky Temperature from MIGHTEE}]{{MIGHTEE:} Deep 1.4 GHz Source Counts and the {Sky Temperature} {Contribution of Star Forming Galaxies and Active Galactic Nuclei} }

\author[C. L. Hale et al.]{C. L. Hale,$^{1}$\thanks{E-mail: Catherine.Hale@ed.ac.uk}
I. H. Whittam,$^{2,3}$
M. J. Jarvis,$^{2,3}$
P. N. Best,$^{1}$
N. L. Thomas,$^{4,5}$
I. Heywood,$^{2,6,7}$
M. Prescott,$^{3,8}$ \newauthor
N. Adams,$^{2, 9}$
J. Afonso,$^{10, 11}$
Fangxia An,$^{12, 8}$
R. A. A. Bowler,$^{2,13}$
J. D. Collier,$^{14, 15, 16}$ 
R. H. W. Cook,$^{17}$\newauthor
R. Dav\'{e},$^{1, 3, 18}$
B. S. Frank,$^{7, 14, 19}$
M. Glowacki,$^{8, 20}$
P. W. Hatfield,$^{2}$
S. Kolwa$^{14, 21}$
C. C. Lovell,$^{22}$
N. Maddox,$^{23}$\newauthor
L. Marchetti,$^{19, 24}$
L. K. Morabito,$^{4, 5}$
E. Murphy,$^{25}$
I. Prandoni,$^{24}$
Z. Randriamanakoto,$^{18, 26}$
A. R. Taylor$^{19, 14, 3}$ 
\\
$^{1}$ School of Physics and Astronomy, Institute for Astronomy, University of Edinburgh, Royal Observatory, Blackford Hill, EH9 3HJ Edinburgh, UK\\
$^{2}$ Sub-Department of Astrophysics, University of Oxford, Keble Road, Oxford OX1 3RH, UK\\
$^{3}$ Department of Physics and Astronomy, University of the Western Cape, Robert Sobukwe Road, 7535 Bellville, Cape Town,
South Africa \\
$^{4}$ Centre for Extragalactic Astronomy, Department of Physics, Durham University, Durham DH1 3LE, UK \\
$^{5}$ Institute for Computational Cosmology, Department of Physics, Durham University, Durham DH1 3LE, UK \\
$^{6}$ Department of Physics and Electronics, Rhodes University, PO Box 94, Makhanda, 6140, South Africa\\
$^{7}$ South African Radio Astronomy Observatory, 2 Fir Street, Black River Park, Observatory, Cape Town, 7925, South Africa \\
$^{8}$ Inter-University Institute for Data Intensive Astronomy, University of the Western Cape, Robert Sobukwe Road, 7535 Bellville, Cape Town, South Africa \\
$^{9}$ Jodrell Bank Centre for Astrophysics, Alan Turing Building, University of Manchester, Oxford Road, Manchester, UK \\
$^{10}$ Instituto de Astrof\'{i}sica e Ci\^{e}ncias do Espa\c co, Universidade de Lisboa, OAL, Tapada da Ajuda, PT1349-018 Lisboa, Portugal\\
$^{11}$ Departamento de F\'{i}sica, Faculdade de Ci\^{e}ncias, Universidade de Lisboa, Edif\'{i}cio C8, Campo Grande, PT1749-016 Lisbon, Portugal \\
$^{12}$ Purple Mountain Observatory, Chinese Academy of Sciences, Nanjing 210023, China \\
$^{13}$ Jodrell Bank Centre for Astrophysics, Department of Physics and Astronomy, School of Natural Sciences, The University of Manchester, Manchester, M13 9PL, UK \\
$^{14}$ The Inter-University Institute for Data Intensive Astronomy, Department of Astronomy, University of Cape Town, Private Bag X3, Rondebosch, 7701, South Africa \\
$^{15}$ School of Science, Western Sydney University, Locked Bag 1797, Penrith, NSW 2751, Australia \\
$^{16}$ CSIRO Astronomy and Space Science, PO Box 1130, Bentley, WA, 6102, Australia \\
$^{17}$ International Centre for Radio Astronomy Research (ICRAR), University of Western Australia, Crawley, WA 6009, Australia \\
$^{18}$ South African Astronomical Observatories, P.O. Box 9, Observatory, Cape Town, 7935, South Africa \\
$^{19}$ Department of Astronomy, University of Cape Town, Private Bag X3, Rondebosch 7701, South Africa \\
$^{20}$ International Centre for Radio Astronomy Research, Curtin University, Bentley, WA 6102, Australia\\
$^{21}$Physics Department, University of Johannesburg, 5 Kingsway Ave, Rossmore, Johannesburg, 2092, South Africa \\
$^{22}$ Centre for Astrophysics Research, School of Physics, Astronomy and Mathematics, University of Hertfordshire, College Lane, Hatfield, Hertfordshire AL10 9AB, UK \\
$^{23}$ University Observatory, Faculty of Physics, Ludwig-Maximilians-Universit\"at, Scheinerstr. 1, 81679 Munich, Germany \\
$^{24}$ INAF-Istituto di Radioastronomia, Via P. Gobetti 101, 40129, Bologna, Italy \\
$^{25}$ National Radio Astronomy Observatory, 520 Edgemont Road, Charlottesville, VA 22903, USA \\
$^{26}$ Department of Physics, University of Antananarivo, P.O. Box 906, Antananarivo, Madagascar 
}
\date{Accepted XXX. Received YYY; in original form ZZZ}

\pubyear{2022}

\begin{document}
\label{firstpage}
\pagerange{\pageref{firstpage}--\pageref{lastpage}}
\maketitle

\begin{abstract}
We present deep 1.4 GHz source counts from {$\sim$5 deg$^2$ of} the {{continuum Early Science}} data release of the MeerKAT International Gigahertz Tiered Extragalactic Exploration (MIGHTEE) survey down to {{$S_{1.4\textrm{GHz}}\sim$}}15 $\muup$Jy. Using observations over two extragalactic fields (COSMOS and XMM-LSS){{,}} we provide a comprehensive {investigation into} correcting the incompleteness of the raw source counts within the survey to understand the true underlying {{source count population.}} {We use} a {{variety}} of simulations {{that}} account for{:} {errors in} source detection and characterisation, {clustering}, and variations in the assumed source model used to simulate sources within the field {and characterise source count incompleteness}. We present these deep source count distributions and {use them to} investigate the {contribution of extragalactic sources to the sky background temperature at} {1.4 GHz using} a relatively large {sky} {area.} We then use the wealth of ancillary data covering {a subset of the COSMOS field to} investigate the specific contributions from both active galactic nuclei {{(AGN)}} and star forming galaxies {{(SFGs)}} to the {source counts and} sky background {{temperature.}} We find, similar to previous deep studies, that we are unable to reconcile the {sky} temperature observed by the ARCADE 2 experiment. We show that AGN provide the majority contribution to the sky {temperature contribution from radio sources}, but the relative contribution of SFGs rises sharply below 1 mJy, reaching an approximate {15}-25\% contribution to the {total {sky} background temperature} {{($T_b\sim$100 mK)}} at $\sim$15 $\muup$Jy. \\

\end{abstract}




\section{Introduction}
\label{sec:intro}
{As radio astronomers head towards the era of the Square Kilometre Array Observatory (SKAO)\footnote{{\url{https://www.skao.int}}}, a combination of SKAO precursor and pathfinder telescopes are transforming the ability to observe galaxies to sub-mJy and even to $\muup$Jy sensitivities at radio frequencies {of} tens of MHz to several GHz and these facilities combine both fast survey speeds with large area observations. This includes surveys from precursor facilities such as the Meer Karoo Array Telescope \citep[MeerKAT;][]{MeerKAT1, MeerKAT2} which is located at the SKAO site in South Africa and pathfinder facilities which span the frequencies of the proposed SKAO. These pathfinder facilities include mid frequency ($\sim$GHz) observations with facilities such as {the} Australian Square Kilometre Array Pathfinder \citep[ASKAP;][]{ASKAP1, ASKAP2, ASKAP3} and low frequency ($\sim$10-200 MHz) observations with the LOw Frequency ARray \citep[LOFAR;][]{LOFAR} as well as those radio facilities which span both low and mid frequencies such as the {{Upgraded Giant Metrewave Radio Telescope \citep[u-GMRT;][]{ugmrt}}} and {the upgraded} Karl G. Jansky Very Large Array \citep[VLA;][]{VLA}. These telescopes allow observations of radio populations at incredibly deep sensitivities, detecting a wealth of previously undetected radio sources, enabling more in depth studies of galaxy evolution, and studies to higher redshifts.}

{Within these deep extragalactic radio surveys, the sources are typically classifed into two populations: star forming galaxies (SFGs) and active galactic nuclei (AGN). {The radio emission from both of these populations} (at $\sim$1 GHz) is {{dominated by synchrotron radiation \citep{Condon1992}, though free-free emission may be important for SFGs and becomes more important at higher rest-frame frequencies \citep[see e.g.][]{Tabatabaei2017, Galvin2018}}}. In {the synchrotron} mechanism, radiation is emitted when electrons, that are moving at relativistic speeds, spiral in magnetic fields. For SFGs, the {relativistic electrons} are generated in supernova remnants, and so this radio emission acts as a proxy for star formation within a galaxy. This leads to relations as in the works of \cite{Bell2003, Garn2009, Jarvis2010, Davies2017, Delhaize2017, Gurkan2018,Delvecchio2021} and \cite{Smith2021}, which link radio luminosity to star formation rates (SFRs) and also to their infrared emission through the infrared radio correlation. For {AGN, the} {relativistic electrons spiral in the} jets associated with the {accreting supermassive black holes}. Historically, those AGN which exhibit jets are often further classified based on their morphology \citep{FanaroffRiley} and more recently AGN have been classified on their accretion mechanisms \citep[see e.g.][]{Best2012,Heckman2014,Whittam2018, Williams2018}. For faint surveys, with the telescopes described above, a substantial population of radio quiet AGN will also become important within the sources observed.}

One way in which we can investigate the contribution of different extragalactic radio populations to the radio source {{landscape}} is by looking at the distribution of radio sources as a function of flux density. This is typically {done} through investigating the source counts of radio sources \citep[see e.g.][]{Owen2008, Ibar2009, deZotti2010, Vernstrom2016, Mandal2021, Matthews2021, vandervlugt2021}. At high flux densities, the dominant radio source populations are powerful AGN \citep[see e.g.][]{Mauch2007,Padovani2016,Smolcic2017b} and this is {therefore} reflected in various simulated catalogues of radio sources \citep[][]{Wilman2008,Bonaldi2019}.  However, {with sensitive surveys such as those described in {\cite{Smolcic2017, Shimwell2019, Heywood2022, Tasse2021, Sabater2021}} and \cite{Norris2021} we are} able to detect significant numbers of the faint radio extragalactic populations. These include SFGs as well as the faint, radio quiet AGN populations {{\citep{Padovani2015, White2015, White2017}}}. The contribution of these sources {is responsible} for the flattening in the source counts distribution at $\lesssim$mJy flux densities at 1.4 GHz \citep[see e.g.][]{Jarvis2004, Smolcic2017b}. 

These faint source counts have been investigated using the new, sensitive surveys from LOFAR \citep{Mandal2021}, VLA \citep{Smolcic2017b, vandervlugt2021} {and} {{GMRT \citep{Ocran2020}}}. The recent source counts from MeerKAT DEEP2 observations \citep{Mauch2020, Matthews2021} covered 1.04 deg$^2$ and used both the source counts from {catalogues} as well as inferred sub-threshold source counts from probability of deflection, P(D), analysis \citep[][]{Matthews2021}. Previous deep sub-$\muup$Jy source counts {have been inferred} with both P(D) analysis \cite[see e.g.][]{Condon2012, Vernstrom2016} as well as using Bayesian stacking \citep[see e.g.][]{Zwart2015}. {These have} produced the best constraints on source counts at sub-$\muup$Jy levels to date. {These deep observations are typically restricted to small areas, whilst at low frequencies the LOFAR surveys have constructed source counts over relatively large areas \citep[$\sim$25 deg$^2$][]{Mandal2021} {{to $\sim$200 $\muup$Jy at 144 MHz ($\sim$40 $\muup$Jy at 1.4 GHz)}}. {For the deepest observations at GHz frequencies, the {surveyed} areas are {small, including the deepest source counts available from \cite{vandervlugt2021} and \cite{Algera2020} which covers 350 arcmin$^2$ and {so is} limited by sample variance \citep[e.g.][]{Heywood2013}}.}}

Knowledge of the source {counts} distribution at {{faint flux densities}} is also essential for understanding the integrated sky background temperature. {This provides the information necessary to model the contributions of faint extra-galactic} sources to the background emission at radio frequencies. The radio sky background is especially interesting to investigate at faint flux densities due to the large sky temperature excess found by the ARCADE 2 experiment \citep{Fixsen2011}. {{In their work, \cite{Fixsen2011} used radiometers to measure the sky temperature between 3-90 GHz at seven frequency values. This was combined with literature values \citep[such as][at 1.4 GHz]{Reich1986} to create a model for the {total} {sky background temperature} in the range 22 MHz - 10 GHz. However this work has been shown to be in disagreement with work from the catalogues of radio surveys.}} Whilst one explanation for this large sky temperature could have been an excess of faint (${\sim}\muup$Jy) radio sources, recent work by \cite{Vernstrom2011, Murphy2018,Hardcastle2021} and \cite{Matthews2021b} have indicated that it is not possible to explain the ARCADE 2 measurement using deep radio surveys.

One deep, relatively large area radio survey which also benefits from a vast wealth of ancillary multi-wavelength data is the MeerKAT International Giga Hertz Tiered Extragalactic Exploration (MIGHTEE) survey \citep{Jarvis2016, Heywood2022}. When completed, these observations will cover a total {{area}} of 20 deg$^2$, covering four extragalactic fields (COSMOS, E-CDFS, ELAIS-S1 and XMM-LSS). This should allow a range of different environments (e.g. clusters, voids etc.) to be observed and investigated, mitigating the effect of sample variance. {{The continuum Early Science data release of the MIGHTEE survey \citep{Heywood2022} covers {a fraction of} two of the four fields: COSMOS and XMM-LSS. This release consists of both a lower ($\sim$8\arcsec) and higher ($\sim$5\arcsec) resolution image}}. In total these {observations} cover $\sim5$ deg$^2$ to a typical thermal noise of $\sim$2 $\muup$Jy beam$^{-1}$ in the lower resolution image and $\sim$6 $\muup$Jy beam$^{-1}$ in the higher resolution image.

Importantly, MIGHTEE's survey strategy targets those fields with some of the best multi-wavelength ancillary data. This spans the vast ranges of the electromagnetic spectrum, and a non-exhaustive list of these observations include those from the X-ray \citep[see e.g.][]{Hasinger2007, Chen2018, Ni2021}, optical \citep[see e.g.][]{Davies2018, Aihara2018, Davies2021}, near-IR \citep[see e.g.][]{McCracken2012, Jarvis2013, Laigle2016}, {{mid-IR \citep[see e.g.][]{Lonsdale2003, Mauduit2012},}} far-IR \citep[see e.g.][]{Oliver2012, Ashby2013} and {{radio \citep[see e.g.][]{Bondi2003, Tasse2007, Smolcic2017, Hale2019, Heywood2020}}} {wavelengths}. This produces a wealth of information to help characterise source types (e.g. SFG or AGN) and also the properties of the host galaxies (e.g. star formation rate, SFR, and stellar mass, $M_{*}$) through methods such as spectral energy distribution (SED) fitting.

In this paper we investigate the deep source counts distribution from the {{continuum Early Science}} data release of the MIGHTEE survey in the COSMOS and XMM-LSS fields. We then make use of the {classifications which use the} large amounts of ancillary data within the MIGHTEE fields to consider the contribution to the integrated background sky temperature from AGN and SFGs separately. Using radio observations at these depths and investigating the sky temperature contribution from AGN and SFG respectively is something which benefits from surveys such as MIGHTEE where depth, area, and multi-wavelength information {are all} combined. 

The layout of this paper is as follows{{:}} in Section \ref{sec:data} we describe the data used for this analysis before we then outline the methods used for {calculating the incompleteness of the source counts in Section \ref{sec:methods}. Using the measurements of source count completeness we determine the corrected source counts which we present in Section \ref{sec:results} before using these corrected source counts to determine the integrated sky background temperature contribution of AGN and SFGs. We then discuss these results in Section \ref{sec:discussion}, before drawing conclusions in Section \ref{sec:conclusions}. }

\section{Data}
\label{sec:data}
In this section we give a brief overview of the continuum data from the MIGHTEE continuum {{Early Science}} data release {\citep{Heywood2022}} that {are} used in this paper. {Furthermore, we also use the} {{catalogues generated from}} {{cross-matching}} {(Prescott et al. subm.) and {{further classified by their source type \citep{Whittam2022}}, which are used to}} investigate the contribution of AGN and SFGs. Further information and details on the MIGHTEE {{Early Science continuum}} data release can be found in \cite{Heywood2022}, where {information} on data access can also be found.

\subsection{MIGHTEE Continuum Data}
The images used for this work are taken from the {{Early Science}} data {{release in}} the MIGHTEE survey, which cover the COSMOS {{($\sim$1.6 deg$^2$)}} and XMM-LSS fields {{(${\sim}3.5$ deg$^2$)}}. For the COSMOS field, a total of 17.45 hours of observations (on target) were {taken} over a single field of view centered at RA: {10$^{\textrm{h}}$00$^{\textrm{m}}$28.6$^{\textrm{s}}$, Dec: +02\degree12\arcmin21\arcsec}. Three observations of the field were taken in April 2018, May 2018 and April 2020 respectively. For {{XMM-LSS, 3}} pointings were used to construct the mosaicked image of the field, with individual field centres of {(02$^{\textrm{h}}$17$^{\textrm{m}}$51$^{\textrm{s}}$, $-$04\degree49\arcmin59\arcsec), (02$^{\textrm{h}}$20$^{\textrm{m}}$42$^{\textrm{s}}$, $-$04\degree49\arcmin59\arcsec) and (02$^{\textrm{h}}$23$^{\textrm{m}}$22$^{\textrm{s}}$, $-$04\degree49\arcmin59\arcsec)}. Each pointing was observed twice during October 2018 with $\sim$12.4 hours {on each field centre}.

Data reduction is described {comprehensively} in \cite{Heywood2022} and used a combination of both direction-independent and direction-dependent calibration. \texttt{CASA} \citep{CASA} was used to determine gain solutions from the primary and secondary calibrators and {these} were applied to the target data which {were} subsequently flagged using \texttt{TRICOLOUR}\footnote{\url{https://github.com/ska-sa/tricolour}}. Direction-independent imaging and {self-}calibration of the target data set was performed using a combination of \texttt{WSCLEAN} \citep{Offringa2014} and the \texttt{CASA GAINCAL} task. Direction-dependent calibration was then calculated and {the fields were then imaged using} a combination of \texttt{KILLMS} \citep{Smirnov2015} and \texttt{DDFACET} \citep{Tasse2018}. 

Final images were constructed using two Briggs' weighting values \citep{Briggs1995}: 0.0 and $-$1.2. The first Briggs' weighting of 0.0 was optimised to improve the sensitivity of the image ({{thermal noise}} $\sim$2 $\muup$Jy beam$^{-1}$, {though observed noise} {in the central regions} {is $\sim$4-5 $\muup$Jy beam$^{-1}$ due to confusion}), however this compromised the resolution and led to 8.6\arcsec \ (8.2\arcsec) resolution for COSMOS (XMM-LSS) field. A second Briggs' weighting of $-$1.2 instead prioritized resolution over depth of the image and resulted in images with 5.0\arcsec \ resolution but with poorer sensitivity ({{thermal noise}} $\sim$6 $\muup$Jy beam$^{-1}$). {For the work in this paper{{,}} we only make use of the low resolution images{{,}} to probe the source counts and sky background  temperature {{to faintest flux densities possible.}} {{However, this does mean our images are more {likely to be} affected by confusion.}}}

{ Source catalogues were {generated} {by running} the Python Blob Detector and Source Finder \citep[\texttt{PyBDSF;}][]{PyBDSF} using the default source extraction parameters. \texttt{PyBDSF} produces both a source catalogue (\texttt{srl}) file as well as a list of the Gaussian components (\texttt{gaul}) that are used to model the radio emission above 3$\sigma$ of the {{local}} sky background. The respective advantages of these two catalogues will be described further in Section \ref{sec:src_comp}. Considering the Gaussian component catalogues only, there are a total of 9,915 components in the COSMOS low resolution image {and these were combined into 9,252 sources}. In the XMM-LSS low resolution image there are 20,397 components detected and 19,290 sources. Subsequent visual inspection of these images and catalogue led to a removal of a handful of spurious sources, as described in \cite{Heywood2022}. }

\begin{figure*}
    \centering
    \begin{minipage}[b]{0.39\linewidth}
    \includegraphics[width=0.92\textwidth]{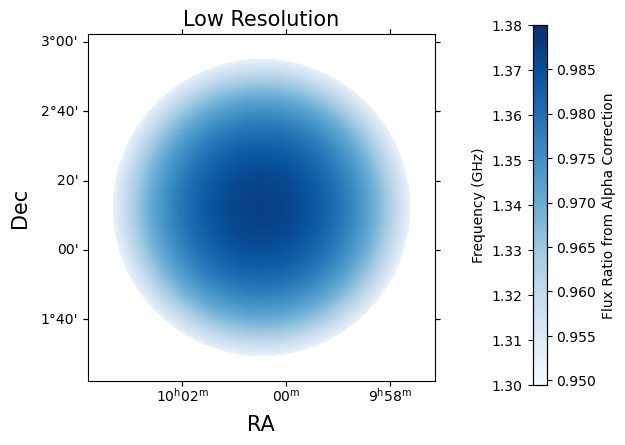}
    \subcaption{COSMOS}
    \end{minipage}%
    \begin{minipage}[b]{0.61\linewidth}
    \includegraphics[width=0.92\textwidth]{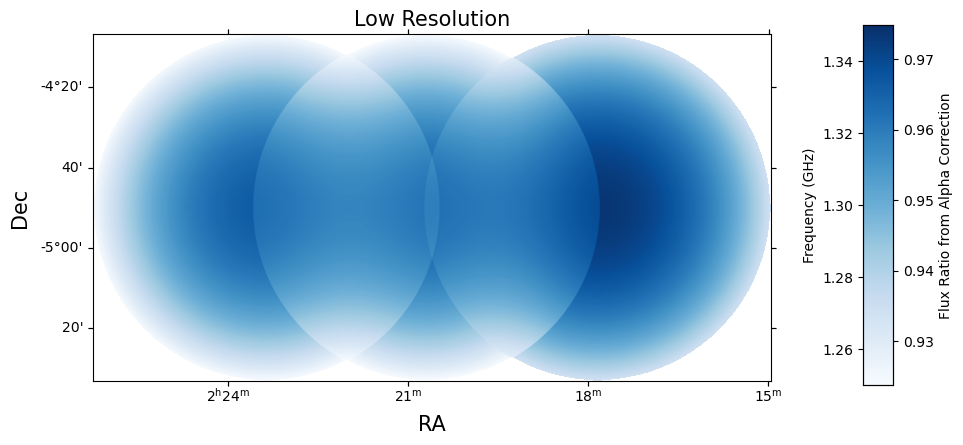}
    \subcaption{XMM-LSS}
    \end{minipage}%
    \caption{Effective frequency map of the COSMOS field (left) and XMM-LSS field (right) {{released in \protect \cite{Heywood2022}}}. The colour bar indicates the effective frequency as well as the correction factor needed to convert flux densities at the given frequency to 1.4 GHz assuming a synchrotron power law.}
    \label{fig:eff_freq}
\end{figure*}

\subsection{Effective Frequency Map}
For each image, an effective frequency map was also constructed in \cite{Heywood2022}. This reflects the changing nature of the effective frequency at each location within the image {{due to}} the response of the primary beam of MeerKAT being both a function of position within a pointing as well as frequency. {Observations were taken across a} {wide frequency band, $\sim$900-1600 MHz, and} {factors such as the flagging of the raw data, the varying response of the primary beam with frequency and the mosaicing of data means the effective frequency is not a constant value across the image.}

The effective frequency maps {{that were created and released with \cite{Heywood2022}}} for the low resolution images in the COSMOS and XMM-LSS fields can be seen in Figure \ref{fig:eff_freq}. Figure \ref{fig:eff_freq}a shows that the effective frequency for the COSMOS field is higher towards the centre of the field {{($\sim$1.4 GHz in the low resolution image)}}, decreasing to lower frequencies at greater distance from the pointing centre. For XMM-LSS (Figure \ref{fig:eff_freq}b), the distribution in effective frequency is more complicated, due to the mosaicking of three pointings that were used to construct the full field. As such there {are higher} values for the effective frequency towards the centre of the east and west-most {{pointings. The}} overlap between the central pointing and the east and west pointings, however, shows slightly lower effective frequencies.

For our work, we {{scale our source counts to a common frequency of 1.4 GHz,}} assuming a synchrotron power law spectrum\footnote{{{$S_{\nu} \propto \nu^{-\alpha}$; where $S_{\nu}$ is the integrated flux density at a given frequency, $\nu$, {and} the spectral index is denoted by $\alpha$. We assume $\alpha=0.7$ throughout this paper unless otherwise stated. A spectral index of 0.7$-$0.8 has been commonly measured, see e.g. \protect \cite{Smolcic2017, CalistroRivera2017, deGasperin2018, An2021} and are commonly assumed values in the literature.}}}. The colour bars {{in}} Figure \ref{fig:eff_freq} therefore not only show the change in frequency, but also the value of the correction for the flux density of sources at each position within the map to ensure a common frequency of 1.4 GHz. Depending on {the} location within these {images, this} correction factor is in the range ${\sim}0.9{-}1.0$. {The effective frequency maps from \cite{Heywood2022} do not have associated errors with the maps and we do not have spectral indices and associated errors for each individual source. Therefore there are likely very small uncertainties on these correction factors. However, given the small frequency corrections, small changes in the spectral index should not contribute significantly to the errors in the source counts presented in Section~\ref{sec:scounts_results}.}

\subsection{AGN and SFG classification of MIGHTEE sources}
\label{sec:sfg_agn_classification}
{The classification of radio sources into AGN and SFGs within the MIGHTEE continuum early science data release is the result of} combined efforts to identify host {{galaxies}} for the objects detected by \texttt{PyBDSF} (described in {Prescott et al., subm.}) and a process of using multiple multi-wavelength diagnostics to {{separate}} AGN from SFGs {\citep[described in][]{Whittam2022}}. {{This identification of host galaxies and classification into AGN and SFGs uses a subset of the MIGHTEE {{Early Science}} continuum data, over $0.8$ deg$^2$ of the COSMOS field.}}

{In Prescott et al., (subm.)}, components within this central region of COSMOS were {{cross-matched}} to {probable} host sources from a compilation of {catalogues which combine optical and near-IR data from a multitude of wavelengths and telescopes{,} such as the Canada-France-Hawaii Telescope Legacy Survey (CFHTLS), Hyper Suprime Cam (HSC), Visible and Infrared Survey Telescope for Astronomy (VISTA) and {the} \textit{Spitzer} space telescope \citep[for more information on these compilation catalogues see][]{Bowler2020,Adams2020,Adams2021}}. An updated version of the \texttt{XMATCHIT} code \citep[see ][]{Prescott2018} was used for visual {{host galaxy}} identification, using composite images for each source that combined UltraVISTA \citep{McCracken2012} $K_S$-band images with radio contours from MIGHTEE and from the VLA 3 GHz COSMOS survey \citep{Smolcic2017} overlaid on the image. These images were visually inspected by members of the MIGHTEE team, providing host galaxy identification for $\sim$83\% of {\texttt{PyBDSF} Gaussian components, including those that were in regions masked by the multi-wavelength data. {The remaining components either did not have counterparts assigned or were too confused to assign a host}. This process also identified those Gaussian components which needed to be combined into a single source, {as well as identifying those components which appeared to be from multiple individual host sources}. Exact details of the number of components which are classified as multi-component sources, have no counterpart or are confused, can be found in {Prescott et al. (subm.)}.}

Source classifications into AGN or SFGs were subsequently made using the wealth of multi-wavelength data and the knowledge of the host from the cross-matched catalogue {as} described in {\cite{Whittam2022}}. The combined multiple diagnostics are summarised here. Firstly, diagnostics from X-ray emission were used to identify AGN, with $L_{X}{>}10^{42}$ {{erg\,s$^{-1}$}}. Secondly, excess radio emission was identified using the infrared-radio correlation from \cite{Delvecchio2021} where sources with radio {emission ${>}2{\sigma}$ above} the correlation {were} defined to be AGN. Moreover, AGN were identified from their mid-infrared colours using the colour cut described in \cite{Donley2012}. Finally, sources that are found to be point-like at optical wavelengths (using Hubble ACS I-band data) were described to be optical AGN. The remaining sources were assumed to be SFGs {if they failed all of these four criteria and probable SFGs if they had $z{>}0.5$ but satisfied all the non-AGN criteria {\citep[due to X-ray observation limitations, see][]{Whittam2022}}}. {For the sources which were cross matched to a host galaxy {in Prescott et al. (subm.)} $\sim$88\% of sources are associated as either an AGN, SFG or probable SFG. This represents ${\sim}73\%$ of the total sources, including sources within masked regions. It is with these classifications that we will investigate the respective contribution of SFG and AGN to the background sky temperature. }

{{As mentioned, this only uses the classifications across the $\sim$0.8 deg$^2$ central area of the COSMOS field. Therefore any assumptions on the fraction of AGN/SFGs for the larger COSMOS region or for the XMM-LSS field are made assuming the ratio from the $\sim$0.8 deg$^2$ COSMOS region.}}

\section{{{Calculation of Source Counts and Incompleteness}}}
\label{sec:methods} 
In this section we discuss the methods to determine the source counts for the catalogue of radio sources and to subsequently calculate the background sky temperature for these data. We also discuss our methods to calculate the incompleteness within these images and to correct for this to understand the intrinsic source count distribution.

\subsection{Calculation of Source Counts}
\label{sec:methods_calc}
Source counts quantify the number of sources ($N$) within a flux density ($S_{\nu}$) bin (i.e. $\frac{dN}{dS_{\nu}}$) per unit steradian observed on the sky (combined to give $n(S_{\nu})$). Typically, the counts are Euclidean normalised and so the Euclidean normalised source counts are denoted by $n(S_{\nu})S_{\nu}^{2.5}$. {{We first calculate the raw source counts using the \texttt{PyBDSF} catalogues of \cite{Heywood2022} corrected to a frequency of 1.4 GHz using the effective frequency map. However,}} these {{observed raw source counts will decrease at faint flux densities due to incompleteness from varying sensitivity across the image}}. {Therefore in order to calculate the intrinsic source counts distribution we must first determine the appropriate completeness corrections to account for underestimations in the raw source counts.}

\subsection{Source vs. Component catalogues}
\label{sec:src_comp}
As described in Section \ref{sec:data}, \texttt{PyBDSF} produces both a source and component catalogue. The component catalogue describes the property of each Gaussian component used to model emission within the image, whilst the source catalogue describes the properties of sources where Gaussian components, which are believed by the algorithm to be associated with the same source, have been combined together\footnote{{{See}} \url{https://www.astron.nl/citt/pybdsf/algorithms.html\#grouping-of-gaussians-into-sources} for further details}. Both of these catalogues have advantages in different regimes and {the decision on which catalogue is appropriate to use will also be dependent on the science {goals}. For images that are close to confusion and where real radio sources may appear close together on the sky, it may be more appropriate to use the Gaussian component catalogue, at the faintest flux densities, to avoid combining different true extragalactic radio {{sources}} into a single source. However, using a Gaussian component catalogue will mean that resolved jetted AGN or nearby SFGs may be split into many Gaussian components, which typically affects brighter flux densities.  }

\begin{figure*}
    \centering
    \includegraphics[width=0.65\textwidth]{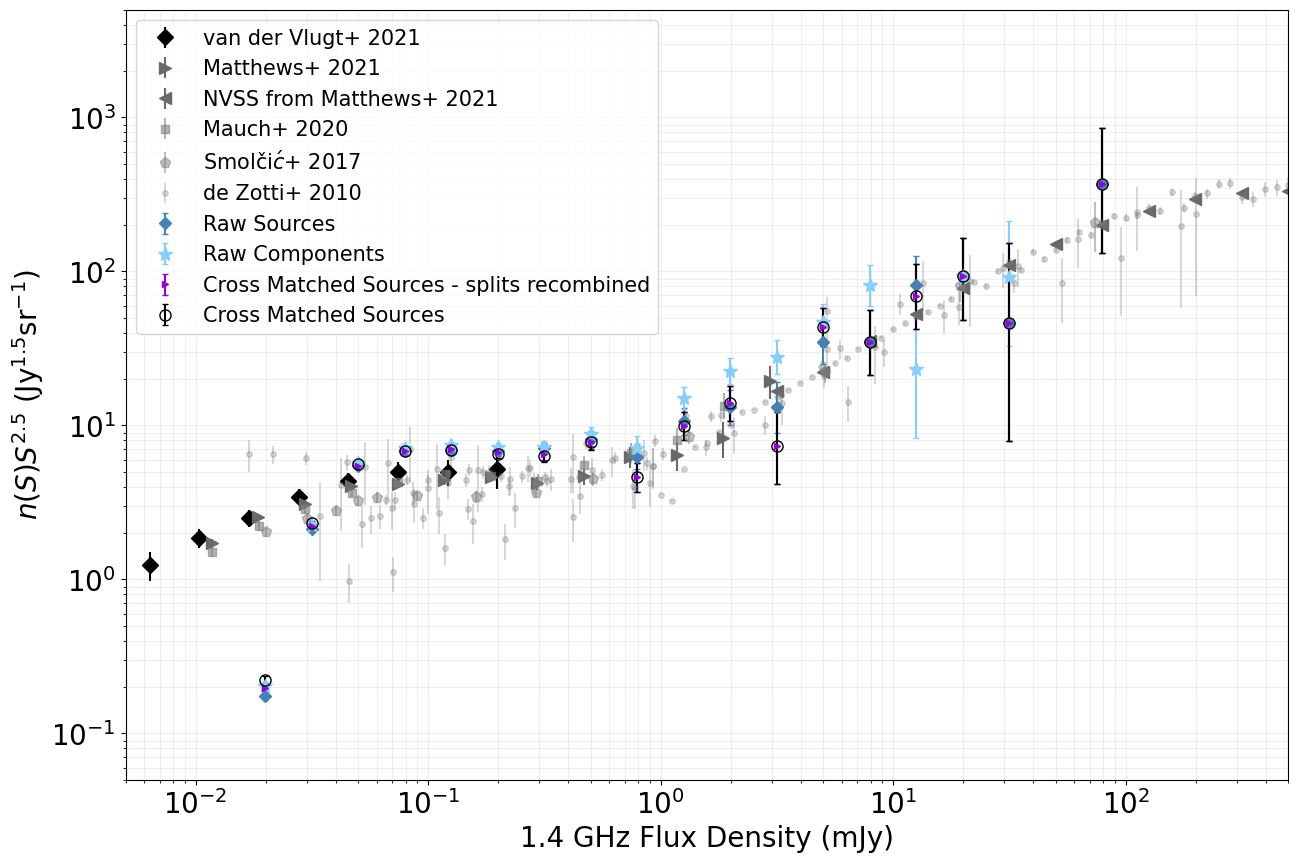}
    \caption{{Euclidean normalised source counts at 1.4 GHz using the raw \texttt{PyBDSF} components (light blue stars) and source (steel blue diamonds) catalogues over the central $\sim0.8$ deg$^2$ of the COSMOS field, compared to the {{cross-matched}} catalogue (black open circles triangles) and the {{cross-matched}} catalogue where any {{components which were split based on their 3 GHz flux densities}} have been recombined (purple triangles). {Also plotted in grey to highlight the results from previous observational data are also shown for 1.4 GHz source counts from \protect\cite{deZotti2010} (dots), \protect\cite{Smolcic2017b} (pentagons), \protect\cite{Mauch2020} (squares), \protect\cite{Matthews2021} (triangles) {and \protect\cite{vandervlugt2021} (diamonds). For data at other frequencies, these are scaled to 1.4 GHz assuming $\alpha=0.7$.}} }}
    \label{fig:sc_match}
\end{figure*}

To investigate what is the best catalogue to use for our specific science goal, we consider which source counts appear most appropriate for the data using knowledge of the source counts from the cross-matched catalogue{. We} show, in Figure \ref{fig:sc_match}, the difference between the {raw source counts (i.e. not corrected for incompleteness)} using the \texttt{PyBDSF} source and Gaussian {component} catalogues over the $\sim$0.8 deg$^2$ cross-matched area and compare this to the {source counts of the} cross-matched catalogue of {Prescott et al. (subm.)}. 

Figure~\ref{fig:sc_match} shows the effect of combining associated components using the 0.8 deg$^2$ {COSMOS} cross-matched region. Above 1 mJy, these source counts differ significantly from the counts from the \texttt{PyBDSF} {Gaussian} component (\texttt{gaul}) catalogue, and are more similar to the counts from the \texttt{PyBDSF} source (\texttt{srl}) catalogue. This relates to large, bright, multi-component AGN within the field such as those with Fanaroff Riley Type I and II morphologies \citep{FanaroffRiley}. At fainter flux densities (${\sim}50 \ \muup$Jy{-}1 mJy), there is less variation between the cross-matched catalogue source counts and those from the raw source and component catalogues. {Below ${\sim}50 \ \muup$Jy, again there {{are}} discrepancies between the cross-matched catalogue source counts and those from the raw source and component catalogues, but in this flux density range this is a consequence of splitting {objects detected as single sources in \texttt{PyBDSF} which are in-fact multiple sources which are confused}. This can be seen by the source count distribution where any split sources have been recombined. {Below ${\sim}50 \ \muup$Jy, the cross-matched source counts seem to slightly better reflect those of the component catalogue. This is probably because the fainter population of sources are more often single component objects, {and therefore} the source counts based on the {\texttt{PyBSDF}} source catalogue are instead underestimated compared to the cross-matched catalogue. This would be due to sources being incorrectly combined with other nearby sources into multi-component objects and is expected due to the effect of confusion within the low resolution MIGHTEE images. }}

 As can be seen in Figure \ref{fig:sc_match}, the source catalogue from \texttt{PyBDSF} provides {more comparable agreement} to the source counts from the cross-matched catalogue {over a wide range of flux densities}, {compared to those from the Gaussian catalogue}. {As such, we proceed with this work by making use of the raw data source catalogues to calculate the source counts and calculate the source counts completeness corrections using the simulated and recovered source catalogues from our simulated images}. This should help provide an understanding of the source counts distribution across a large flux density range of $\sim$0.01-100 mJy. {For bright sources, which are rare and are less well sampled in the area of the MIGHTEE Early Science data, these {are better constrained, across a range of frequencies,} from the catalogues of larger area sky surveys \citep[such as][]{Condon1998, Shimwell2019, Hale2021}.}

\subsection{Simulations to Determine Incompleteness}
\label{sec:compsims}
In order to understand the intrinsic source counts distribution, {{we use simulations to quantify the incompleteness in these source counts, which we then correct for}}. For {{these simulations}}, we use realistic mock radio catalogues which reflect the radio sky to investigate the detection of sources across the image. {{These simulations allow us to consider the combined incompleteness seen due to the effects of source finder incompleteness, resolution bias and sensitivity variations across the image. We use three different radio sky models {from simulations} in order to investigate the completeness, which shall be discussed separately in Sections \ref{sec:skads}, \ref{sec:skadsmod} and \ref{sec:simba}. For each of these different input source models, we}} {{follow the approach of many previous works \citep[see e.g.][]{Williams2018,Hale2019,Hale2021,Williams2021,Shimwell2022} and inject simulated sources into images of the corresponding field and determine how successful source detection is}}.

{{For our work, it is important to understand both which image we should inject our simulated sources into, as well as how many sources to inject into the given image. Due to the confusion within the image, it is challenging to inject a large number of sources into the image itself. Alternatively, sources can be injected into the residual image, which is the observed image with the modelled Gaussian components subtracted. In the residual image, a much larger number of simulated sources can be injected into the image. However, as discussed above, the MIGHTEE images suffer from confusion so there will still be a large number of faint sources in the residual image that were previously unable to be detected above 5$\sigma$. Due to confusion, the rms (root mean square, or noise) will be affected by the {sources (both number and flux density) within the image. With no bright sources in the residual image, the intrinsic rms of this residual image will likely be lower than the rms calculated for the original image; this will therefore affect the measured completeness as a function of flux density.} Similarly, if too many simulated sources are injected, the rms may be much larger than measured for the original image. This choice of which image to inject sources into and how many simulated sources to inject will be dependent on the simulation used. We therefore discuss these details further for each simulation respectively in Sections \ref{sec:skads}-\ref{sec:simba}}}.

\subsubsection{SKADS}
\label{sec:skads}
{Firstly, we created simulated sources across the image using the radio sources from the Square Kilometre Array Design Study simulations} \citep[SKADS;][]{Wilman2008, Wilman2010}. To do this, we take the SKADS components catalogue covering 100 deg$^2$ of simulated sky to a minimum source flux density of 5 $\muup$Jy at 1.4 GHz. Each source is constructed using components, which have an individual flux density, a simulated size and a simulated position. For some sources, such as SFGs, these can be constructed using single SKADS components. For other sources, such as Fanaroff-Riley type AGN \citep{FanaroffRiley}, these consist of multiple components to {represent} the core and lobes of the source. The input source counts distribution for the SKADS simulation can be seen in Figure \ref{fig:scmodel}{{, and appears to}} underestimate the source counts at faint flux densities ($S_{1.4 \ \textrm{GHz}} \lesssim$0.1 mJy) compared to recent measured source counts distributions {\citep[see e.g.][]{Smolcic2017, Prandoni2018, Mauch2020, Matthews2021}}. Therefore we will also consider a modified version of this input distribution, which is discussed in Section \ref{sec:skadsmod}.

In order to construct simulated images from which we can estimate the completeness, we choose {{locations randomly distributed over the sky within the field of view to inject simulated sources}}. A {simulated source is then generated in the following manner, following the method of e.g. \citep{Hale2021}}. After randomly selecting a source from the SKADS catalogue, {each SKADS component is modelled} as an elliptical disk {or a point source depending on the source size}. Each component is then convolved with a 2D Gaussian kernel which has the same FWHM as the restoring beam of the radio image and scaled to retain the integrated flux density of the component {{(scaled to the effective frequency at the position of the source)}}. Each component for a given source is combined together to make a model for the entire source. {This model is then injected into the image at the random location for the source. }

{{As we want to understand the completeness within the image, we choose to inject a small number of simulated sources into the image} {itself. For each} {simulation we inject 1,000 sources for the COSMOS field and 2,000 sources in the larger XMM-LSS. We repeat these simulations {1000} times on each image in order to build up better statistics of the completeness.}}

\subsubsection{Modified SKADS Source Model}
\label{sec:skadsmod}

As {{described in Section \ref{sec:skads},}} there is growing evidence that the SKADS model underestimates the observed source counts at faint flux densities ($S_{1.4 \ \textrm{GHz}} \lesssim 0.1$mJy). To ensure that underestimations of the source counts model from SKADS is not affecting our derived completeness, we also use a modified version of the SKADS catalogue in which the SFG sample within the SKADS catalogue have been doubled. {{This difference in source population may affect the measured completeness. For example, if the additional SFGs have a different source size distribution to the AGN at these flux densities, this then could affect the impact of resolution bias on completeness.}} Doubling this {{population creates a raw}} source count distribution which {is in much better agreement with recent observations of source counts at the faintest fluxes}. We use this new input catalogue in the same way as described in Section \ref{sec:skads} to produce {{1000} simulations again with the same number of injected sources.}

\subsubsection{\textsc{SIMBA} Light Cone}
\label{sec:simba}
Next, we consider the completeness using simulations which account for realistic clustering within the field of view using a 1 deg$^2$ simulated light cone from \textsc{SIMBA}~\citep[see e.g.][]{Simba, Lovell2021}. {\textsc{SIMBA} is a state-of-the-art suite of cosmological hydrodynamic simulations resolving galaxies down to a stellar mass of $M_{\star} = 5.8 \times 10^{8}$ M$_{\odot}$ within a (100 $h^{-1}$Mpc)$^{3}$ box assuming a \citet{Planck2016} concordant cosmology with {{$\Omega_{\textrm{M}}=0.3$}}, $\Omega_{\Lambda}=0.7$, {{$\Omega_{\textrm{b}}= 0.048$}}, $H_{0}=68\ \rm{km \ s^{-1} Mpc^{-1}}$, $\sigma_{8}=0.82$, and $n_{s}=0.97 $. \textsc{\textsc{SIMBA}} is unique in that it models the growth of supermassive black holes via a two mode sub-resolution prescription, namely, Bondi accretion from hot gas and gravitational torque limited accretion from cold gas \citep[see][]{Angles2017}. In addition, \textsc{SIMBA} models the feedback from supermassive black holes motivated by observations~\citep{Heckman2014} including kinetic feedback in the form of bipolar jets. The model employed by \textsc{SIMBA} has shown good agreement with observations of galaxy properties~\citep[e.g.][]{Simba} as well as black hole--galaxy correlations and co-evolution~\citep{Thomas2019}, and reproduces a viable population of radio galaxies~\citep{Thomas2021}. {{Radio luminosities at 1.4 GHz}} for \textsc{SIMBA} galaxies are computed from star formation as well as ongoing jet feedback using the scaling relations detailed in \citet{Thomas2021}.}

Using a realistic light cone is important as, at the sensitivity and resolution of MIGHTEE's lower resolution ($\sim$8\arcsec) images, we are reaching the confusion limit within the survey. As such, the source counts may be affected by confused sources not being correctly identified as separate {sources.} {{Whilst the original SKADS catalogue has large-scale clustering included, \textsc{SIMBA} will more accurately represent both the `1-halo' clustering (within the same dark matter halo) and `2-halo' clustering \citep[within different dark matter haloes, see e.g.][]{Cooray2002, Zehavi2004} as it is based on cosmological simulations.}} {\textsc{SIMBA}} realistically distributes galaxies within a light cone over the redshift range $0<z<6$ and projected over 1 deg$^2$ {{of sky area. We use this to understand the effects of source clustering and large scale structure. Clustering {may} affect source counts measurements both due to the effects of confusion and sample variance (see Section \ref{sec:calc_sc})}.} {This light cone is created by combining together snapshot images of the simulation at different times.}

In order to use this light cone to investigate the effect of {{clustering on completeness, we compare two approaches. In the first, we use the positional information and the flux densities}} of the sources within the light cone simulation\footnote{We convert this from a {{flux density at 1.4 GHz to the effective frequency at the source location}} using an assumed spectral index of $\alpha$=0.7; this is again done to reflect the typical frequency for the image.} We then model each source within the simulation as a point {source using a 2D Gaussian model with the properties of the restoring beam} and inject the source into the {{residual image. The residual image needs to be used in this simulation due to the number of \textsc{SIMBA} sources to be injected. The MIGHTEE image is already close to confusion and so it would not be useful to directly inject these into the image. Injection into the residual image should instead produce an overall source density broadly comparable to that of the data.}} As the simulation only has a 1 deg$^2$ field of view, it will not cover the field in its entirety. Therefore for each realisation we randomly {generate} a central position for the light cone within the field of view and also randomly rotate the simulation within the image. {{For the second approach, we use the same method but instead of using the positions from \textsc{SIMBA}, we use random positions within 1 deg$^2$ of the image. By comparing the completeness using the two approaches we can determine whether the intrinsic clustering plays an important role in affecting the completeness of sources {within the field for this work.} For each of the two \textsc{SIMBA} simulations we create 100 realisations. This is fewer simulations than in Section \ref{sec:skads} and \ref{sec:skadsmod}, however as more sources are injected into the residual image, we maintain good statistics.}} 

In this simulation we make the assumption that we can model each of the simulated \textsc{SIMBA} sources as a point source. {In reality, some of the more extended} sources would be resolved in the MIGHTEE images. {However, as we are primarily using these simulations to make a direct comparison of the completeness with and without clustering, the assumption that the \textsc{SIMBA} sources are unresolved will not be likely to affect the results significantly.} The effect of resolution bias will instead be accounted for in the SKADS simulations. {We also note that the \textsc{SIMBA} simulations may have small {box} edge effects {when generating the light cone} as discussed in \cite{Blaizot2005} (Section 3.2.1) and \cite{Merson2013} (Section 4.1), {but for the small-scale clustering which may be important for completeness, these effects should have a negligible effect.} This simulation also only represents one realisation and so may be affected by sample variance however; we discuss including sample variance in our errors in Section \ref{sec:calc_sc}. } \\ 

\begin{figure}
    \centering
    \includegraphics[width=0.45\textwidth]{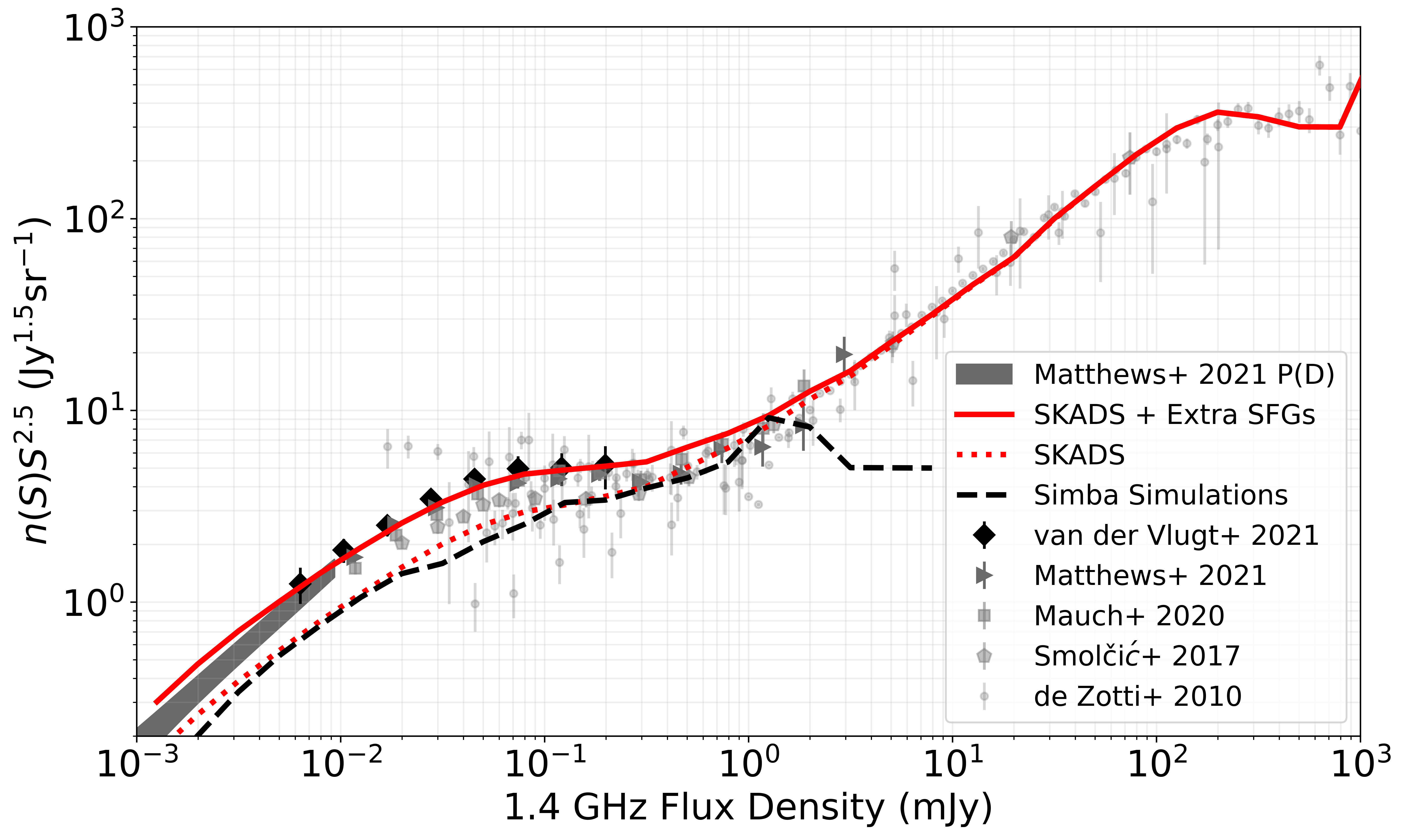}
   \caption{{Euclidean normalised 1.4 GHz source counts models used in this work and compared with previous data and simulations. Simulations shown are from SKADS \protect\citep[red dotted line;][]{Wilman2008}; the modified SKADS model described in \ref{sec:skadsmod} (red solid line); \textsc{SIMBA} (black dashed line). This is compared with P(D) analysis from \protect\cite{Matthews2021} (grey shaded region) and previous observational data from \protect\cite{deZotti2010} (dots), \protect\cite{Smolcic2017b} (pentagons), \protect\cite{Mauch2020} (squares), \protect\cite{Matthews2021} (triangles) {and \protect\cite{vandervlugt2021} (diamonds). For data at other frequencies, these are scaled to 1.4 GHz assuming $\alpha=0.7$.}}}
    \label{fig:scmodel}
\end{figure}
\FloatBarrier

\subsubsection{{Summary of source count models}}

 \noindent All the input source count models used in this work are shown in Figure \ref{fig:scmodel}. As can be seen, the modified SKADS distribution {{appears to more accurately reflect the observed deep source counts compared to both the {original SKADS model and} the \textsc{SIMBA} simulations.}} For \textsc{SIMBA}, as the distribution of galaxies is related to cosmological simulations, this discrepancy could {relate to the calibration chosen between the galaxies observed}, their mass and SFR to the radio flux observations. At bright flux densities {{($>$1 mJy)}}, SKADS models the distribution of source counts well{, however \textsc{SIMBA} cannot constrain the bright source counts due to the small volume size.}

{Once a simulated image was created (using the different models described) \texttt{PyBDSF} was run over the image using the same parameters as used in \cite{Heywood2022}. By using the output catalogues from \texttt{PyBDSF} and comparing this to the input catalogue, it is possible to determine the effects of incompleteness across the field due to the combined effects of rms variations as well as source finder detection issues. Furthermore, this strategy of using simulated sources, including those injected below the nominal 5$\sigma$ detection threshold{,} also allows the effect of Eddington bias \citep{Eddington1913} to be considered. {However they do not account for variations in the source size models.}} {{For each of the simulation methods described in Sections \ref{sec:skads}-\ref{sec:simba}, we repeated the method and generated multiple realisations {to calculate the variation in completeness}, see Section \ref{sec:calccomp}.}}

\subsubsection{High Flux Density Simulations}
{{At the very highest flux densities, the simulations described in Sections \ref{sec:skads}-{\ref{sec:simba}} are limited because the source populations are dominated by faint sources and so fewer sources are injected at bright flux densities. Therefore for the simulations in Sections \ref{sec:skads} and \ref{sec:skadsmod} we conduct additional simulations where we only inject brighter sources ($\geq0.1$ mJy) into our images. }} {{For each high flux density simulation we inject 500 sources in the COSMOS field, and 1,000 sources in the XMM-LSS image and run {1000} realisations}}. {We do not generate the same high flux density simulations for the SIMBA simulations (described in Section \ref{sec:simba}) as these are used to understand the effects of clustering and confusion which primarily affects the faint populations. }

\subsection{{Calculating Source Counts Corrections}}
\label{sec:calccomp}

\subsubsection{Matching Input and Output Catalogues}
\label{sec:calccomp1}
In order to determine {how incomplete\footnote{{Whilst completeness is typically defined as the fraction of sources with an intrinsic given flux density that are detected in the image irrespective of measured flux density, here we define a total source counts completeness correction factor. We define our source counts completeness to be the fraction of sources detected within a flux density bin compared to the number of simulated sources injected within the same flux density bin. This therefore calculates a correction applicable to the source counts as a function of flux density which incorporates both traditionally defined completeness as well as the biases in measuring flux densities {due to} the source finder, {the impact of noise on flux density measurements} and due to confusion.}} our source counts are} we want {{to}} ensure that the sources detected by \texttt{PyBDSF} for each simulation are those same simulated sources injected within the field, and not any {{existing emission within the image prior to adding in the simulated sources. Therefore we compare the output detected source catalogues to those originally within the image (MIGHTEE image or residual image depending on the simulation) before calculating the completeness. We shall call this catalogue the pre-simulation catalogue. For the simulations of Section \ref{sec:skads} and \ref{sec:skadsmod}, where we inject sources into the image itself, these sources would be the catalogue of \cite{Heywood2022}. For Section \ref{sec:simba} where, instead, we inject the simulated sources into the residual image. {Whilst it might be expected that there are no sources in the residual image, with the} $>$5$\sigma$ sources removed from the image, the background emission and rms values within the residual image is lower and therefore some objects now exceed the 5$\sigma$ threshold of \textsc{PyBDSF}. {Whilst some of these new detections may be genuine faint sources, there will also be a contribution of noise artefacts detected}. Therefore, we also run \texttt{PyBDSF} using the same detection parameters as in \cite{Heywood2022} over the residual image to produce a {pre-simulation} source catalogue.}} 

To determine the {source counts incompleteness} for each simulation, we first match both the input simulated catalogues and the output detected catalogues to the \texttt{PyBDSF} {pre-simulation} catalogue as well as matching the output catalogue from the simulated image to both the input simulated source and component catalogues. We remove any sources within either the input or output catalogue that are matched to {the} {pre-simulation} catalogue within a given angular separation. {This angular separation will be discussed further in Section \ref{sec:results_sep} and is chosen to ensure that not only are detected sources correctly associated to an input source, but also that any simulated sources that are associated through the {{cross-matching}} process are not affected by difficulties in determining whether the flux density contribution arises predominately from the input source, {pre-simulation} source or a combination of the two. This is especially important as we are injecting predominantly faint sources, due to the source counts distribution, and so do not want to confuse these faint sources with bright sources which already exist within the image.} Finally, we determine a source to be in our ``detected" catalogue if either the separation between the nearest input source or input component is less than a certain angular separation\footnote{For the \textsc{SIMBA} light cone based simulations we only use the input simulated catalogue source with flux densities $>5 \muup$Jy when matching to the output sources. This is {done} to avoid matching detected sources to a less appropriate faint source due to positional offsets in the source finding process.}{, see Section \ref{sec:results_sep}}. We do this secondary match to ensure that the detected sources are in fact related to the simulated sources.

\subsubsection{Determining the Angular Matching Separation}
\label{sec:results_sep}
\begin{figure*}
    \centering
    \begin{minipage}[b]{0.8\linewidth}
    \includegraphics[width=\linewidth]{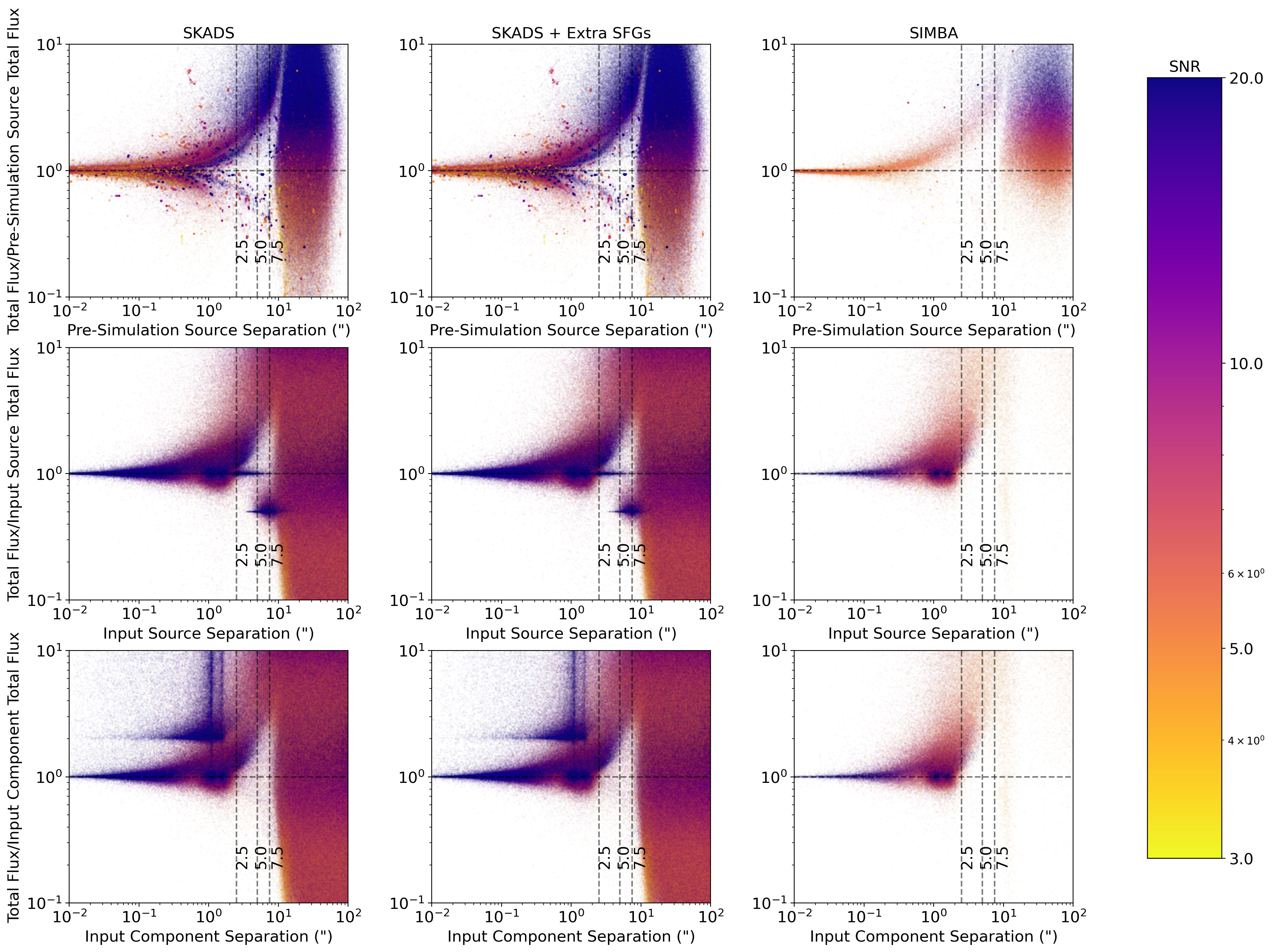}
    \subcaption{COSMOS}
    \end{minipage}%
    \\
    \caption{{Comparison of the ratio of flux densities between a detected source in the simulated image and its nearest pre-simulation source (top row), nearest input source (middle row) and nearest input component (bottom row) as a function of angular separation. This is shown for the SKADS simulations (left, Section \ref{sec:skads}), modified SKADS model simulations (centre, Section \ref{sec:skadsmod}) and \textsc{SIMBA} based simulations (right, Section \ref{sec:simba}) in the COSMOS field. Each point represents a source and is coloured based on its detection signal-to-noise ratio. {{The dashed lines represent 2.5\arcsec, 5\arcsec and 7.5\arcsec separations as a guide only.  {The result for the XMM-LSS field are not presented as the results are very similar to those shown for the COSMOS field.}}}}}
    \label{fig:Angular_separations}
\end{figure*}

First we determine the appropriate matching radius to use to match our detected sources from the simulations to the input sources as well as to mask around any sources {already detectable within the image}. Therefore for each simulated source detected by \texttt{PyBDSF} we determine both its nearest {pre-simulation} catalogue source and nearest input source and input component. In Figure~\ref{fig:Angular_separations} we present the distribution of the ratio of the measured \texttt{PyBDSF} source flux density to the flux density of (a) the nearest {pre-simulation} source, (b) the {nearest input source and (c) the component. These are presented as a function of angular separation for the three simulations described earlier.} There are typically two distinct regions within the distributions of flux {density} ratio {, separated at an angular separation of $\sim$7.5\arcsec. } For the top row of Figure \ref{fig:Angular_separations}, we compare the flux density ratio of the measured source to the nearest pre-simulation source. At very small angular separations ($\lesssim$0.5\arcsec) this flux density ratio is $\sim$1 where we are identifying those sources that are only from the pre-simulation sources already in the image. As the separation increases, this ratio increases to $\sim$2-3 up to $\sim$7.5\arcsec. This is likely a result of pre-simulation sources merging with faint simulated sources. {At separations larger than 7.5\arcsec \ the scatter in the flux ratio distribution increases due to association of an undetected input source with a random nearest neighbour.}

In the middle row of Figure \ref{fig:Angular_separations} we compare to the input source catalogue {{and}} we again see a large number of matches at small angular separations with flux density ratios of $\sim$1, as expected. The scatter around this flux density ratio of 1 increases at large angular separations. {{This increase in flux density ratio (and the scatter around it) appears to be}} due to lower signal to noise sources whose flux density {and positions are} more easily influenced from being on a noise peak or trough. However, this will preferentially be biased to having larger measured flux densities compared to input flux {{densities}} as sources on noise peaks are more likely to be detected by \texttt{PyBDSF} than those on noise troughs.  It is possible to see that there are a small group of sources at larger $\sim$5-10\arcsec \ separation with an input to output flux density ratio of $\sim$0.5. This relates to double lobed AGN within our simulation that have been detected by \texttt{PyBDSF} as two separate sources. Similarly in the bottom row of Figure \ref{fig:Angular_separations}, where we compare to the nearest component, there are now a group of sources with flux density ratios of $\sim$2 at separations $<2.5$\arcsec. These are those sources which were simulated as two components, but \texttt{PyBDSF} only detects a single source.

\begin{figure*}
    \begin{minipage}[b]{0.9\linewidth}
    \includegraphics[width=\textwidth]{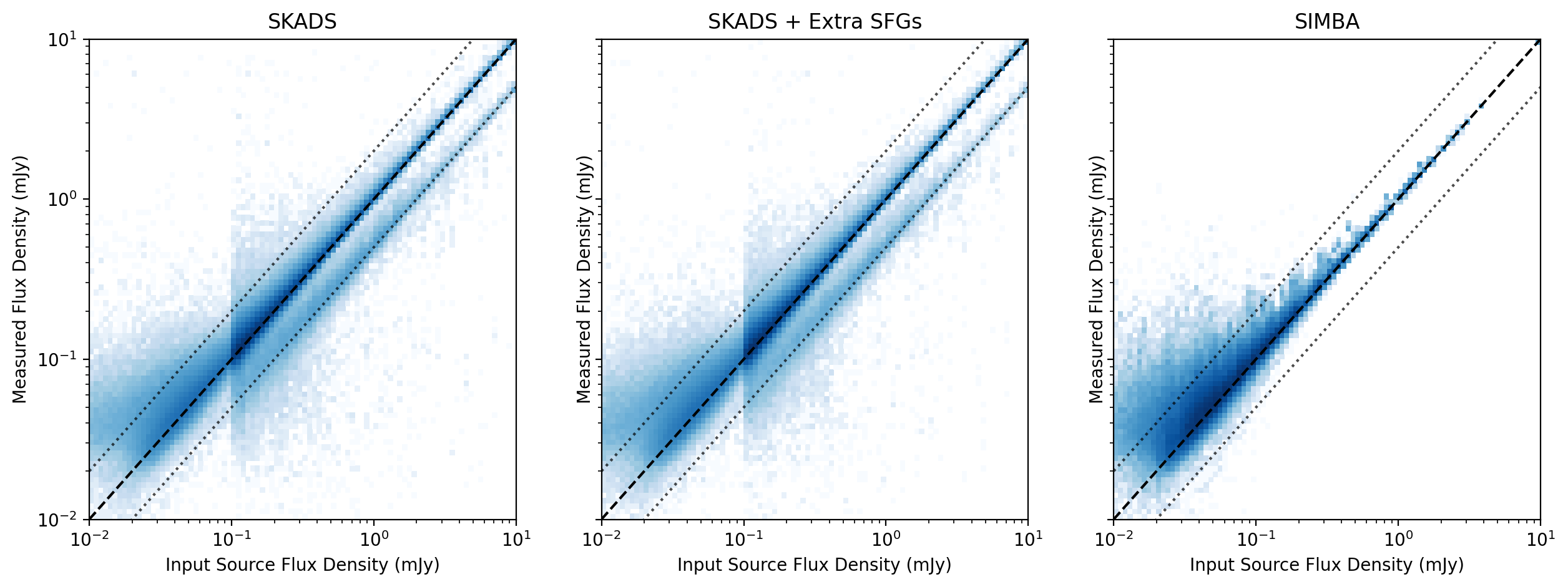}
    \subcaption{COSMOS}
    \end{minipage}%
    \linebreak
    \caption{{Comparison of the input simulated flux density {{(x-axis)}} to the measured flux density {{(y-axis)}} {for the source catalogues} in the {COSMOS field. The results for the XMM-LSS field are not shown, as the results are very similar to those shown for the COSMOS field}. The black {{dashed}} line indicates a 1-to-1 relation and the results are shown for the {combined simulation} results for three simulation models SKADS (left), modified SKADS (centre) and \textsc{SIMBA} light cone (right). The grey dotted lines represent flux density ratios of 0.5 and 2.}}
    \label{fig:fluxflux}
\end{figure*}

{The} dichotomy in sources {{which occurs at}} $\sim$7.5\arcsec \ {leads} us to use this as the matching radius. We do note though that for those multi-component AGN that are detected as two separate sources by \texttt{PyBDSF}{{,}} both detected components will be included in the output catalogue, as opposed to one single input source. However, in our real MIGHTEE images there will also be single sources that \texttt{PyBDSF} detects as multiple components. Therefore where these sources influence the completeness and so the corrected source counts, this will likely be correcting the measured source counts in the catalogues in the same way as necessary for the \texttt{PyBDSF} catalogues from the images. {As discussed in Section \ref{sec:calccomp1}, we therefore remove all simulated input/output sources that are matched to the pre-simulation catalogue within 7.5\arcsec}. After applying this angular separation radius, we present the comparison of the input flux density to the measured flux density {for the COSMOS field} in Figure \ref{fig:fluxflux}.  This is shown for the three input simulation models, {where the high flux simulations can be seen above 0.1 mJy.} 

{As expected}, at high flux densities sources have measured  integrated flux densities in agreement with their simulated flux densities, as the sources are bright and the noise is comparatively low. However, for fainter simulated sources the noise is more comparable to the flux densities of the sources themselves. For these faint sources, there is a clear excess in flux density {{that}} is important below $\sim0.1$ mJy, leading to an artificial boost in the measured flux density of a simulated source. As discussed previously, sources are both likely to be located on noise troughs as well as peaks, {but} those affected by noise troughs are less likely to be detected by a source finder, due to the reduced peak flux values and hence the reduced signal to noise. {The increase in scatter at $\gtrsim$0.1 mJy reflects the large number of high flux density simulations, where the majority of sources will be detected.} {As can be seen in Figure \ref{fig:fluxflux}, there are a number of sources for the SKADS-based simulations where the measured source flux density of some brighter sources is approximately half of the input flux density. This relates, as discussed above, to those sources with multiple components which have been split into two sources when measured with PyBDSF.} Figure \ref{fig:fluxflux} shows only detected sources, which preferentially have higher flux densities than the injected simulated sources {\citep[Eddington bias; ][]{Eddington1913}}. This bias towards measuring larger flux densities than were simulated is also notable in Figure \ref{fig:completeness1} where the {source count} completeness appears larger {{than}} 1 {at certain flux densities}. 

\subsubsection{Quantifying Source Counts Completeness and the Associated Errors}
The {flux density-dependent} {source count} completeness for each field is then determined by comparing the binned flux density distribution of input sources {{scaled to 1.4 GHz based on the input source position}} (excluding those within a certain angular separation to the original image {{catalogue) to}} that of output measured flux density of the sources in the detected catalogue{{, again scaled to 1.4 GHz (and excluding those matched to the original image catalogue)}}. We then compare the full input and output flux density distributions using logarithmically spaced flux density bins. {As such, the completeness {{(as defined in this work)}} can be found to be greater than {1. This} {can occur} when predominately faint sources are boosted to higher {flux densities (although they may also decrease in flux densities}). However these differences may also relate to any measurement errors when using the source finder, {\texttt{PyBDSF}}. This is} less likely to affect bright sources.

{The combined average source counts completeness value is determined using the ratio of the detected binned flux distribution to input source flux density distribution across all the simulations. To determine the uncertainty, we used the modified SKADS model to estimate the expected number of sources in each flux density bin considered. Using these numbers of sources, we construct random samples from out simulated sources which have the expected number in each flux density bin. By comparing the input flux density distribution of these sources to the flux density distribution of their measured counterparts (if they exist) a measurement of the completeness can be made. This process is then repeated a number of times ($N_\textrm{resamp}$) and the standard deviation of these realisations is used to quantify the error. $N_\textrm{resamp}$ is the approximate the number of independent samples we can consider and is calculated by determining the median number of independent samples across the flux density bins. This is $\sim$20 samples for the low flux density simulations, rising to $\sim$200 samples when we use the high flux density bins. The completeness errors are independently determined for the standard and high flux density simulations. The completeness and its errors above 0.5 mJy are constructed from these high flux density simulations. }

At the very highest flux densities, the {accuracy of the completeness estimates from these simulations may still be limited. Therefore, we set the completeness to 1 (and the completeness error to 0) above a flux density limit of 10 mJy. At these flux densities the number of sources in a field are small and so the errors will be dominated by small number statistics in these bins.}

\subsection{Resulting Source Count Corrections}
\label{sec:calc_sc}
\begin{figure*}
    \centering
   \includegraphics[width=0.75\textwidth]{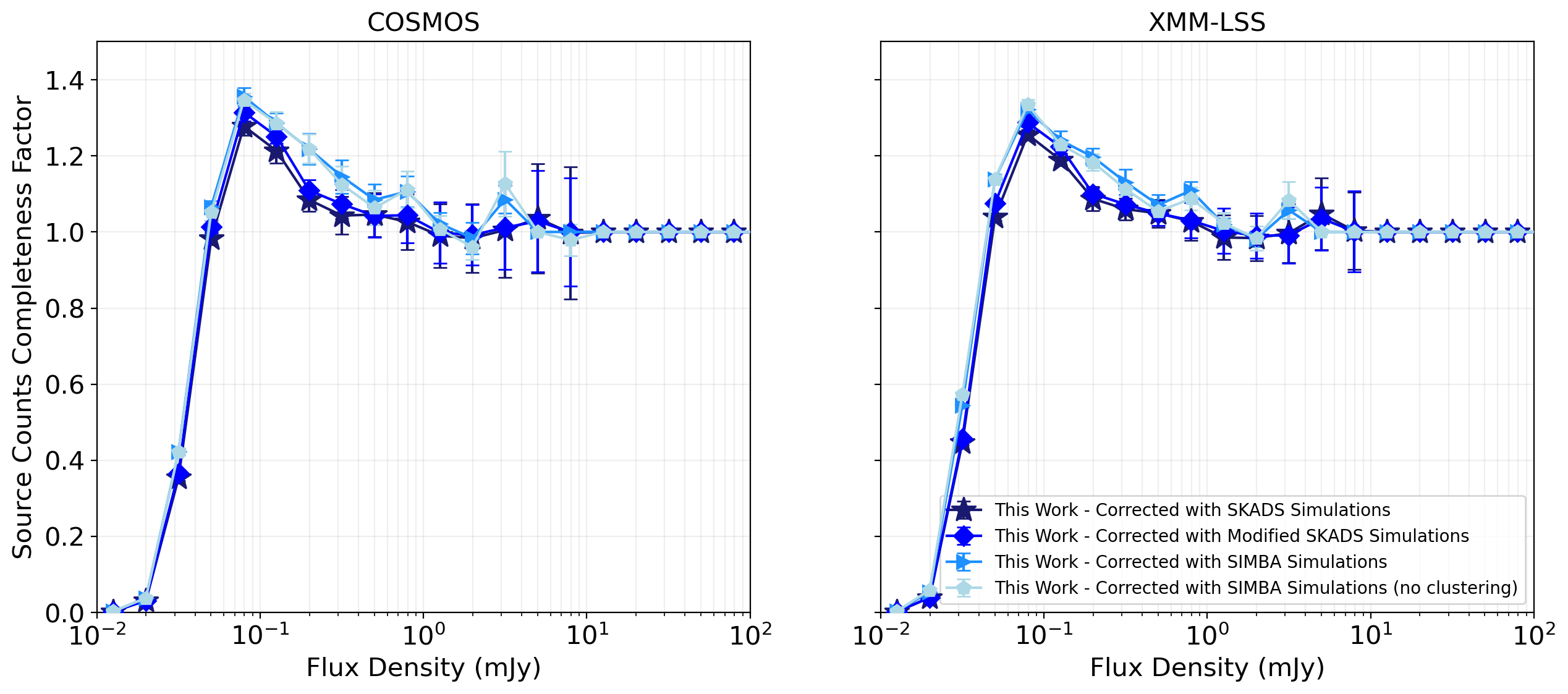}    
    \caption{{Source counts correction factor as a function of flux density within the COSMOS (left) and XMM-LSS (right) fields for the simulations using the SKADS (navy stars), modified SKADS (blue diamonds) and \textsc{SIMBA} light cone both with (blue triangles) and without (light blue pentagons) clustering effects. }}
    \label{fig:completeness1}
\end{figure*}

We present the results from investigating the {source count} completeness as a function of flux density in Figure \ref{fig:completeness1}. This is shown using the distribution of input source flux density to detected source flux density and for the three different simulation models. The completeness increases from 0 at ${\sim}10 \ \muup$Jy to a value larger than 1, before declining back down to a value of 1 at flux densities \ $\gtrsim$1 mJy. As discussed earlier, this increase above a value of 1 is not unexpected, and reflects the differences between the input simulated flux density and the flux density recovered when detected. 

{{The underlying} {intrinsic source counts were calculated by dividing the raw source counts (scaled to 1.4 GHz) by the source counts completeness calculated above.}} The associated errors are determined by combining, in quadrature, the {errors on the counts \citep[from Equations 9 and 12 of][]{Gehrels1986}} as well as the standard deviation derived from the completeness simulations and finally the error due to sample variance {from} \cite{Heywood2013}. 

\section{Results}
\label{sec:results}

In this section we present the results from investigating both the {corrected source counts and the integrated sky background temperature from AGN and SFGs}. 

\subsection{{Corrected} Source Counts}
\label{sec:scounts_results}
\begin{figure*}
    \centering
    \begin{minipage}[b]{0.8\linewidth}
    \includegraphics[width=\textwidth]{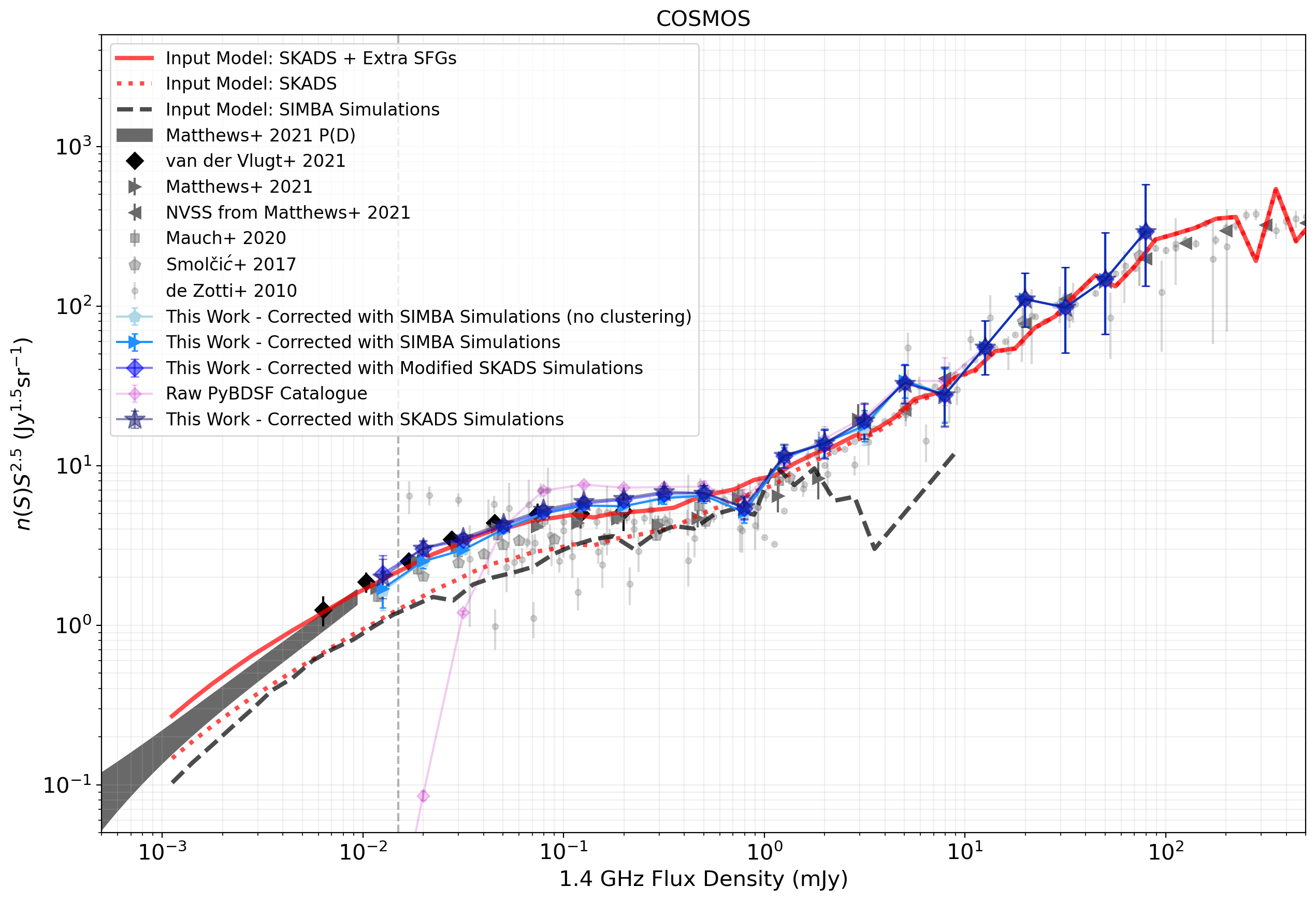}
    \end{minipage}%
    \linebreak
    \begin{minipage}[b]{0.8\linewidth}
    \includegraphics[width=\textwidth]{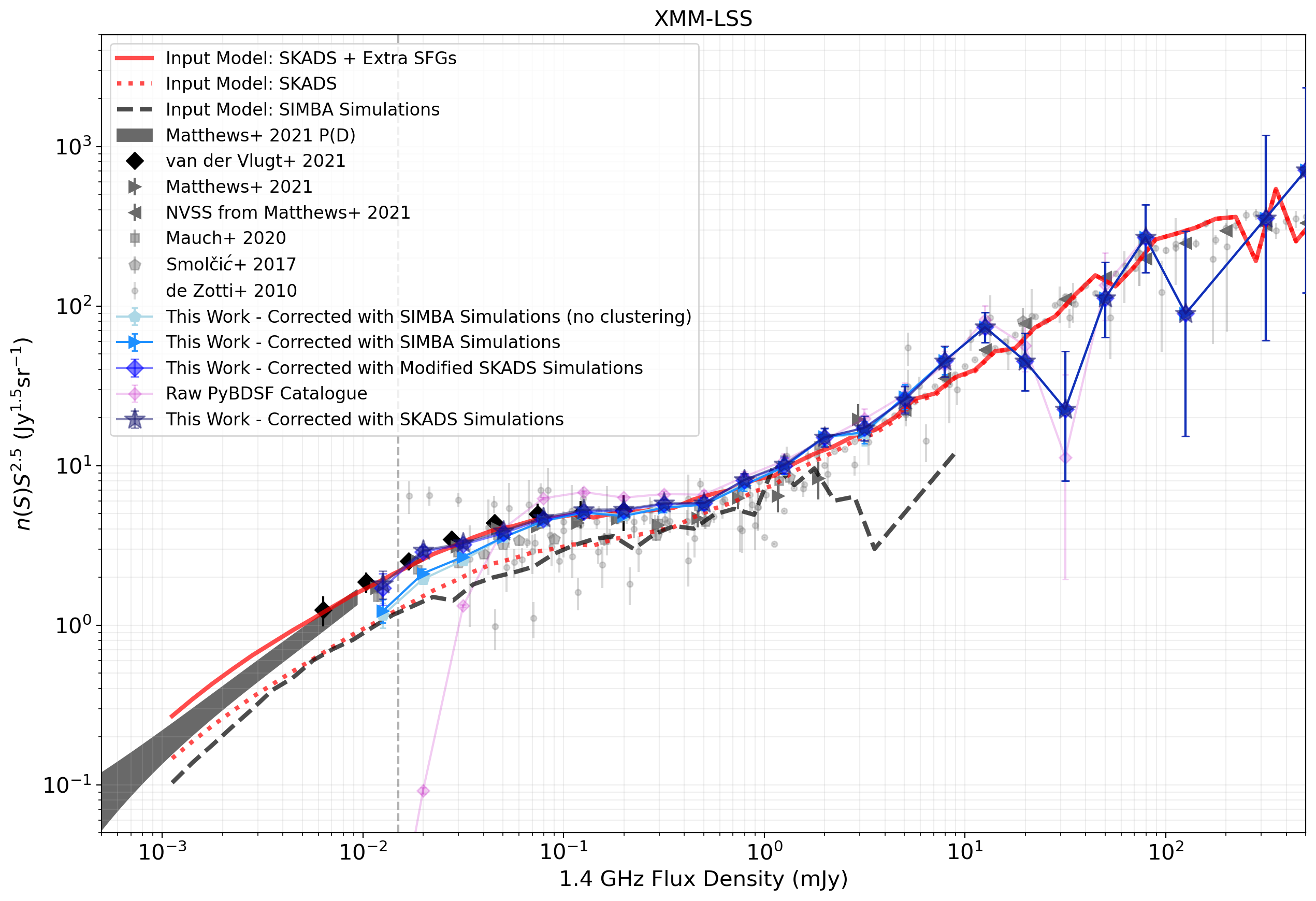}
    \end{minipage}
    \caption{{The 1.4 GHz Euclidean source counts for the MIGHTEE {{Early Science}} fields compared to previous models and observations for the COSMOS (upper) and XMM-LSS (lower) fields. Presented are the raw counts ({{pink}} diamonds), corrected source counts using the simulations with SKADS {{(dark blue stars, Section \ref{sec:skads})}}, {{modified SKADS (blue right diamonds, Section \ref{sec:skadsmod})}} {{and \textsc{SIMBA} light cone simulation both with clustering {{(blue right facing triangles, Section \ref{sec:simba})}} and without (light blue pentagons)}}. This is compared to observational data from the compilation by \protect \cite{deZotti2010} (grey dots); from the 3GHz VLA COSMOS survey \protect \cite[1.4 GHz source counts from][grey pentagons]{Smolcic2017b}; MeerKAT DEEP-2 observations by \protect \cite[][grey squares and right facing triangles]{Mauch2020, Matthews2021}; NVSS source counts \protect \cite[as given in][grey left facing triangles]{Matthews2021} {and deep COSMOS-XS observations \citep[][black diamonds]{vandervlugt2021}}. Also plotted {are} simulated models from SKADS \protect \cite[][red dotted line]{Wilman2008, Wilman2010}, the modified SKADS model described in \ref{sec:skadsmod} (red solid line), the \textsc{SIMBA} simulations \protect \cite[][black dashed line, also see Section \ref{sec:simba}]{Simba} and finally the sub-$\muup$Jy models from P(D) analysis from \protect \cite[][grey shaded region]{Matthews2021}. {For data at other frequencies, these are scaled to 1.4 GHz assuming $\alpha=0.7$.} The {vertical} grey dashed line indicates a value of 15 $\muup$Jy, which is 3$\times$ the minimum flux density used in our simulations. At fainter flux densities, our assumed minimum source flux may be affecting our work. Therefore whilst these fainter flux densities are included here to indicate our agreement with previous work, results below $\sim$15 $\muup$Jy should not be used for future comparisons and are therefore omitted from Tables \ref{tab:sc_cosmos} - \ref{tab:sc_xmm}. }}
    \label{fig:sc_corrected}
\end{figure*}

We present the corrected {Euclidean} {normalised} source counts from the combined results of the simulations using both the COSMOS (upper) and XMM-LSS (lower) fields in Figure \ref{fig:sc_corrected}. We present the corrected source counts using each of the three models described in Sections \ref{sec:skads} - \ref{sec:simba} as well as a comparison to the source counts from the raw (uncorrected) MIGHTEE source counts. We further present comparisons to input simulated models from \cite{Wilman2008}, and the simulated light cone from \textsc{SIMBA} as well as previous observational data from \cite{deZotti2010, Smolcic2017, Mauch2020, Matthews2021} and \cite{vandervlugt2021}. {The source counts from \cite{deZotti2010} are a compilation of {1.4 GHz} source counts from the literature from the work of \cite{Bridle1972, White1997, Ciliegi1999, Gruppioni1999, Richards2000, Hopkins2003, Fomalont2006, Bondi2008, Owen2008, Kellermann2008, Seymour2008}.} {Our derived} source counts are also presented in {Tables \ref{tab:sc_cosmos} and \ref{tab:sc_xmm}} for both the COSMOS and XMM-LSS fields {respectively, in each case giving the raw and} the corrected source counts from the three simulation methods\footnote{As we use logarithmic binning, the quoted flux density mid-point is taken using the mid-point of the logarithmic flux density bin. {We also note that although we include the SIMBA corrected source counts, these are underestimated due to the fact sources are injected into the residual image, see text. The SKADS or modified SKADS source counts should be used for future comparisons. }}. The source counts are shown for flux densities $\gtrsim$10 $\muup$Jy however, given the 5 $\muup$Jy flux density limit on simulated sources (and hence potential residual Eddington bias effects) as well as the increasing risk of systemic errors in completeness calculations at the lowest flux densities (for example, due to the effects of source size distributions on resolution bias), we recommend that strong conclusions are only drawn above 3-4$\sigma$ (i.e. $\sim$15$-$20 $\muup$Jy) in order to properly account for Eddington bias. {This is indicated by the dashed line at 15 $\muup$Jy within Figure \ref{fig:sc_corrected}. Our tabulated results only include the source counts above 15\,$\muup$Jy.}

{As can be seen in Figure \ref{fig:sc_corrected}, the four simulation variations each produce corrected source counts in good agreement with one another in {the COSMOS field source counts and in XMM-LSS the two SKADS simulations agree with each other and two SIMBA simulations are in good agreement with each other, though with corrected source counts slightly lower than the SKADS based source counts}; this will be discussed further in Section \ref{sec:discuss_sc}. Therefore for future discussions of the integrated background sky temperature we will only use the results for the model described in Section \ref{sec:skadsmod} for the modified SKADS simulations. }

\begin{sidewaystable*}
    \begin{minipage}[b]{\linewidth}
    \centering
    \caption{{Euclidean normalised 1.4 GHz source counts (at $>15 \muup$Jy) from the COSMOS field. Shown are the mid-point flux density, flux density range, number of sources within the flux density bin, N, raw source counts (Raw SC) {and the associated errors due to Poissonian statistics only from \protect \cite{Gehrels1986}.{ We then present the corrected source counts from the SKADS simulations (Section \ref{sec:skads}), modified SKADS simulations (Section \ref{sec:skadsmod}{{)}} and \textsc{SIMBA} simulations (with source clustering, Section \ref{sec:simba}) and the errors which have been determined by combining the Poissonian errors, completeness simulation errors and cosmic variance, see Section \protect \ref{sec:calc_sc} for details.}} Also included are the source counts split into AGN and SFGs as described in Section \protect \ref{sec:sc_sf_agn} for the modified SKADS {model, again with the errors from Poissonian errors, completeness simulation errors and cosmic variance included, as well as the errors from resampling the AGN/SFG fractions, see Section \protect \ref{sec:sc_sf_agn}}. These are presented for the two cases where the unclassified sources are assumed to be SFG (the SC$_{\textrm{AGN}}$ and SC$_{\textrm{SFG+Unc}}$ columns) and when the unclassified sources are assumed to be a mix of SFGs and AGN based on the flux density ratio of classified sources (given here with the subscript `ratio'). {We note that as discussed in Section \ref{sec:calc_temp}, whilst we present here the raw and corrected source counts above 10 mJy for the full sample, these are not used for the calculation of {sky background temperature contribution from extragalactic sources, due to the smaller number statistics}. As such, when we consider the source counts for the different source types (AGN vs SFGs), these are not included above 10 mJy where the source counts from NVSS are used and classified using the source ratio from MIGHTEE. Source counts are quoted to 3 significant figures.}}}
        \label{tab:sc_cosmos}
    \begin{tabular}{ccccccccccccc}
    $S$ & $S_{\textrm{min}} - S_{\textrm{max}}$ & $N$ & Raw SC & SC$_{\textrm{SKADS}}$ & SC$_{\textrm{mod. SKADS}}$ & SC$_{\textrm{\textsc{SIMBA}}}$ & SC$_{\textrm{AGN}}$ & SC$_{\textrm{SFG+Unc}}$  & SC$_{\textrm{AGN, ratio}}$ & SC$_{\textrm{SFG, ratio}}$  \\ 
       {{[$\muup$Jy] }} &{{[$\muup$Jy] }} & & {{[Jy$^{1.5}$sr$^{-1}$]}}& {{[Jy$^{1.5}$sr$^{-1}$] }}& {{[Jy$^{1.5}$sr$^{-1}$] }}& {{[Jy$^{1.5}$sr$^{-1}$] }} & {{[Jy$^{1.5}$sr$^{-1}$] }} & {{[Jy$^{1.5}$sr$^{-1}$] }} & {{[Jy$^{1.5}$sr$^{-1}$] }} & {{[Jy$^{1.5}$sr$^{-1}$] }}  \\ \hline \hline
20 & 16 - 25 & 249$^{+17}_{-16}$ & 0.966 $^{+0.007}_{-0.006}$ & 3.03 $^{+0.35}_{-0.35}$ & 3.02 $^{+0.32}_{-0.31}$ & 2.52 $^{+0.25}_{-0.25}$ & 0.620 $^{+0.091}_{-0.084}$ & 2.39 $^{+0.25}_{-0.26}$ & 0.874 $^{+0.126}_{-0.117}$ & 2.13 $^{+0.24}_{-0.24}$ \\
32 & 25 - 40 & 1617$^{+41}_{-40}$ & 1.25 $^{+0.03}_{-0.03}$ & 3.54 $^{+0.20}_{-0.20}$ & 3.42 $^{+0.19}_{-0.19}$ & 2.96 $^{+0.16}_{-0.16}$ & 0.712 $^{+0.051}_{-0.049}$ & 2.70 $^{+0.16}_{-0.15}$ & 0.980 $^{+0.069}_{-0.067}$ & 2.44 $^{+0.14}_{-0.14}$ \\
50 & 40 - 63 & 2751$^{+53}_{-52}$ & 4.25 $^{+0.08}_{-0.08}$ & 4.33 $^{+0.23}_{-0.23}$ & 4.19 $^{+0.22}_{-0.22}$ & 3.99 $^{+0.21}_{-0.20}$ & 1.02 $^{+0.06}_{-0.06}$ & 3.17 $^{+0.17}_{-0.17}$ & 1.40 $^{+0.09}_{-0.08}$ & 2.80 $^{+0.15}_{-0.15}$ \\
79 & 63 - 100 & 2200$^{+48}_{-47}$ & 6.78 $^{+0.15}_{-0.14}$ & 5.31 $^{+0.28}_{-0.28}$ & 5.16 $^{+0.28}_{-0.28}$ & 5.00 $^{+0.27}_{-0.27}$ & 1.58 $^{+0.10}_{-0.10}$ & 3.58 $^{+0.21}_{-0.19}$ & 2.15 $^{+0.14}_{-0.13}$ & 3.01 $^{+0.18}_{-0.17}$ \\
126 & 100 - 158 & 1174$^{+35}_{-34}$ & 7.22 $^{+0.22}_{-0.21}$ & 5.95 $^{+0.36}_{-0.36}$ & 5.77 $^{+0.33}_{-0.32}$ & 5.62 $^{+0.33}_{-0.32}$ & 2.13 $^{+0.16}_{-0.16}$ & 3.64 $^{+0.23}_{-0.23}$ & 2.86 $^{+0.21}_{-0.20}$ & 2.91 $^{+0.21}_{-0.20}$ \\
200 & 158 - 251 & 552$^{+25}_{-23}$ & 6.77 $^{+0.30}_{-0.29}$ & 6.25 $^{+0.43}_{-0.42}$ & 6.11 $^{+0.42}_{-0.41}$ & 5.56 $^{+0.40}_{-0.39}$ & 2.76 $^{+0.23}_{-0.22}$ & 3.34 $^{+0.27}_{-0.25}$ & 3.63 $^{+0.29}_{-0.27}$ & 2.48 $^{+0.23}_{-0.21}$ \\
316 & 251 - 398 & 292$^{+18}_{-17}$ & 7.15 $^{+0.44}_{-0.42}$ & 6.86 $^{+0.62}_{-0.60}$ & 6.66 $^{+0.56}_{-0.55}$ & 6.24 $^{+0.54}_{-0.52}$ & 3.68 $^{+0.39}_{-0.36}$ & 2.96 $^{+0.34}_{-0.31}$ & 4.65 $^{+0.47}_{-0.44}$ & 1.98 $^{+0.29}_{-0.27}$ \\
501 & 398 - 631 & 144$^{+13}_{-12}$ & 7.04 $^{+0.64}_{-0.59}$ & 6.73 $^{+0.78}_{-0.74}$ & 6.75 $^{+0.78}_{-0.74}$ & 6.49 $^{+0.70}_{-0.66}$ & 4.45 $^{+0.61}_{-0.56}$ & 2.28 $^{+0.42}_{-0.38}$ & 5.36 $^{+0.69}_{-0.63}$ & 1.38 $^{+0.34}_{-0.30}$ \\
794 & 631 - 1000 & 58$^{+9}_{-8}$ & 5.65 $^{+0.84}_{-0.74}$ & 5.51 $^{+0.94}_{-0.86}$ & 5.42 $^{+0.92}_{-0.84}$ & 5.11 $^{+0.82}_{-0.73}$ & 4.10 $^{+0.75}_{-0.66}$ & 1.30 $^{+0.39}_{-0.33}$ & 4.70 $^{+0.82}_{-0.71}$ & 0.696 $^{+0.293}_{-0.221}$ \\
1259 & 1000 - 1585 & 59$^{+9}_{-8}$ & 11.5 $^{+1.7}_{-1.5}$ & 11.6 $^{+2.0}_{-1.9}$ & 11.5 $^{+2.0}_{-1.8}$ & 11.2 $^{+1.8}_{-1.6}$ & 9.62 $^{+1.86}_{-1.72}$ & 1.75 $^{+0.98}_{-0.75}$ & 10.6 $^{+1.9}_{-1.8}$ & 0.695 $^{+0.794}_{-0.399}$ \\
1995 & 1585 - 2512 & 35$^{+7}_{-6}$ & 13.6 $^{+2.7}_{-2.3}$ & 13.8 $^{+3.1}_{-2.7}$ & 13.7 $^{+3.0}_{-2.6}$ & 13.8 $^{+2.9}_{-2.5}$ & 11.9 $^{+2.7}_{-2.4}$ & 1.49 $^{+1.31}_{-0.80}$ & 13.2 $^{+2.8}_{-2.6}$ & 0.175 $^{+0.864}_{-0.175}$ \\
3162 & 2512 - 3981 & 25$^{+6}_{-5}$ & 19.4 $^{+4.7}_{-3.8}$ & 19.3 $^{+5.3}_{-4.6}$ & 19.1 $^{+5.2}_{-4.4}$ & 17.8 $^{+4.4}_{-3.7}$ & 13.5 $^{+1.7}_{-1.8}$ & 3.24 $^{+1.62}_{-1.40}$ & 16.2 $^{+1.3}_{-1.4}$ & 0.223 $^{+1.243}_{-0.223}$ \\
5012 & 3981 - 6310 & 22$^{+6}_{-5}$ & 34.0 $^{+8.9}_{-7.2}$ & 32.8 $^{+9.8}_{-8.4}$ & 33.1 $^{+9.8}_{-8.3}$ & 34.0 $^{+9.0}_{-7.4}$ & 17.2 $^{+2.6}_{-2.9}$ & 4.99 $^{+2.88}_{-2.39}$ & 21.5 $^{+1.3}_{-2.2}$ & 0.307 $^{+2.498}_{-0.307}$ \\
7943 & 6310 - 10000 & 9$^{+4}_{-3}$ & 27.7 $^{+12.7}_{-9.1}$ & 27.8 $^{+13.6}_{-10.4}$ & 27.7 $^{+13.3}_{-10.0}$ & 27.7 $^{+12.7}_{-9.1}$ & 24.0 $^{+12.1}_{-8.7}$ & 2.45 $^{+4.84}_{-1.99}$ & 26.1 $^{+12.7}_{-9.2}$ & 0.383 $^{+3.581}_{-0.383}$ \\
12589 & 10000 - 15849 & 9$^{+4}_{-3}$ & 55.4 $^{+25.2}_{-18.1}$ & 55.4 $^{+25.4}_{-18.3}$ & 55.4 $^{+25.4}_{-18.3}$ & 55.4 $^{+25.4}_{-18.3}$ & {-} &  {-} & {-}  & {-}  \\ 
19953 & 15849 - 25119 & 9$^{+4}_{-3}$ & 110 $^{+50}_{-36}$ & 110 $^{+51}_{-36}$ & 110 $^{+51}_{-36}$ & 110 $^{+51}_{-36}$ & {-} &  {-} & {-}  & {-} \\ 
31623 & 25119 - 39811 & 4$^{+3}_{-2}$ & 98.0 $^{+77.2}_{-46.7}$ & 98.0 $^{+77.4}_{-47.0}$ & 98.0 $^{+77.4}_{-47.0}$ & 98.0 $^{+77.4}_{-47.0}$ & {-} &  {-} & {-}  & {-}   \\
50119 & 39811 - 63096 & 3$^{+3}_{-2}$ & 147 $^{+142}_{-80}$ & 147 $^{+142}_{-80}$ & 147 $^{+142}_{-80}$ & 147 $^{+142}_{-80}$ & {-} &  {-} & {-}  & {-}   \\
79433 & 63096 - 100000 & 3$^{+3}_{-2}$ & 292 $^{+284}_{-159}$ & 292 $^{+284}_{-159}$ & 292 $^{+284}_{-159}$ & 292 $^{+284}_{-159}$ & {-} &  {-} & {-}  & {-}    \\ \hline 
    \end{tabular}
    \end{minipage}%
\end{sidewaystable*}

\begin{sidewaystable*}
    \begin{minipage}[b]{\linewidth}
    \centering
    \caption{{Euclidean normalised source counts table as in \protect Table \ref{tab:sc_cosmos} but for the XMM-LSS field.}}
    \label{tab:sc_xmm}
    \begin{tabular}{ccccccccccccc}
    $S$ & $S_{\textrm{min}} - S_{\textrm{max}}$ & $N$ & Raw SC & SC$_{\textrm{SKADS}}$ & SC$_{\textrm{mod. SKADS}}$ & SC$_{\textrm{\textsc{SIMBA}}}$ & SC$_{\textrm{AGN}}$ & SC$_{\textrm{SFG+Unc}}$  & SC$_{\textrm{AGN, ratio}}$ & SC$_{\textrm{SFG, ratio}}$  \\ 
       {{[$\muup$Jy] }} &{{[$\muup$Jy] }} & & {{[Jy$^{1.5}$sr$^{-1}$]}}& {{[Jy$^{1.5}$sr$^{-1}$] }}& {{[Jy$^{1.5}$sr$^{-1}$] }}& {{[Jy$^{1.5}$sr$^{-1}$] }} & {{[Jy$^{1.5}$sr$^{-1}$] }} & {{[Jy$^{1.5}$sr$^{-1}$] }} & {{[Jy$^{1.5}$sr$^{-1}$] }} & {{[Jy$^{1.5}$sr$^{-1}$] }}  \\ \hline \hline
20 & 16 - 25 & 648$^{+26}_{-25}$ & 0.116 $^{+0.005}_{-0.005}$ & 2.94 $^{+0.19}_{-0.18}$ & 2.89 $^{+0.20}_{-0.20}$ & 2.11 $^{+0.14}_{-0.14}$ & 0.595 $^{+0.074}_{-0.068}$ & 2.29 $^{+0.18}_{-0.17}$ & 0.840 $^{+0.101}_{-0.094}$ & 2.04 $^{+0.17}_{-0.16}$ \\
32 & 25 - 40 & 4094$^{+65}_{-64}$ & 1.46 $^{+0.02}_{-0.02}$ & 3.27 $^{+0.14}_{-0.14}$ & 3.20 $^{+0.14}_{-0.14}$ & 2.68 $^{+0.12}_{-0.12}$ & 0.667 $^{+0.040}_{-0.039}$ & 2.53 $^{+0.11}_{-0.11}$ & 0.919 $^{+0.054}_{-0.053}$ & 2.28 $^{+0.10}_{-0.10}$ \\
50 & 40 - 63 & 5686$^{+76}_{-75}$ & 4.05 $^{+0.05}_{-0.05}$ & 3.89 $^{+0.16}_{-0.16}$ & 3.76 $^{+0.16}_{-0.16}$ & 3.55 $^{+0.15}_{-0.15}$ & 0.918 $^{+0.050}_{-0.049}$ & 2.85 $^{+0.12}_{-0.12}$ & 1.25 $^{+0.07}_{-0.07}$ & 2.51 $^{+0.11}_{-0.11}$ \\
79 & 63 - 100 & 4189$^{+66}_{-65}$ & 5.95 $^{+0.09}_{-0.09}$ & 4.74 $^{+0.20}_{-0.20}$ & 4.62 $^{+0.20}_{-0.20}$ & 4.50 $^{+0.19}_{-0.19}$ & 1.42 $^{+0.08}_{-0.07}$ & 3.21 $^{+0.14}_{-0.14}$ & 1.93 $^{+0.10}_{-0.10}$ & 2.69 $^{+0.13}_{-0.12}$ \\
126 & 100 - 158 & 2220$^{+48}_{-47}$ & 6.29 $^{+0.14}_{-0.13}$ & 5.29 $^{+0.24}_{-0.24}$ & 5.14 $^{+0.24}_{-0.24}$ & 5.07 $^{+0.25}_{-0.25}$ & 1.89 $^{+0.13}_{-0.12}$ & 3.24 $^{+0.17}_{-0.17}$ & 2.55 $^{+0.16}_{-0.15}$ & 2.59 $^{+0.16}_{-0.16}$ \\
200 & 158 - 251 & 1023$^{+33}_{-32}$ & 5.78 $^{+0.19}_{-0.18}$ & 5.32 $^{+0.31}_{-0.31}$ & 5.26 $^{+0.27}_{-0.27}$ & 4.83 $^{+0.26}_{-0.26}$ & 2.37 $^{+0.17}_{-0.17}$ & 2.88 $^{+0.19}_{-0.18}$ & 3.12 $^{+0.21}_{-0.20}$ & 2.13 $^{+0.17}_{-0.17}$ \\
316 & 251 - 398 & 545$^{+24}_{-23}$ & 6.15 $^{+0.27}_{-0.26}$ & 5.80 $^{+0.38}_{-0.37}$ & 5.73 $^{+0.37}_{-0.36}$ & 5.44 $^{+0.36}_{-0.35}$ & 3.17 $^{+0.28}_{-0.26}$ & 2.55 $^{+0.25}_{-0.24}$ & 4.01 $^{+0.33}_{-0.31}$ & 1.71 $^{+0.23}_{-0.22}$ \\
501 & 398 - 631 & 271$^{+17}_{-16}$ & 6.10 $^{+0.39}_{-0.37}$ & 5.81 $^{+0.48}_{-0.47}$ & 5.82 $^{+0.47}_{-0.46}$ & 5.70 $^{+0.45}_{-0.43}$ & 3.84 $^{+0.41}_{-0.39}$ & 1.97 $^{+0.32}_{-0.30}$ & 4.61 $^{+0.44}_{-0.43}$ & 1.19 $^{+0.27}_{-0.24}$ \\
794 & 631 - 1000 & 186$^{+15}_{-14}$ & 8.35 $^{+0.66}_{-0.61}$ & 8.13 $^{+0.82}_{-0.78}$ & 8.11 $^{+0.80}_{-0.77}$ & 7.53 $^{+0.68}_{-0.64}$ & 6.14 $^{+0.76}_{-0.70}$ & 1.95 $^{+0.51}_{-0.44}$ & 7.03 $^{+0.80}_{-0.73}$ & 1.04 $^{+0.41}_{-0.31}$ \\
1259 & 1000 - 1585 & 112$^{+12}_{-11}$ & 10.0 $^{+1.0}_{-0.9}$ & 10.2 $^{+1.3}_{-1.2}$ & 10.0 $^{+1.3}_{-1.2}$ & 9.84 $^{+1.10}_{-1.02}$ & 8.39 $^{+1.30}_{-1.22}$ & 1.53 $^{+0.83}_{-0.64}$ & 9.26 $^{+1.30}_{-1.23}$ & 0.607 $^{+0.691}_{-0.346}$ \\
1995 & 1585 - 2512 & 83$^{+10}_{-9}$ & 14.8 $^{+1.8}_{-1.6}$ & 15.1 $^{+2.1}_{-2.0}$ & 15.0 $^{+2.1}_{-2.0}$ & 15.1 $^{+2.0}_{-1.8}$ & 13.1 $^{+2.1}_{-2.1}$ & 1.64 $^{+1.40}_{-0.86}$ & 14.5 $^{+2.1}_{-2.0}$ & 0.194 $^{+0.955}_{-0.194}$ \\
3162 & 2512 - 3981 & 48$^{+8}_{-7}$ & 17.1 $^{+2.8}_{-2.5}$ & 17.2 $^{+3.2}_{-2.9}$ & 17.3 $^{+3.2}_{-2.9}$ & 16.2 $^{+2.8}_{-2.4}$ & 13.5 $^{+1.7}_{-1.8}$ & 3.24 $^{+1.62}_{-1.41}$ & 16.2 $^{+1.3}_{-1.4}$ & 0.224 $^{+1.243}_{-0.224}$ \\
5012 & 3981 - 6310 & 38$^{+7}_{-6}$ & 27.0 $^{+5.1}_{-4.4}$ & 25.8 $^{+5.5}_{-4.9}$ & 26.1 $^{+5.5}_{-4.8}$ & 27.0 $^{+5.2}_{-4.5}$ & 17.3 $^{+2.6}_{-2.9}$ & 4.98 $^{+2.88}_{-2.38}$ & 21.5 $^{+1.2}_{-2.2}$ & 0.306 $^{+2.502}_{-0.306}$ \\
7943 & 6310 - 10000 & 32$^{+7}_{-6}$ & 45.4 $^{+9.5}_{-8.0}$ & 45.3 $^{+10.7}_{-9.4}$ & 45.4 $^{+10.8}_{-9.5}$ & 45.4 $^{+9.7}_{-8.2}$ & 39.5 $^{+11.1}_{-10.2}$ & 4.19 $^{+7.27}_{-3.35}$ & 42.7 $^{+11.1}_{-10.3}$ & 0.640 $^{+5.844}_{-0.640}$ \\
12589 & 10000 - 15849 & 26$^{+6}_{-5}$ & 73.7 $^{+17.5}_{-14.3}$ & 73.7 $^{+17.7}_{-14.6}$ & 73.7 $^{+17.7}_{-14.6}$ & 73.7 $^{+17.7}_{-14.6}$ & {-} &  {-} & {-}  & {-}   \\
19953 & 15849 - 25119 & 8$^{+4}_{-3}$ & 45.2 $^{+22.3}_{-15.6}$ & 45.2 $^{+22.3}_{-15.7}$ & 45.2 $^{+22.3}_{-15.7}$ & 45.2 $^{+22.3}_{-15.7}$ & {-} &  {-} & {-}  & {-}   \\
31623 & 25119 - 39811 & 2$^{+3}_{-1}$ & 22.6 $^{+29.6}_{-14.5}$ & 22.6 $^{+29.7}_{-14.6}$ & 22.6 $^{+29.7}_{-14.6}$ & 22.6 $^{+29.7}_{-14.6}$ & {-} &  {-} & {-}  & {-}   \\
50119 & 39811 - 63096 & 5$^{+3}_{-2}$ & 113 $^{+76}_{-48}$ & 113 $^{+76}_{-49}$ & 113 $^{+76}_{-49}$ & 113 $^{+76}_{-49}$ & {-} &  {-} & {-}  & {-}   \\
79433 & 63096 - 100000 & 6$^{+4}_{-2}$ & 269 $^{+161}_{-107}$ & 269 $^{+161}_{-107}$ & 269 $^{+161}_{-107}$ & 269 $^{+161}_{-107}$ & {-} &  {-} & {-}  & {-}   \\
125893 & 100000 - 158489 & 1$^{+2}_{-1}$ & 89.6 $^{+205.0}_{-74.2}$ & 89.6 $^{+205.0}_{-74.3}$ & 89.6 $^{+205.0}_{-74.3}$ & 89.6 $^{+205.0}_{-74.3}$ & {-} &  {-} & {-}  & {-}   \\
316228 & 251189 - 398107 & 1$^{+2}_{-1}$ & 357 $^{+816}_{-296}$ & 357 $^{+816}_{-296}$ & 357 $^{+816}_{-296}$ & 357 $^{+816}_{-296}$ & {-} &  {-} & {-}  & {-}   \\
501187 & 398107 - 630957 & 1$^{+2}_{-1}$ & 712 $^{+1628}_{-590}$ & 712 $^{+1628}_{-590}$ & 712 $^{+1628}_{-590}$ & 712 $^{+1628}_{-590}$ & {-} &  {-} & {-}  & {-}   \\ \hline 
    \end{tabular}
    \end{minipage}%
\end{sidewaystable*}

\subsubsection{{The effect of input source model}}

{Whilst the consistency of our results give us confidence in our completeness corrections, in this section we examine the effect of using very different input source models. To do this we use two parameterised models for the source counts which allow a great deal of variation, including an uptick in the Euclidean normalised source counts at the faintest flux densities. Using these input source models we use the random simulations from the models of Section \ref{sec:skads} to determine what the ``observed source counts" from a given input model may be, which can be compared to the raw source counts to test what limits of an input source model could be assumed and still reconcile observations.  We use the assumed source size distributions from SKADS as this should give us a good estimate of the input model size distribution, assuming these are approximately correct.}

{The first model is a broken power law model of the form:}
\begin{equation}
\frac{dN}{dS}S^{2.5}   = \begin{cases} C \left( \frac{S}{S_0}\right)^{\alpha} &  (S\leq S_0)\\
C \left( \frac{S}{S_0}\right)^{\beta}  & (S>S_0)  \\
\end{cases}
\label{eq:doublepl}
\end{equation}

\noindent {and the second is a quadratic polynomial of the form:}
\begin{equation}
\log_{10} \left( \frac{dN}{dS}S^{2.5} \right)   =  \sum_{i=0}^{2} a_i \times \textrm{log}_{10}(S_{1.4 \ \textrm{GHz}})^i .
\label{eq:quadratic}
\end{equation}

{We sample the parameters $C$, $S_0$, $\alpha$ and $\beta$ (for the broken power law model) and three parameters for the polynomial ($a_i$ for i=0,1,2) to determine an input source counts model. Using the random simulated sources described in Section \ref{sec:skads} (without the additional high flux density simulations), we obtain an input simulated source ``catalogue", for a given input source model, and using the detected flux densities for these sources to determine the ``observed" source counts. These model ``observed" source counts are then compared to the {measured raw MIGHTEE source counts} using \texttt{emcee} \citep{emcee} to sample the posterior likelihood space, assuming a $\chi^2$  log likelihood function ($\ln\mathcal{L} = -\chi^2/2$) fit over the flux density range: $S_{1.4}: 1.5 \times 10^{-5} - 1.0 \times 10^{-3}$Jy. {For each run, 50 walkers with a chain length of 2000 steps are used to build up our samples. We then repeat this 100 times for each field using and each model. In order to ease computation for the polynomial models, extreme models (with $>10^7$ sources in a flux density bin were excluded). } } {The range of parameters used for this fitting with Equation \ref{eq:doublepl} were: $\log_{10}(S_0)$ = [-5.5, -3.5]; $\log_{10}$(C) = [0, 2]; $\alpha$ = [-0.5, 1.5] and $\beta$ = [-1, 1]. For the model with Equation \ref{eq:quadratic}, instead the ranges used were: $a_0$ = [-3.0, 1.0]; $a_1$ = [-3.0, 1.0] and $a_2$ = [-0.5, 0.5].}

\begin{figure*}
    \centering
    \includegraphics[width=0.85\textwidth]{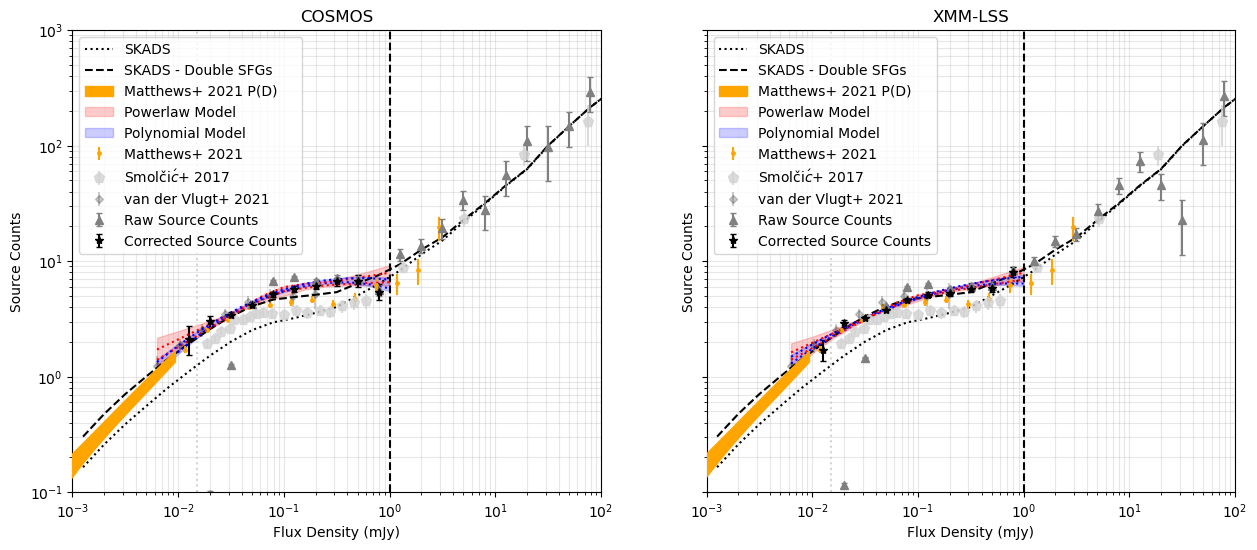}
    \caption{{Results for the power law (red) and polynomial (blue) models for the COSMOS (left) and XMM-LSS field (right). The filled region shows the range of values for the final step in the chain of those walkers not stuck in an poor fit and are given by the 5$^{\textrm{th}}$ and 95$^{\textrm{th}}$ percentiles. The dotted lines show the 16$^{\textrm{th}}$ and 84$^{\textrm{th}}$ percentiles from these fits. Also shown are the SKADS source counts from \protect\cite{Wilman2008} (black dotted) and modified SKADS model (black dashed) and observed source counts from \protect\cite{Matthews2021} (gold dots), \protect\cite{Smolcic2017} (grey pentagons) and \protect\cite{vandervlugt2021} (grey diamonds). The P(D) analysis from \protect\cite{Matthews2021} is also included (gold shaded region). The raw source counts in each field (grey triangles) and corrected source counts in the field (black stars, using the modified SKADS source counts model) are also shown.}}
    \label{fig:sc_vary_input}
\end{figure*}

{As the sampling code will randomly sample an input random catalogue distribution based on the model, it is the case that a different likelihood value can be obtained despite using the same model parameters. As such, chains were able to become stuck in a value where the likelihood for that given model and that given random sample was high. To avoid over weighting these particular parameter/randoms combinations, we use the last chain for each walker when comparing the models produced by the sampler. The results from the final chains of each of the source model simulations are shown in Figure \ref{fig:sc_vary_input}. We note that for the quadratic polynomial model, the large parameter range which is probed by the walkers results in a number of walkers appearing to be stuck in likelihood values that have not optimised. For the majority of walkers ($\sim 60-75\%$ of walkers on average, closer to 100\% for the broken power law model) the final chains have log likelihood values $>-50$, whilst the remaining walkers have anomalous log likelihood values (large negative values), as such, these are removed from the samples plotted as it is clear that these are poor fits to the data. Figure \ref{fig:sc_vary_input} shows the range in models only for those final chains with  $\ln\mathcal{L}>=-100$. The 5$^{\textrm{th}}$, 16$^{\textrm{th}}$, 84$^{\textrm{th}}$ and 95$^{\textrm{th}}$ percentiles for these chains are indicated on the plot. Figure \ref{fig:sc_vary_input} shows that both models agree well with both the data of \cite{Matthews2021} below $\sim70$ $\muup$Jy as well as with the modified SKADS model. The errors associated with the corrected source counts calculated in Section \ref{sec:scounts_results} are comparable or larger than those associated with the 16$^{\textrm{th}}$ and 84$^{\textrm{th}}$ percentiles, due to the restricted parameterisation.}

{Our source count models and measured source counts are in good agreement with one another, in general, and the two parameterisations agree well, though there are increasing discrepancies at the faintest flux densities, where we are less able to constraining our model, due to the 5$\muup$Jy limit for our simulated sources. The fact that our corrected source counts are in good agreement with these models indicates that the assumed source count model simulations that were used in order to calculate the corrected source counts in Section \ref{sec:calc_sc} are not substantially affecting the corrected source count models that we determine. We do note that our models are slightly higher than the P(D) results from \cite{Matthews2021} below 10 $\muup$Jy, but as these are in the flux density ranges below where we fit our data and our simulated sources had flux densities $\geq$5 $\muup$Jy, a discrepancy here is not necessarily unexpected. }

\subsection{Sky Background Temperature}

\subsubsection{Calculation of Sky Temperatures}
\label{sec:calc_temp}
The corrected source count distributions {{can then be}} used to calculate the background sky temperature at 1.4 GHz. Following the procedure of \cite{Hardcastle2021} and \cite{Matthews2021b} we estimate the sky background {{temperature, $T_b$}}, at a given frequency, $\nu$, through the equations relating {the thermodynamic temperature}, $T$, and spectral radiance, $I_{\nu}$, for a {{blackbody (Planck's law):}}
\begin{equation}
I_{\nu} = \frac{2 h \nu^3}{c^2} \frac{1}{e^{\frac{h\nu}{ k_B T}} - 1},
\end{equation}

\noindent where $h$ is the Planck's constant, $k_B$ the Boltzmann's constant and $c$ is the speed of light in {{a vacuum}}.

The spectral radiance is a measure of the flux density per unit solid angle, at a given frequency, with standard units Wm$^{-2}$sr$^{-1}$Hz$^{-1}$. Given that at radio frequencies we are in the Rayleigh-Jeans regime ($\frac{h\nu}{ k_B T}<< 1$){, then given a spectral radiance measurement at a given frequency,} this can be simplified {to give the expression for the brightness temperature, $T_B$, as:}
\begin{equation}
T_B = \frac{I_{\nu} c^2}{2 k_B \nu^2}.
\end{equation}

In order to determine the integrated spectral radiance from the data, and specifically from the {contribution of individual sources}, we must sum the contribution of the flux density of sources {observed} within our image, and normalise for the solid angle subtended. Following the methods of \cite{Hardcastle2021}, we use the equation:

\begin{equation}
 \begin{array}{l}
I_{\nu}(\geq S_{\nu}) \ = \int_{S_{\nu}}^{\infty} {S'}_{\nu} n({S'}_{\nu}) {dS'}_{\nu}  \\
\indent \indent \indent \indent  = \int_{S_{\nu}}^{\infty} {S'}_{\nu}^{-1.5} n({S'}_{\nu}) S_{\nu}^{\prime 2.5} {dS'}_{\nu}  \\
 \end{array}
\end{equation}

\noindent where $n(S'_{\nu})$ is the non-Euclidean source counts described in Section \ref{sec:methods_calc} and $S'_{\nu}$ the flux density of the source at a given frequency, $\nu$. {The contribution of individual extragalactic sources to the integrated sky brightness temperature above a given flux density, $T_b$,} is calculated by:

\begin{equation}
T_b(\geq S_{\nu}) = \frac{c^2}{2 k_B \nu^2} \int_{S_{\nu}}^{\infty} {S'}_{\nu}^{-1.5} n({S'}_{\nu}) {S'}_{\nu}^{2.5} {dS'}_{\nu}.
\label{eq:skytemp}
\end{equation}

\noindent We note that for the rest of this paper, we omit the subscript for frequency ($\nu$) notation in our description of source counts and temperature for simplicity, however they are evaluated at 1.4 GHz.

We use Equation \ref{eq:skytemp} with the corrected source counts derived in Section \ref{sec:scounts_results}. However, whilst the area of these observations ($\sim$5 deg$^2$) is relatively large for such deep observations {\citep[$\sim 5\times$ and $\sim 50\times$ larger than used in the works of][respectively]{Matthews2021, vandervlugt2021}}, it is still limited in observing the brightest, rarest sources. These sources can only be observed in large numbers using surveys that cover large fractions of the sky such as NVSS \citep{Condon1998}, TGSS-ADR \citep{Intema2017}, LoTSS \citep{Shimwell2019} and RACS \citep{racs, Hale2021}. {These bright sources can have a significant contribution to the background sky temperature, and so the poor statistics in small areas can lead to a large amount of Poisson noise.} Therefore, we follow the method of \cite{Matthews2021}, and combine the source counts from MIGHTEE with the source counts from NVSS at high flux densities \citep[from Table 6 of][]{Matthews2021}. We use the source counts from MIGHTEE below a flux density of 10 mJy. Whilst we have data up to $\sim$100 mJy, as can be seen in Figure \ref{fig:sc_corrected}, the source counts in the 10-100 mJy flux density range are more variable, especially in the XMM-LSS field. This is likely a result of two contributions: (1) sample variance and (2) multi component bright AGN which have not been combined into a single object.

\subsubsection{Contribution of AGN and SFG to the Sky Temperature}
As discussed in Section \ref{sec:intro}, one of the key benefits of the MIGHTEE survey is the wealth of ancillary data within the fields being observed. This information from across the electromagnetic spectrum can be combined using multiple diagnostics in order to distinguish those radio sources which are AGN dominated, compared to those dominated by star formation. This, therefore, allows for direct measurement of the {contribution of these extragalactic SFGs and AGN} to the {integrated sky background} temperature. 
The relative contribution to the sky temperature has been inferred recently by \cite{Matthews2021b} through linking the source counts distribution to an evolving luminosity function from the local {{radio luminosity functions}}. This, therefore, does not use direct measurements of the proportion of AGN and SFGs within the population to classify into a certain source type. For this work, though, the wealth of ancillary data in the MIGHTEE fields provides an excellent opportunity to directly use the AGN and SFG fractional contributions to the source counts in order to determine their {separate} contribution to the sky temperature. 

The fraction of AGN and SFGs within the MIGHTEE data as a function of flux density can {be} calculated from the catalogue {{produced}} in {\cite{Whittam2022}}. {In this work MIGHTEE sources were classified} into AGN (as well as sub-categories of AGN), SFGs {{and}} probable SFGs {(which we consider here to be SFGs)}, however there also remained {{a subset of}} sources which could not be classified or those which could not be cross-matched, either due to a lack of multi-wavelength source or due to the radio source being confused (see {Prescott et al., subm.}). {For this work we consider three potential options for these unmatched or unclassified sources in order to understand how their lack of classification may affect our measurement of the contribution of SFGs and AGN to the source counts and background sky temperature. The first possibility is that all these undetected/unclassified sources are dusty SFGs which are not detected at other wavelengths due to attenuation of their emission. The second possibility is that these sources are AGN which are predominately at high redshift. This may be the case for the unclassified sources, which \cite{Whittam2022} find to predominantly be at higher redshifts. However, the most likely option is that the unclassified/unmatched sources are a combination of SFGs and AGN as both have selection biases which may affect how easily a host galaxy could be detected or, for those with a host, how easily these could be classified. Therefore, we also use the case where the unclassified/unmatched sources are assumed to have the same split in SFGs to AGN as the classified sources at the given flux density. By considering these cases, we are able to better ascertain the spread in classified source counts. } 

\begin{figure}
    \begin{minipage}[b]{\linewidth}
    \begin{center}
    \centering
    \begin{subfigure}{0.9\textwidth}
    \includegraphics[width=\textwidth]{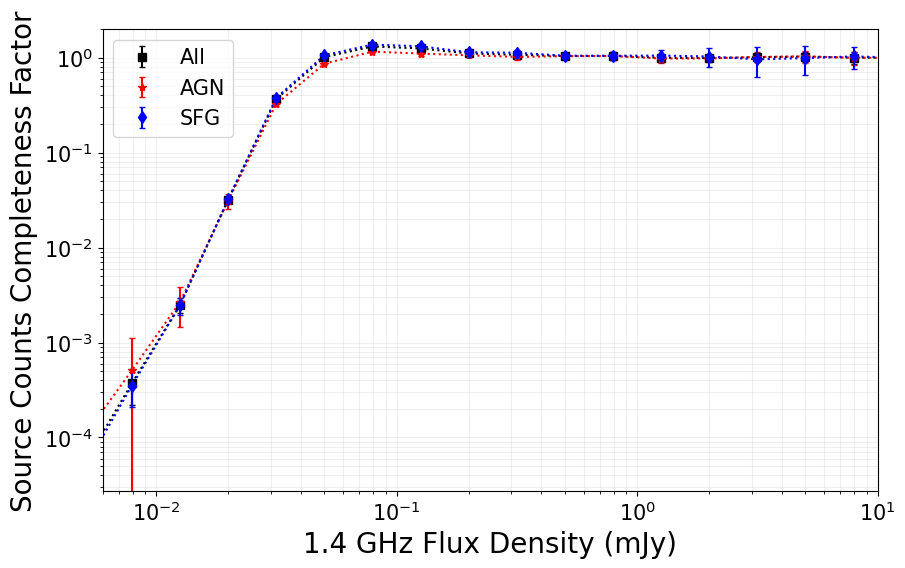}    
     \end{subfigure}%
      \end{center}
     \end{minipage}
    \caption{{{Source counts completeness of SFGs (blue), AGN (red) and for all sources (black) in the COSMOS field for simulations using the modified SKADS model {(Section} \ref{sec:skadsmod}).}}}
    \label{fig:comp_sfg_agn}
\end{figure}

As discussed {previously, we also note that the} AGN/SFG classifications are only {available} over the central $\sim$0.8 deg$^2$ of COSMOS where the \texttt{PyBDSF} Gaussian component catalogue has been combined together and cross-matched to ancillary data. Therefore the exact AGN/SFG fraction across both fields (COSMOS and XMM-LSS) may be different to that used here from just this smaller region. This will be further improved with the completion of MIGHTEE observations, and the associated source classifications, across all the four fields (COSMOS, E-CDFS, ELAIS-S1 and XMM-LSS). In this work, we make the assumption that the completeness of SFGs and AGN {(as a function of flux density)} agree with one another, even at the faintest flux densities. Therefore, even though we are incomplete at the faintest flux densities, the ratio of AGN to SFGs represents the true ratio of sources if we were complete. {We test this with the SKADS-based simulations, which have source type information so we can compare the completeness of AGN to SFGs{. From} Figure \ref{fig:comp_sfg_agn}, it can be seen that completeness of SFGs and AGN for the SKADS and modified SKADS simulations agree with each other within the errors at the brightest and faintest flux densities. There are small differences in the measured source count completeness values of AGN and SFGs in the range 0.05-0.2 mJy, however these differences are small and are unlikely to significantly impact our results. }

In this work, {in order to} determine the number of AGN, SFGs (including probable SFGs) and unclassified/unmatched sources as a function of flux density {we use} coarser logarithmic binning (using 15 bins between $\textrm{log}_{10}(S_{1.4 \textrm{GHz}})$ of -5.2 to -1) than used to investigate the source counts. We {can then} interpolate from this binned distribution to determine a function and from this we then calculate the fraction of different source types for each flux density bin we evaluate the source counts at. {This will be used to help resample our data to allow the contribution of AGN and SFGs to the 1.4 GHz source counts and the associated errors on that, as described below}. We made the assumption at bright flux densities ($>1$ mJy where there are no sources within the flux density bin), that the fraction of AGN in our sample would {go to} 1{, this is consistent with assumed ratios in works such as \cite{Wilman2008, deZotti2010,Bonaldi2019}}. 

\subsection{Uncertainties on the Sky Background Temperature}
\label{sec:sc_sf_agn}
To calculate the uncertainty on the background sky temperature accounting for Poissonian statistics, completeness, sample variance and the fractional contribution of AGN/SFGs we take the following approach. {First, we produce 1000 source count realisations by randomly sampling a normal distribution centred on the corrected source counts value within each flux density bin and with errors from the combined errors described in Section \ref{sec:calc_sc}. Due to the asymmetric errors we use 50\% of these samples with the positive and negative errors respectively.} 

{We then further attempt to model the uncertainty associated with the split in AGN and SFGs for both the classified source counts and sky temperature contribution}. This is challenging, as it is hard to distinguish the error in classification using multiple diagnostics as well as the error from the AGN/SFG fractional contributions due to the fact that $\sim$0.8 deg$^2$ of COSMOS was used to calculate these contributions, not the full $\sim$5 deg$^2$. {Therefore, we try to understand how the SFG/AGN split may be affecting the background sky temperature contributions by using resampling to make 1000 more realisations of the already resampled source counts to determine the fractional AGN and SFG contributions}. We {therefore use resampling to recalculate the number of SFGs, AGN and unclassified sources in {the coarser flux density bins that for the AGN/SFG fractions, as discussed above}. We then use these to} calculate a new fraction of SFGs and AGN within each of the coarse flux density bins based on the fraction of resampled each respective population compared to the sum of the resampled SFG, AGN and unclassified populations. Again, we then interpolate from these distributions to evaluate the fraction of {SFGs and AGN} at the flux density bins that the source counts are evaluated at. {As discussed,} we assume at the brightest flux densities that the AGN fractions can be assumed to be 1 {and hence 0 for SFGs}. This led to a total of 1,000,000 realisations each for the intrinsic source counts distributions for SFGs and AGN {for each of the respective models where we make the assumptions for the consistency of the unclassified sources}. {This method to determine errors is limited, as it does not allow for systematic classification errors in the diagnostics used in \cite{Whittam2022}, however these are difficult to properly account for, and we note that this may lead to an underestimation of the uncertainties.}

\subsubsection{{Contribution of AGN and SFGs to the {Source Counts}}}

\begin{figure*}
    \centering
    \begin{minipage}[b]{0.8\linewidth}
    \includegraphics[width=\textwidth]{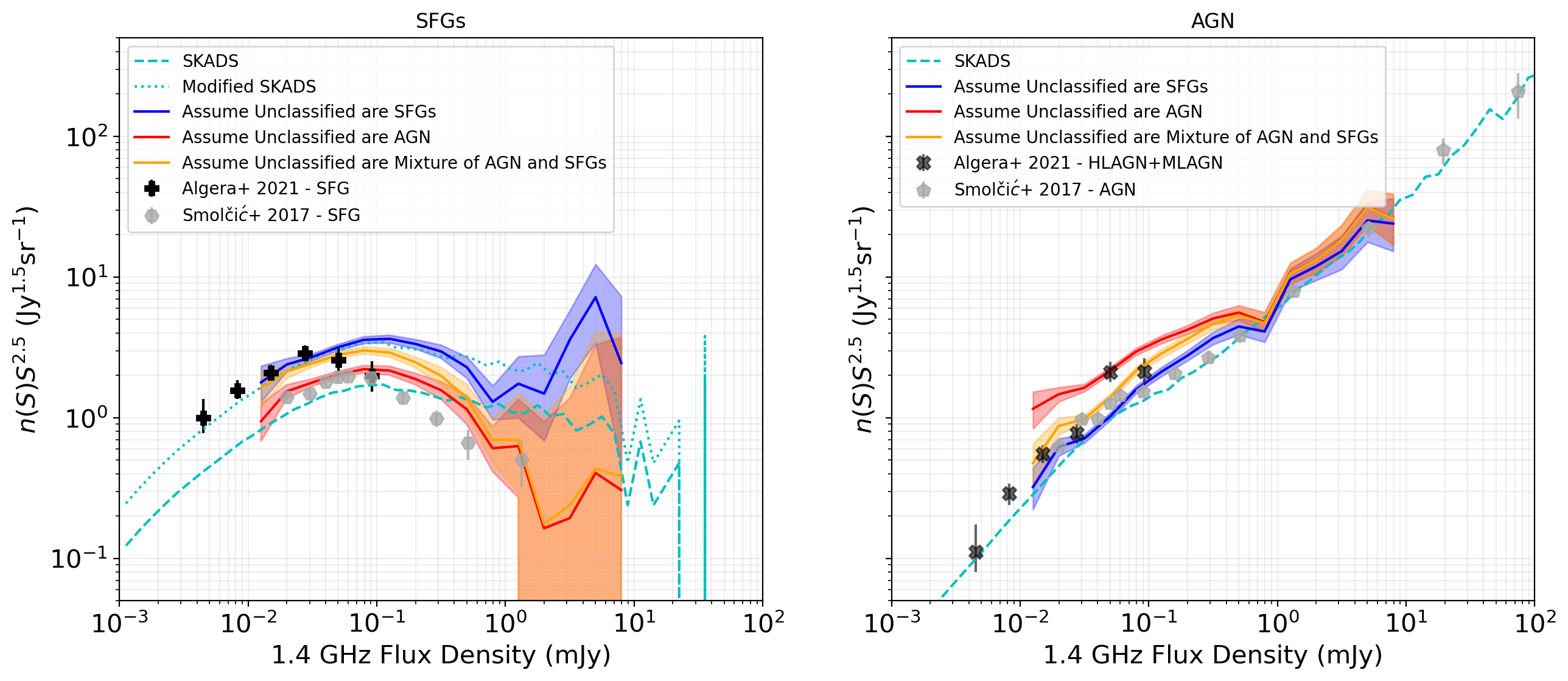}
    \subcaption{COSMOS}
    \end{minipage}%
    \\
    \begin{minipage}[b]{0.8\linewidth}
    \includegraphics[width=\textwidth]{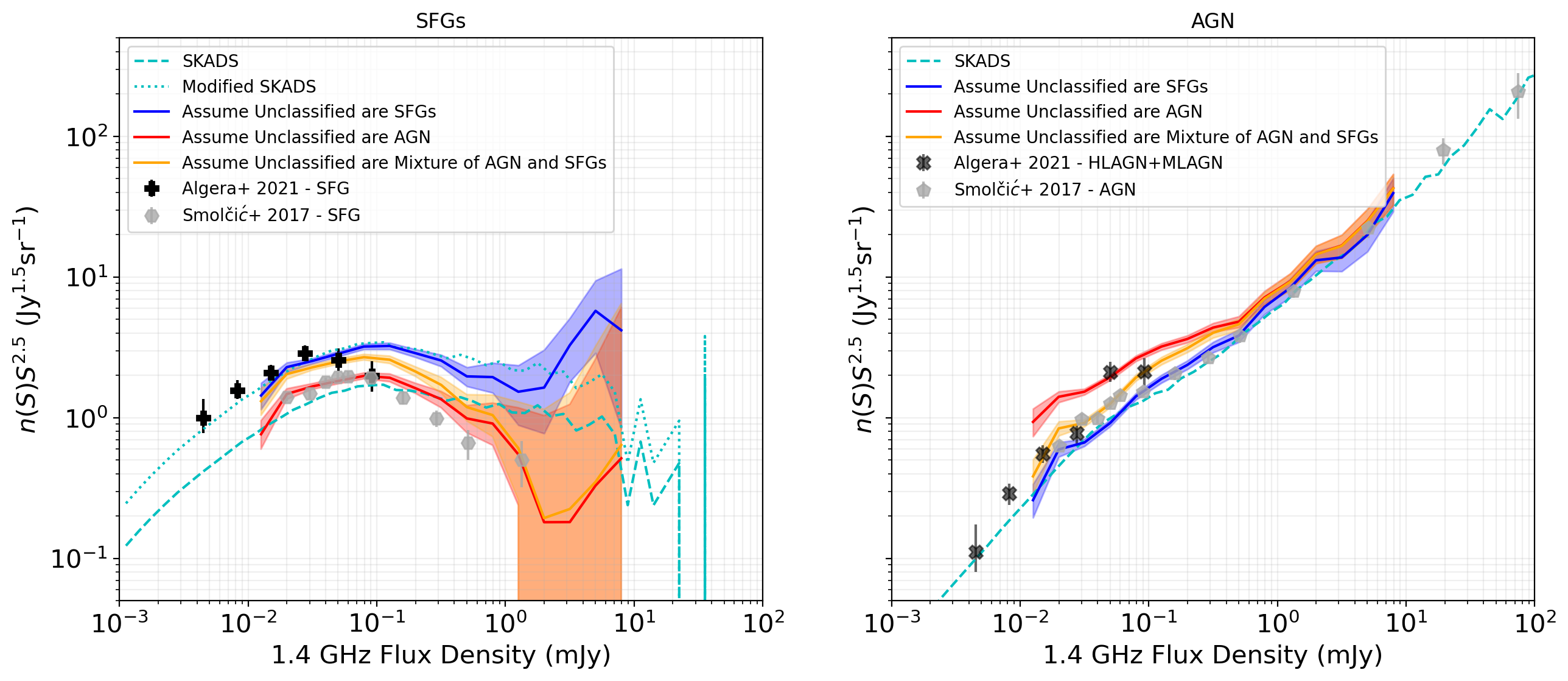}
    \subcaption{XMM-LSS}
    \end{minipage}%
    \caption{{{{The 1.4 GHz Euclidean source counts for the COSMOS (upper) and XMM-LSS field (lower) split into SFGs (left) and AGN (right) using the modified SKADS simulations (Section \ref{sec:skadsmod}). The shaded colourful regions indicate the three different assumptions on the unclassified population: the unclassified sources are SFGs (blue), AGN (red) and a mixture of AGN and SFGs dependent on the classified ratio for that flux bin (yellow). Also included are the SFG and AGN source counts from \protect\cite{Smolcic2017b} (grey hexagons for SFGs, grey pentagons for AGN) and from \protect\cite{Algera2020} (black plus for SFGs, black crosses for AGN). Also plotted are the respective SFG or AGN source models from SKADS \protect\cite[][cyan dashed line]{Wilman2008, Wilman2010} and the modified SKADS model described in \ref{sec:skadsmod} (cyan dotted line). {For data at other frequencies, these are scaled to 1.4 GHz assuming $\alpha=0.7$.}}}}}
    \label{fig:sc_split}
\end{figure*}

{We present the source counts generated using this resampling process as a function of source type in Figure {\ref{fig:sc_split} for both the COSMOS and XMM-LSS fields. The differences in the assumptions for the unclassified/unmatched sources affects the flux density at which SFGs appear to become the significant population. For example, if the unclassified sources (which includes the unmatched sources and we now on refer to solely as unclassified) are all assumed to be SFGs, then the SFG population becomes a significant fraction of the source population at flux densities $S_{\textrm{1.4 \ GHz}} \lesssim$0.3 mJy. If instead the unclassified sources are assumed to be AGN then the source counts for these two populations {show} similar behaviour below $S_{\textrm{1.4 \ GHz}} \lesssim$0.05 mJy. Finally, if we assume these unclassified sources have the same flux density ratio as to the classified sources, then the SFGs do dominate below $S_{\textrm{1.4 \ GHz}} \lesssim$0.1 mJy. We also include the source counts of SFGs and AGN from the previous works of \cite{Smolcic2017b} and \cite{Algera2020}. }}{Using these source counts models for the different source types, we then use these to calculate the integrated background sky temperature above a given flux density limit. From these samples, we then quantify the integrated background sky temperature by determining the median temperature contributions for the two populations and report the uncertainties from the 16$^{\textrm{th}}$ and 84$^{\textrm{th}}$ percentiles of the samples.}

\subsubsection{Sky Temperature Results}
\begin{figure*}
    \centering
    \begin{minipage}[b]{0.45\linewidth}
    \includegraphics[width=0.85\textwidth]{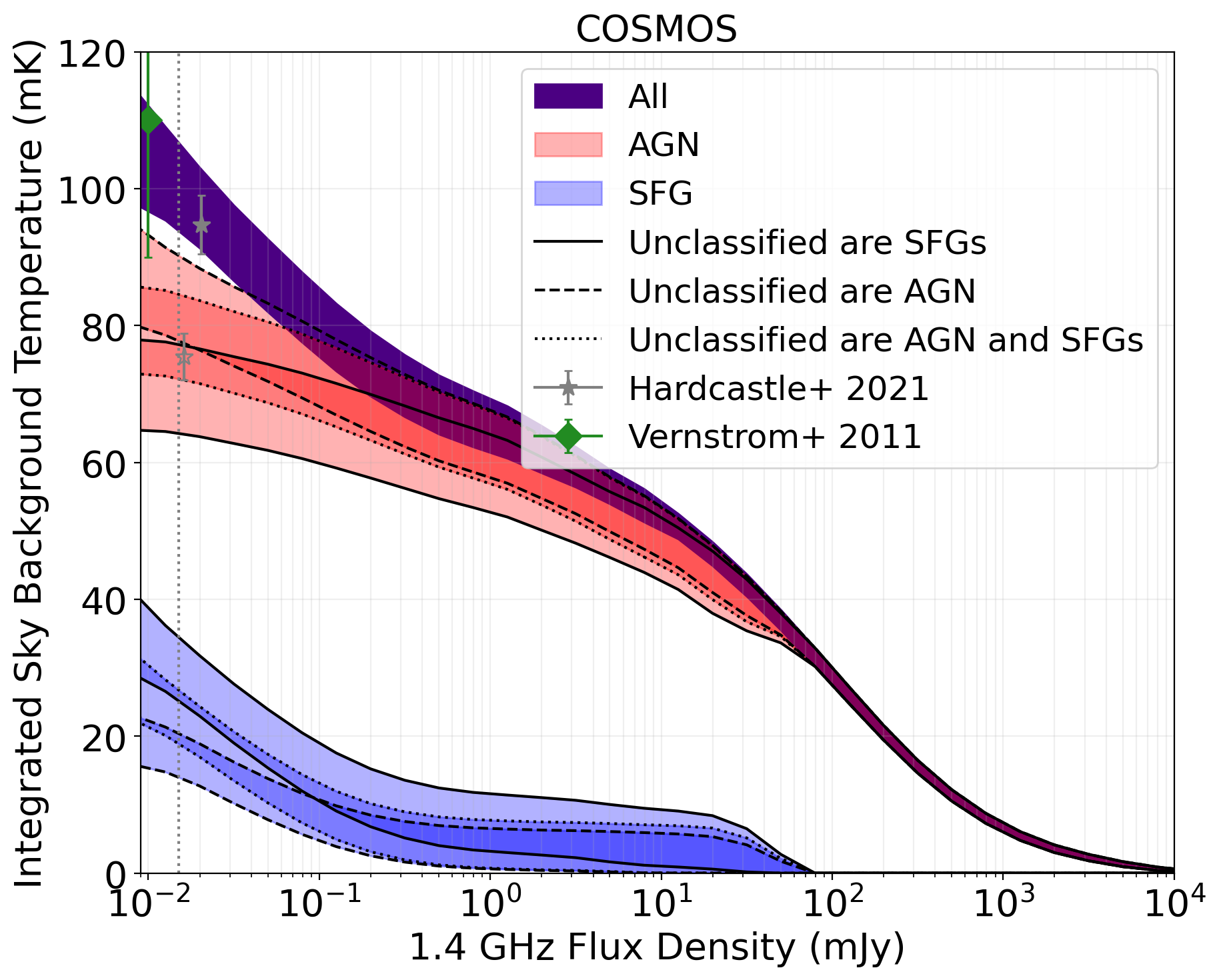}
    \end{minipage}%
     \begin{minipage}[b]{0.45\linewidth}
    \includegraphics[width=0.85\textwidth]{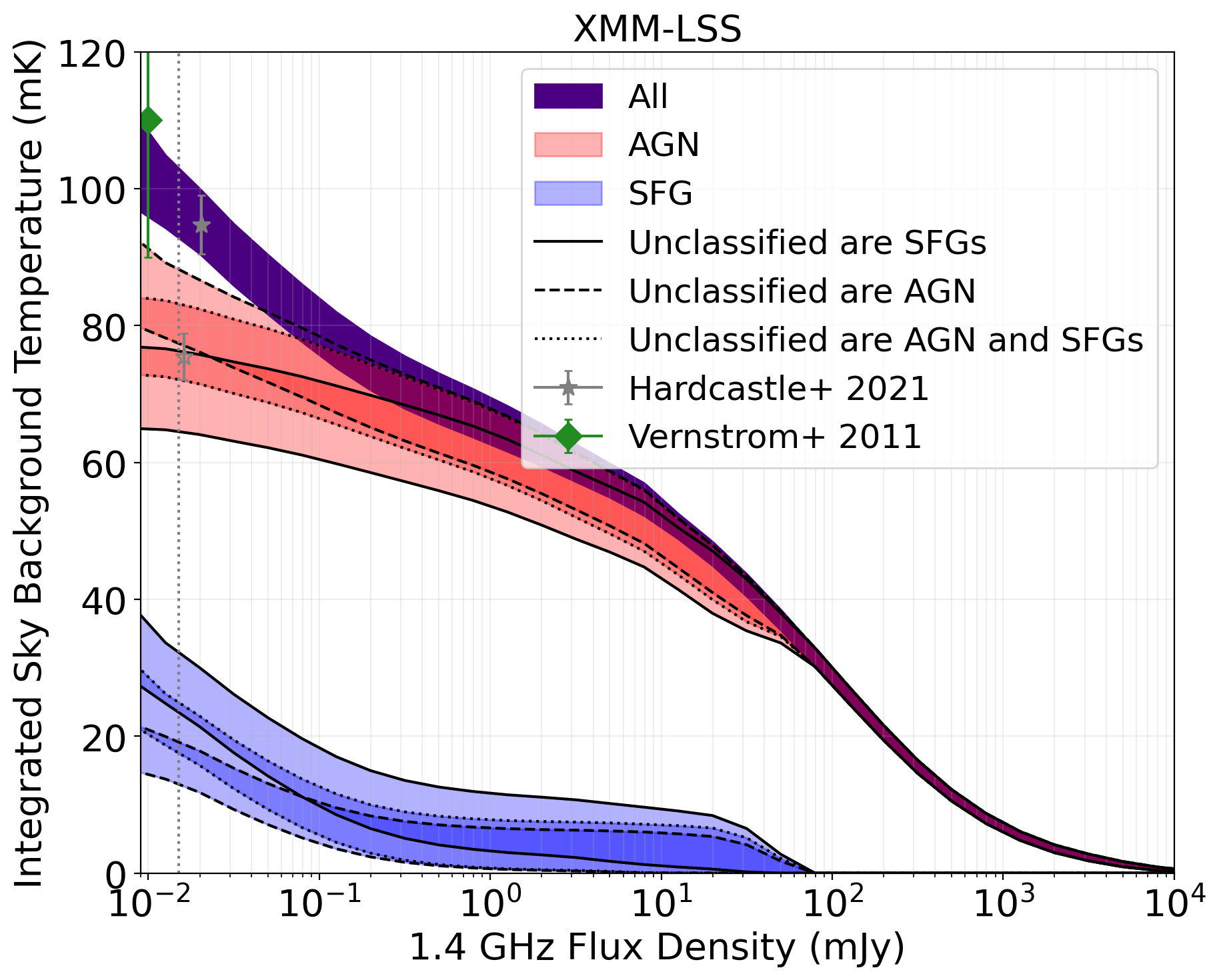}
    \end{minipage}%
    \caption{{Contribution of AGN (red) and SFGs (blue) {to} the total (purple) integrated background sky temperature indicated by a filled region between their 16$^{\textrm{th}}$ and 84$^{\textrm{th}}$ percentiles. This is shown for the COSMOS (left) and XMM-LSS (right) fields for the simulations using the modified SKADS model (Section \ref{sec:skadsmod}). The three assumptions for what makes up the unclassified sources are indicated by the black lines. Solid black lines indicate where all unclassified sources are assumed to be SFGs, dashed black lines indicate where the unclassified sources are assumed to be SFGs and, finally, black dotted lines indicate where unclassified sources are assumed to be a mixture of SFGs with their ratio the same as for the classified sources in the given flux density bin. Also shown is {the {sky background temperature contribution from extragalactic sources}} from \protect \cite{Vernstrom2011} (green diamond) and \protect \cite{Hardcastle2021} (grey star; {filled marker assuming $\alpha=0.7$ and as an indicative example of the effect of spectral index this is also show with $\alpha$ = 0.8,} {{open marker}}) scaled to 1.4 GHz using equation \ref{eq:temp_conv}. The grey dotted vertical lines indicates a $\sim$15 $\muup$Jy {flux density} cut.}}
    \label{fig:skytemp_split}
\end{figure*}

Finally, we present the integrated sky background temperature as a function of flux density from both AGN, SFGs and unmatched/unclassified sources in the COSMOS and XMM-LSS fields in Figure \ref{fig:skytemp_split}. {As mentioned previously, this uses the corrections based on the modified SKADS simulations described in Section \ref{sec:skadsmod}.} {We show the contribution to the integrated sky background temperature from AGN and SFGs using the three assumptions of what the unclassified sources could be. We find that the contribution to the sky temperature from extragalactic sources to be $T_b \sim 100$ mK at $\sim$ 15 $\muup$Jy. In Figure \ref{fig:skytemp_split} we compare this to the integrated background sky temperature measured in both \cite{Vernstrom2011} and \cite{Hardcastle2021}. \cite{Vernstrom2011} used a compilation of data from surveys at 150\,MHz to 8.4\,GHz \citep[see references in Table 1 of][]{Vernstrom2011}. They evaluated the sky temperature {contribution from all sources above 10 $\muup$Jy at} 1.4\,GHz and found $T_{b} = 110 \pm 20$ mK. \cite{Hardcastle2021} used data from the LOFAR deep fields \citep[see e.g.][]{Tasse2021,Sabater2021} to calculate the total sky background temperature at 144 MHz and found $T_{b} = 44 \pm 2$ K above 100 $\muup$Jy at 144 MHz}. 

To convert the {measurements of background sky temperatures at other frequencies} to 1.4 GHz we follow the method {used in \cite{Hardcastle2021} and} convert the temperatures using:
\begin{equation}
T_{b} = T_{\nu} \times \left(\frac{\nu \left(\textrm{GHz}\right)}{1.4}\right)^{\beta}.
\label{eq:temp_conv}
\end{equation}

\noindent {For our definition of spectral index convention $\beta=2+\alpha$, as in \cite{Hardcastle2021}. As we assumed $\alpha=0.7$}, {we use $\beta=2.7$ for this frequency conversion. We plot the value from \cite{Hardcastle2021} also including the limiting flux density used, converted to 1.4 GHz. {However, we also present the results of \cite{Hardcastle2021} scaled to 1.4 GHz assuming $\alpha=0.8$ (and $\beta$ = 2.8). Whilst a difference in spectral index of 0.1 will not make much difference to the conversion of the source counts at 1.4 GHz (as the frequency of MIGHTEE is close to 1.4 GHz, see Figure \ref{fig:eff_freq}) {it can be seen to} have an important impact on the conversion of temperatures from {144 MHz}.} }

\section{Discussion}
\label{sec:discussion}
{We now discuss the corrected source counts and the integrated background sky temperature based on the MIGHTEE Early Science data.}

\subsection{Source Counts}
\label{sec:discuss_sc}

In Figure \ref{fig:sc_corrected} we present our corrected source counts, as well as comparisons to previous studies. As can be seen in Figure \ref{fig:sc_corrected}, at the faintest flux densities the completeness corrections are able to correct the underestimated raw source counts to values in better agreement (compared to the raw counts) to those previously measured from \cite{Smolcic2017b, Mauch2020,Matthews2021} and \cite{vandervlugt2021}. {However, the corrected source counts, using all three models, are typically higher than \cite{Matthews2021} in the range $\sim0.1-2$ mJy. At flux densities $\sim 0.05-0.2$ mJy,} what is most striking is the contrast between the raw source counts from the COSMOS and XMM-LSS fields to the corrected source counts. In this regime, the raw source counts are notably higher than those which are corrected. This suggests that whilst on average the simulated and measured (recovered) flux densities follow a 1-to-1 line (see Figure \ref{fig:fluxflux}), there are small offsets between the measured distribution of sources by \texttt{PyBDSF} compared to any input simulation. This is seen in the source counts completeness plots of Figure \ref{fig:completeness1}, {where the source counts completeness can be larger than 1, as we are combining completeness with the measurement of the recovered sources with boosted flux density. This leads to a downwards correction of the raw source counts especially where these values were found to be} greatly in excess of most previous observations \cite[although with some overlap with source counts from the compilation by][]{deZotti2010} {{becoming}} in better agreement with observations from e.g. \cite{Matthews2021}.

{There are some small discrepancies, though, at faint flux densities  {{($\lesssim 100 \muup$Jy)}} between the observations from \cite{Smolcic2017b}, the MeerKAT DEEP-2 observations \citep{Matthews2021}, the COSMOS-XS observations \citep{vandervlugt2021} and the work presented here. At these flux densities, the source counts from \cite{Smolcic2017b} are lower than those observed with MeerKAT (both with DEEP2 and MIGHTEE) but also to VLA observations at 3 GHz from \cite{vandervlugt2021}}. These differences could arise from several reasons such as field to field variation due to sample variance and the relatively small field sizes observed in these surveys {as well as differences in the assumptions used to calculate completeness}. {Furthermore, \cite{Prandoni2018}, have shown that comparisons of the same fields can lead to differences in source counts measurements at the faintest flux densities, which is what we find in our COSMOS field source counts compared to that of \cite{Smolcic2017b}. These differences could be attributed to assumptions on the spectral index made in scaling the source counts from 3 GHz to 1.4 GHz or could be attributed to the increased number of SFGs at faint flux densities \citep[$\sim$100 $\muup$Jy, see e.g.][]{Wilman2008, Smolcic2017b, Bonaldi2019}.} {Furthermore, if these SFGs are resolved, it is possible that due to the baseline configuration of the VLA used for the VLA 3GHz COSMOS project which produced images at very high resolution \citep[0.75\arcsec \ resolution][]{Smolcic2017}, then extended emission may be more difficult to observe with the VLA}. This may result in an under-prediction of the source counts even in regions where completeness is high unless these extended sources are included in simulations {(see Appendix \ref{sec:appendix1})}. {In their work, \cite{Smolcic2017} did include resolution bias, but this could be underestimated for the most nearby and extended sources. In their work, \cite{vandervlugt2021} explain differences between their counts and that of \cite{Smolcic2017} as a combination of resolution bias and field-to-field variation, as they show the \cite{Smolcic2017} observations over the same area, which are in better agreement. In, this work we probe a larger area than the 350 arcmin$^2$ of \cite{vandervlugt2021}, and for both fields our work shows larger source counts than that of \cite{Smolcic2017b}. This therefore suggests that spectral index assumptions (converting from 3 GHz to 1.4 GHz) and resolution bias may also play an important role.}

At bright flux densities, $S_{1.4 \textrm{GHz}}\gtrsim$ 1 mJy, the results from COSMOS and XMM-LSS source counts are in roughly good agreement with e.g. the counts from NVSS in \cite{Matthews2021} and from the source counts {compilation} {{of}} \cite{deZotti2010}, although there is a lot of scatter. For example, at $\sim 10 - 50$ mJy, the source counts appear to be lower {in the XMM-LSS field} compared to previous measurements. This likely arises from the need to combine together multiple components of bright extended AGN manually, as in {Prescott et al. (subm.)}, that have not been combined together by \texttt{PyBDSF}. {{At faint flux densities ($\sim$0.02-0.05 mJy), the corrected source counts from the XMM-LSS and COSMOS fields are in good agreement with one another as well as being in good agreement ($\lesssim$0.05 mJy)}} with previous deep measurements from \cite{Mauch2020, Matthews2021} and \cite{vandervlugt2021}. {Our source counts should only be trusted above $\sim$15 $\muup$Jy, however we note that the source counts in our faintest flux density bin are {in} good agreement with \cite{Matthews2021} and {\cite{vandervlugt2021}}.}

{Comparing the different SKADS models we find that they provide corrected source counts that are in good agreement. When comparing to the two \textsc{SIMBA} simulations to compare the results with and without realistic clustering invoked, we find that for the COSMOS field the two \textsc{SIMBA} models are in excellent agreement both with each other and with the corrected source counts from the SKADS models. For the XMM-LSS field, small discrepancies can be seen between the \textsc{SIMBA} source counts and those from the SKADS simulations in the two lowest flux density bins below 30 $\muup$Jy. However the discrepancies between the \textsc{SIMBA} model with and without clustering invoked within the simulations are consistent with each other. Combining these two fields this suggests that the effect of clustering on completeness appears small and will not have a significant impact on our results moving forwards. The difference between the SKADS based models and that of \textsc{SIMBA} is therefore likely a result of a combination of resolution bias, which is not included in the \textsc{SIMBA} simulations, and any differences in completeness due to the effect of injecting SIMBA sources into the residual (as opposed to restored) image. Therefore, despite different methodology and different assumptions in the input source models, we can be confident that the corrected source counts measured here, using the SKADS based {corrections, represent} the true underlying source model. }

{Finally, we discuss our results for the source counts split by source type (using the modified SKADS based corrections), as presented in Figure \ref{fig:sc_split}. In Figure \ref{fig:sc_split} we show the comparison of our source counts to those of \cite{Smolcic2017b} (AGN and Clean SFGs, as presented in their Table 2) and \cite{Algera2020} (combining HLAGN and MLAGN), who both use observations over the COSMOS field to determine the contribution of AGN and SFGs. As can be seen in Figure \ref{fig:sc_split}, these are considered for the three possible assumptions about the unclassified sources, which we shall discuss now individually. Firstly, if all unclassified sources are AGN, the results for the source counts for the AGN populations appear to, in general, be much larger than found by either \cite{Smolcic2017b} or \cite{Algera2020} below $\sim$0.3 mJy. For SFGs, the source counts model has good agreement with that of \cite{Smolcic2017b}, but significantly under predicts the counts of SFGs compared to \cite{Algera2020}. Secondly, in the case where the unclassified sources are assumed to be SFGs, there is good agreement between the AGN source counts presented here with, in general, both the work of \cite{Smolcic2017b} and \cite{Algera2020}. For SFGs, there is good agreement with the work of \cite{Algera2020} below 0.05 mJy, but {the SFG source counts presented here are} higher than \cite{Smolcic2017b}. Finally, if we consider the unclassified sources to have the same fraction of AGN/SFGs as in the classified sample then, again, there is relatively good agreement with the AGN source counts from both works, and agrees significantly better with the results of \cite{Algera2020} than for \cite{Smolcic2017b}. This could reflect the fact that the source counts in \cite{Smolcic2017b} are for ``Clean SFGs", and so this may underestimate the true SFG population in \cite{Smolcic2017b}.}

{Our work demonstrates that the choice of classification for the sources that do not have a robust classification can significantly affect the contribution of AGN and SFGs to the measured source counts. Therefore, further investigations into deep multi-wavelength fields, using many multi-wavelength diagnostics, are important to help understand the contribution of SFGs and AGN to the source counts. This will be improved with the full MIGHTEE survey \citep[see][]{Jarvis2016}. However, our work does suggest that, in order to agree with the previous work of \cite{Algera2020} that these unclassified sources in our sample must be either SFG dominated or a flux-weighted ratio of AGN and SFGs and cannot be dominated by AGN, we therefore do not include the source counts from AGN and SFGs using these assumptions in Tables \ref{tab:sc_cosmos}-\ref{tab:sc_xmm}.}

\subsection{Integrated Background Sky Temperature}
\label{sec:discuss_skytemp}
With our source counts in good agreement with each other and previous measurements, {{we now discuss the results from the integrated sky background temperature contributions from AGN and SFGs}}. {As discussed, given the results from the SKADS} {simulations are in good agreement and clustering (from the \textsc{SIMBA} simulations) does not appear to have a strong effect} therefore, we only use the modified SKADS simulations (Section \ref{sec:skadsmod}) to investigate the integrated sky background temperature in the COSMOS to XMM-LSS fields. We choose the modified SKADS simulation given its close agreement between its source counts model to that of observed data. Using these, the results from the two fields are in very good agreement with each other and consistent within the errors{, although we note that we again use the same AGN/SFG split from the 0.8 deg$^2$ of the COSMOS field. However as the corrected source counts are calculated separately for each fields, there will be differences between the temperature contributions from the two fields.}

{If we consider the contribution of AGN and SFGs to the background sky temperature, it is important to note that the bright sources (which are generally AGN) have a large influence on the sky background temperature even though they are fewer in numbers. As can be seen from Figure \ref{fig:skytemp_split}, the temperature contribution of SFGs becomes {a more significant fraction of the total temperature} below $\sim$0.2-1 mJy, depending on the assumption of the split of AGN and SFGs in the unclassified sources. This leads to a contribution to the sky background temperature at $\sim$15 $\muup${Jy} in the range of $\sim$15-30 mK from SFGs, as seen in Figures \ref{fig:skytemp_split}(a) and (b). However, the previous discussion on source counts suggests that we are unlikely to be in the regime in which the unclassified sources are dominated by AGN. If we only consider the possibilities where the unclassified sources are all SFGs or a mixture of SFGs and AGN with the same fractional split as the classified data, then the contribution of SFGs to the background sky temperature at $\sim$15 $\muup${Jy} is $\sim${15}-{25} mK. Given the total integrated background temperature at the flux density limit is $\sim$100 mK, this suggests that at these faint flux densities SFGs only contribute {{$\sim${15-25}\%}} of the integrated background temperature, whereas they contribute $\sim$50\% of the sources.}

Comparing to previous results, our total sky background temperature estimate is in good agreement within the uncertainties and when frequency differences are accounted for with both the work of \cite{Vernstrom2011} and \cite{Hardcastle2021}, assuming $\alpha$=0.7. Indeed both fields are in excellent agreement with the measurement from \cite{Hardcastle2021} of $44\pm2$ K at 144 MHz at a flux density limit of $S_{144 \textrm{MHz}} \sim100$ $\muup$Jy. This measurement from \cite{Hardcastle2021} is equivalent to $\sim97$ {mK} at 1.4 GHz at $\sim$20 $\muup$Jy. {However if, instead, $\alpha=0.8$ is considered to convert the work of \cite{Hardcastle2021}, then the temperature, $T_b(\gtrsim10\muup$Jy), is closer to 75 {mK}. This would suggest that a low frequency spectral index of $\alpha=0.8$ is too steep when comparing between 1.4 GHz and 144 MHz and that $\alpha=0.7$, as assumed in this work throughout, is a more appropriate value.} {Our models also extrapolate to those from \cite{Vernstrom2011} at $\sim$10 $\muup$Jy, though this is below the flux density threshold for this work. The results} from \cite{Vernstrom2011, Hardcastle2021} and our observations of T$_b\sim$100 mK at $\sim$15$\muup$Jy, however, are a factor of $\sim4-5$ lower than measured with the ARCADE 2 experiment \citep{Fixsen2011}, where {the total integrated background temperature was estimated to be $\sim$500 mK} at 1.4 GHz. This suggests that there is {{no such population}} of faint extragalactic sources {to these sensitivities} that could explain such {difference in} temperature. 

The relative contribution of AGN and SFGs to the background sky temperature is something {{which can only be}} investigated with modern radio surveys, where the faint SFG population are detected in large numbers. Therefore, only recent studies such as \cite{Matthews2021b} have been able to look at the fractional contribution of AGN and SFGs to the background temperature. As discussed earlier, in \cite{Matthews2021b} the fractional contribution of AGN and SFGs to the sky background temperature was determined through evolving local {{radio luminosity functions}} in order to reproduce the total source counts when integrated over redshift. This work, on the other hand, {uses classifications of MIGHTEE sources to estimate} the relative contribution of AGN and SFGs. {{As discussed, SFGs contribute approximately {15}-25\% of the background temperature at 15 $\muup$Jy. This}} is compared to $\sim$30\% for the results of \cite{Matthews2021b}, who measure a total temperature of $\sim$90$-$100 mK at 10$-$15 $\muup$Jy. 

This work suggests that an even fainter population of extragalactic sources would need to {{exist}} in order to reconcile {the background temperature with that of} \cite{Fixsen2011}. This will be possible to {{investigate}} with surveys such as those from the future Square Kilometre Array Observatory. {{The SKAO will also have higher angular resolution than MIGHTEE which will aid in avoiding confusion, whilst  retaining surface brightness sensitivity.}} However{{,}} as our source counts seem to extrapolate to the models of \cite{Matthews2021}, it seems improbable that such a numerous faint extragalactic population of galaxies exist but {{are not already}} detected even at sub 5$\sigma$ levels in the deep radio data already available.

\subsection{Model of Background Sky Temperature}
{Finally, we provide a model of the sky background temperature for future comparison. To do this, we fit the models described in Section \ref{sec:calc_temp} using \texttt{numpy polyfit} to model the temperature, $T$, in K as a function of 1.4 GHz flux density, $S_{1.4 \ \textrm{GHz}}$, in Jy as:}
\begin{equation}
    T (>S_{1.4 \ \textrm{GHz}}) = \sum_{i=0}^{6} a_i \times \textrm{log}_{10}(S_{1.4 \ \textrm{GHz}})^i 
\label{eq:poly}
\end{equation}
\noindent {These fits are provided as supplementary material {alongside this work}. {These models are fit where the 16$^{\textrm{th}}$ percentile fits are >0 and temperature values are $>$0.01 mK. As the SFG models are fit over a smaller flux density range, we force $a_6$ to be 0 for these fits. In the supplementary table we provide the field, source type (e.g. AGN-AssumeUnclassAreSFG is the AGN model where unclassified sources are {considered to be} SFGs), percentile being fit (e.g. median, 16$^{\textrm{th}}$) and maximum flux density (in Jy) the fit can be used up to, above which it oscillates around 0 mK. }}

\section{Conclusions}
\label{sec:conclusions}
The MIGHTEE survey is an exciting new radio astronomy survey with MeerKAT, which will be essential in the study of galaxy evolution due to its depth ({rms$\sim 4-5 \muup$Jy beam$^{-1}$}), large area ($\sim$20 deg$^2$ on completion) and wealth of ancillary data across the four extragalactic fields it will observe. In this paper{{,}} we have investigated the deep source counts to $\sim$15 $\muup$Jy from the two {{Early Science}} fields \citep[$\sim$5 deg$^2$ over the COSMOS and XMM-LSS{{;}}][]{Heywood2022}. We make use of simulations using multiple underlying source population models to account for the incompleteness within the raw data to determine the intrinsic source counts {{distribution}}. By doing this, we account for incompleteness due to confusion, the visible area from RMS variations across the image as well as the detection efficiency and flux density accuracy of the source finding algorithm. Through these methods{{,}} we recover source counts which are in agreement with other recent, deep surveys of {\cite{Mauch2020, Matthews2021, vandervlugt2021}} but using a larger area of observations. {Furthermore we consider how the assumed source model affects the completeness, and thus the corrected source counts. From this we have demonstrated that independent of the input distribution of the underlying source counts, we determine corrected source counts in good agreement with the inferred source models. }

Building upon this, we use the classification of {{{a subset of} sources into AGN and SFGs from {\cite{Whittam2022}}}}, to directly investigate the contribution of SFGs and AGN to the background sky temperature. We show that AGN are dominant in their contribution to the sky temperature, with the contribution from SFGs increasing below 1 mJy, but only having {$\sim${15}-25\%} contribution to the integrated sky background temperature above 15 $\muup$Jy. {We find a total contribution to the sky background temperature from sources} of $\sim$100~mK above 15 $\muup$Jy, which is approximately a factor of 4 smaller than the reported background temperature from \cite{Fixsen2011}. Therefore, despite the sensitivity of these observations, we are unable to reconcile such a large {sky background} temperature in agreement with other previous works \citep[e.g.][]{Vernstrom2011, Hardcastle2021}. Overall, we have shown that MIGHTEE will be an excellent survey for {{developing}} our understanding of the population statistics of $\sim$ $\muup$Jy sources. Using the full 20 deg$^2$ of MIGHTEE will allow these source counts to be better constrained at the faintest flux densities and, when combined with multi-wavelength data over the full area, better constrain the high flux density source counts as well as better constrain the contribution of SFGs and AGN, and not be limited to using the AGN/SFG fraction based on 0.8 deg$^2$, which may be influenced by sample variance.

\section*{Acknowledgements}
{We thank the referee for their helpful comments which helped improve the quality of this manuscript.} The MeerKAT telescope is operated by the South African Radio Astronomy Observatory, which is a facility of the National Research Foundation, an agency of the Department of Science and Innovation. We acknowledge use of the Inter-University Institute for Data Intensive Astronomy (IDIA) data intensive research cloud for data processing. IDIA is a South African university partnership involving the University of Cape Town, the University of Pretoria and the University of the Western Cape. The authors acknowledge the Centre for High Performance Computing (CHPC), South Africa, for providing computational resources to this research project. CLH acknowledges support from the Leverhulme Trust through an Early Career Research Fellowship. This work used cuillin, the IfA's computing cluster (http://cuillin.roe.ac.uk) partially funded by the STFC and ERC. CLH thanks R. Kondapally for installation help with PyBDSF. IHW and MJJ acknowledge support from the Oxford Hintze Centre for Astrophysical Surveys that is funded through generous support from the Hintze Family Charitable Foundation. MJJ and IH are grateful for support from STFC via grant ST/S000488/1. {PNB is grateful for support from the UK STFC via grant ST/V000594/1.} JA acknowledges financial support from the Science and Technology Foundation (FCT, Portugal) through research grants PTDC/FIS-AST/29245/2017, UIDB/04434/2020 and UIDP/04434/2020. RB acknowledges support from an STFC Ernest Rutherford Fellowship [grant number ST/T003596/1]. MG was partially supported by the Australian Government through the Australian Research Council's Discovery Projects funding scheme (DP210102103). {NM} acknowledges the support of the LMU Faculty of Physics. LKM was supported by the Medical Research Council [MR/T042842/1]. {IP acknowledges financial support from INAF through the SKA/CTA PRIN “FORECaST” and the PRIN MAIN STREAM “SAuROS” projects,  and IP and LM acknowledge financial support from the Italian Ministry of Foreign Affairs and International Cooperation (MAECI Grant Number ZA18GR02). IP acknowledges the South African Department of Science and Technology's National Research Foundation (DST-NRF Grant Number 113121) as part of the ISARP RADIOSKY2020 Joint Research Scheme. BSF acknowledges MeerKAT/SARAO and IDIA. The authors acknowledge comments made by S. Randriamampandry on a draft of this paper.} Data in the appendix uses UltraVISTA DR4 observations which are based on data products from observations made with ESO Telescopes at the La Silla Paranal Observatory under ESO programme ID 179.A-2005 and on data products produced by CALET and the Cambridge Astronomy Survey Unit on behalf of the UltraVISTA consortium. For the multi-wavelength catalogues this work is based on data products from observations made with ESO Telescopes at the La Silla Paranal Observatory under ESO programme ID 179.A-2005 (Ultra-VISTA) and ID 179.A- 2006(VIDEO) and on data products produced by CALET and the Cambridge Astronomy Survey Unit on behalf of the Ultra-VISTA and VIDEO consortia. Based on observations obtained with MegaPrime/MegaCam, a joint project of CFHT and CEA/IRFU, at the Canada-France-Hawaii Telescope (CFHT) which is operated by the National Research Council (NRC) of Canada, the Institut National des Science de l’Univers of the Centre National de la Recherche Scientifique (CNRS) of France, and the University of Hawaii. This work is based in part on data products produced at Terapix available at the Canadian Astronomy Data Centre as part of the Canada-France-Hawaii Telescope Legacy Survey, a collaborative project of NRC and CNRS. The Hyper Suprime-Cam (HSC) collaboration includes the astronomical communities of Japan and Taiwan, and Princeton University. The HSC instrumentation and soft- ware were developed by the National Astronomical Observatory of Japan (NAOJ), the Kavli Institute for the Physics and Mathematics of the Universe (Kavli IPMU), the University of Tokyo, the High Energy Accelerator Research Organization (KEK), the Academia Sinica Institute for Astronomy and Astrophysics in Taiwan (ASIAA), and Princeton University. Funding was contributed by the FIRST program from Japanese Cabinet Office, the Ministry of Education, Culture, Sports, Science and Technology (MEXT), the Japan Society for the Promotion of Science (JSPS), Japan Science and Technology Agency (JST), the Toray Science Foundation, NAOJ, Kavli IPMU, KEK, ASIAA, and Princeton University. This work made use of Python and specifically the Python packages: \texttt{astropy} \citep{astropy1, astropy2}, \texttt{aplpy} \citep{aplpy1, aplpy2}, \texttt{numpy} \citep{numpy1, numpy2}, \texttt{scipy} \citep{scipy}, \texttt{matplotlib} \citep{matplotlib}, \texttt{tqdm} \citep{tqdm}. We also made use of \texttt{ds9} \citep{ds9_1, ds9_2} and \texttt{Topcat} \citep{topcat1, topcat2}.  For the purpose of Open Access, the author has applied a CC BY public copyright licence to any Author Accepted Manuscript version arising from this submission.

\section*{Data Availability}

{The MIGHTEE {{Early Science}} data used for this work is discussed in depth in \cite{Heywood2022} and information on the data release is described there. The cross-matched catalogue is described in the work of {{Prescott et al. (subm.)}} and the classification of radio galaxies into SFGs and AGN is described in {\cite{Whittam2022}}. The catalogues will be released in accompaniment with their work.} {The derived data produced in this work can be found in the article and supplementary material, or can be shared upon reasonable request to the corresponding author.}



\bibliographystyle{mnras}
\bibliography{MIGHTEE-SourceCounts} 

\begin{thebibliography}{}
\makeatletter
\relax
\def\mn@urlcharsother{\let\do\@makeother \do\$\do\&\do\#\do\^\do\_\do\%\do\~}
\def\mn@doi{\begingroup\mn@urlcharsother \@ifnextchar [ {\mn@doi@}
  {\mn@doi@[]}}
\def\mn@doi@[#1]#2{\def\@tempa{#1}\ifx\@tempa\@empty \href
  {http://dx.doi.org/#2} {doi:#2}\else \href {http://dx.doi.org/#2} {#1}\fi
  \endgroup}
\def\mn@eprint#1#2{\mn@eprint@#1:#2::\@nil}
\def\mn@eprint@arXiv#1{\href {http://arxiv.org/abs/#1} {{\tt arXiv:#1}}}
\def\mn@eprint@dblp#1{\href {http://dblp.uni-trier.de/rec/bibtex/#1.xml}
  {dblp:#1}}
\def\mn@eprint@#1:#2:#3:#4\@nil{\def\@tempa {#1}\def\@tempb {#2}\def\@tempc
  {#3}\ifx \@tempc \@empty \let \@tempc \@tempb \let \@tempb \@tempa \fi \ifx
  \@tempb \@empty \def\@tempb {arXiv}\fi \@ifundefined
  {mn@eprint@\@tempb}{\@tempb:\@tempc}{\expandafter \expandafter \csname
  mn@eprint@\@tempb\endcsname \expandafter{\@tempc}}}

\bibitem[\protect\citeauthoryear{{Adams}, {Bowler}, {Jarvis},
  {H{\"a}u{\ss}ler}, {McLure}, {Bunker}, {Dunlop}  \& {Verma}}{{Adams}
  et~al.}{2020}]{Adams2020}
{Adams} N.~J.,  {Bowler} R.~A.~A.,  {Jarvis} M.~J.,  {H{\"a}u{\ss}ler} B.,
  {McLure} R.~J.,  {Bunker} A.,  {Dunlop} J.~S.,   {Verma} A.,  2020, \mn@doi
  [\mnras] {10.1093/mnras/staa687}, \href
  {https://ui.adsabs.harvard.edu/abs/2020MNRAS.494.1771A} {494, 1771}

\bibitem[\protect\citeauthoryear{{Adams}, {Bowler}, {Jarvis}, {H{\"a}u{\ss}ler}
   \& {Lagos}}{{Adams} et~al.}{2021}]{Adams2021}
{Adams} N.~J.,  {Bowler} R.~A.~A.,  {Jarvis} M.~J.,  {H{\"a}u{\ss}ler} B.,
  {Lagos} C.~D.~P.,  2021, \mn@doi [\mnras] {10.1093/mnras/stab1956}, \href
  {https://ui.adsabs.harvard.edu/abs/2021MNRAS.506.4933A} {506, 4933}

\bibitem[\protect\citeauthoryear{{Aihara} et~al.,}{{Aihara}
  et~al.}{2018}]{Aihara2018}
{Aihara} H.,  et~al., 2018, \mn@doi [\pasj] {10.1093/pasj/psx081}, \href
  {https://ui.adsabs.harvard.edu/abs/2018PASJ...70S...8A} {70, S8}

\bibitem[\protect\citeauthoryear{{Algera} et~al.,}{{Algera}
  et~al.}{2020}]{Algera2020}
{Algera} H.~S.~B.,  et~al., 2020, \mn@doi [\apj] {10.3847/1538-4357/abb77a},
  \href {https://ui.adsabs.harvard.edu/abs/2020ApJ...903..139A} {903, 139}

\bibitem[\protect\citeauthoryear{{An} et~al.,}{{An} et~al.}{2021}]{An2021}
{An} F.,  et~al., 2021, \mn@doi [\mnras] {10.1093/mnras/stab2290}, \href
  {https://ui.adsabs.harvard.edu/abs/2021MNRAS.507.2643A} {507, 2643}

\bibitem[\protect\citeauthoryear{{Angl{\'e}s-Alc{\'a}zar}, {Dav{\'e}},
  {Faucher-Gigu{\`e}re}, {{\"O}zel}  \& {Hopkins}}{{Angl{\'e}s-Alc{\'a}zar}
  et~al.}{2017}]{Angles2017}
{Angl{\'e}s-Alc{\'a}zar} D.,  {Dav{\'e}} R.,  {Faucher-Gigu{\`e}re} C.-A.,
  {{\"O}zel} F.,   {Hopkins} P.~F.,  2017, \mn@doi [\mnras]
  {10.1093/mnras/stw2565}, \href
  {https://ui.adsabs.harvard.edu/abs/2017MNRAS.464.2840A} {464, 2840}

\bibitem[\protect\citeauthoryear{{Ashby} et~al.,}{{Ashby}
  et~al.}{2013}]{Ashby2013}
{Ashby} M.~L.~N.,  et~al., 2013, \mn@doi [\apj] {10.1088/0004-637X/769/1/80},
  \href {https://ui.adsabs.harvard.edu/abs/2013ApJ...769...80A} {769, 80}

\bibitem[\protect\citeauthoryear{{Astropy Collaboration} et~al.,}{{Astropy
  Collaboration} et~al.}{2013}]{astropy1}
{Astropy Collaboration} et~al., 2013, \mn@doi [\aap]
  {10.1051/0004-6361/201322068}, \href
  {https://ui.adsabs.harvard.edu/abs/2013A&A...558A..33A} {558, A33}

\bibitem[\protect\citeauthoryear{{Astropy Collaboration} et~al.,}{{Astropy
  Collaboration} et~al.}{2018}]{astropy2}
{Astropy Collaboration} et~al., 2018, \mn@doi [\aj] {10.3847/1538-3881/aabc4f},
  \href {https://ui.adsabs.harvard.edu/abs/2018AJ....156..123A} {156, 123}

\bibitem[\protect\citeauthoryear{{Bell}}{{Bell}}{2003}]{Bell2003}
{Bell} E.~F.,  2003, \mn@doi [\apj] {10.1086/367829}, \href
  {https://ui.adsabs.harvard.edu/abs/2003ApJ...586..794B} {586, 794}

\bibitem[\protect\citeauthoryear{{Best} \& {Heckman}}{{Best} \&
  {Heckman}}{2012}]{Best2012}
{Best} P.~N.,  {Heckman} T.~M.,  2012, \mn@doi [\mnras]
  {10.1111/j.1365-2966.2012.20414.x}, \href
  {https://ui.adsabs.harvard.edu/abs/2012MNRAS.421.1569B} {421, 1569}

\bibitem[\protect\citeauthoryear{{Blaizot}, {Wadadekar}, {Guiderdoni},
  {Colombi}, {Bertin}, {Bouchet}, {Devriendt}  \& {Hatton}}{{Blaizot}
  et~al.}{2005}]{Blaizot2005}
{Blaizot} J.,  {Wadadekar} Y.,  {Guiderdoni} B.,  {Colombi} S.~T.,  {Bertin}
  E.,  {Bouchet} F.~R.,  {Devriendt} J. E.~G.,   {Hatton} S.,  2005, \mn@doi
  [\mnras] {10.1111/j.1365-2966.2005.09019.x}, \href
  {https://ui.adsabs.harvard.edu/abs/2005MNRAS.360..159B} {360, 159}

\bibitem[\protect\citeauthoryear{{Bonaldi}, {Bonato}, {Galluzzi}, {Harrison},
  {Massardi}, {Kay}, {De Zotti}  \& {Brown}}{{Bonaldi}
  et~al.}{2019}]{Bonaldi2019}
{Bonaldi} A.,  {Bonato} M.,  {Galluzzi} V.,  {Harrison} I.,  {Massardi} M.,
  {Kay} S.,  {De Zotti} G.,   {Brown} M.~L.,  2019, \mn@doi [\mnras]
  {10.1093/mnras/sty2603}, \href
  {https://ui.adsabs.harvard.edu/abs/2019MNRAS.482....2B} {482, 2}

\bibitem[\protect\citeauthoryear{{Bondi} et~al.,}{{Bondi}
  et~al.}{2003}]{Bondi2003}
{Bondi} M.,  et~al., 2003, \mn@doi [\aap] {10.1051/0004-6361:20030382}, \href
  {https://ui.adsabs.harvard.edu/abs/2003A&A...403..857B} {403, 857}

\bibitem[\protect\citeauthoryear{{Bondi}, {Ciliegi}, {Schinnerer},
  {Smol{\v{c}}i{\'c}}, {Jahnke}, {Carilli}  \& {Zamorani}}{{Bondi}
  et~al.}{2008}]{Bondi2008}
{Bondi} M.,  {Ciliegi} P.,  {Schinnerer} E.,  {Smol{\v{c}}i{\'c}} V.,  {Jahnke}
  K.,  {Carilli} C.,   {Zamorani} G.,  2008, \mn@doi [\apj] {10.1086/589324},
  \href {https://ui.adsabs.harvard.edu/abs/2008ApJ...681.1129B} {681, 1129}

\bibitem[\protect\citeauthoryear{{Booth}, {de Blok}, {Jonas}  \&
  {Fanaroff}}{{Booth} et~al.}{2009}]{MeerKAT2}
{Booth} R.~S.,  {de Blok} W.~J.~G.,  {Jonas} J.~L.,   {Fanaroff} B.,  2009,
  arXiv e-prints, \href {https://ui.adsabs.harvard.edu/abs/2009arXiv0910.2935B}
  {p. arXiv:0910.2935}

\bibitem[\protect\citeauthoryear{{Bowler}, {Jarvis}, {Dunlop}, {McLure},
  {McLeod}, {Adams}, {Milvang-Jensen}  \& {McCracken}}{{Bowler}
  et~al.}{2020}]{Bowler2020}
{Bowler} R.~A.~A.,  {Jarvis} M.~J.,  {Dunlop} J.~S.,  {McLure} R.~J.,  {McLeod}
  D.~J.,  {Adams} N.~J.,  {Milvang-Jensen} B.,   {McCracken} H.~J.,  2020,
  \mn@doi [\mnras] {10.1093/mnras/staa313}, \href
  {https://ui.adsabs.harvard.edu/abs/2020MNRAS.493.2059B} {493, 2059}

\bibitem[\protect\citeauthoryear{{Bridle}, {Davis}, {Fomalont}  \&
  {Lequeux}}{{Bridle} et~al.}{1972}]{Bridle1972}
{Bridle} A.~H.,  {Davis} M.~M.,  {Fomalont} E.~B.,   {Lequeux} J.,  1972,
  \mn@doi [\aj] {10.1086/111301}, \href
  {https://ui.adsabs.harvard.edu/abs/1972AJ.....77..405B} {77, 405}

\bibitem[\protect\citeauthoryear{{Briggs}}{{Briggs}}{1995}]{Briggs1995}
{Briggs} D.~S.,  1995, in American Astronomical Society Meeting Abstracts. p.
  112.02

\bibitem[\protect\citeauthoryear{{Calistro Rivera} et~al.,}{{Calistro Rivera}
  et~al.}{2017}]{CalistroRivera2017}
{Calistro Rivera} G.,  et~al., 2017, \mn@doi [\mnras] {10.1093/mnras/stx1040},
  \href {https://ui.adsabs.harvard.edu/abs/2017MNRAS.469.3468C} {469, 3468}

\bibitem[\protect\citeauthoryear{{Chen} et~al.,}{{Chen}
  et~al.}{2018}]{Chen2018}
{Chen} C. T.~J.,  et~al., 2018, \mn@doi [\mnras] {10.1093/mnras/sty1036}, \href
  {https://ui.adsabs.harvard.edu/abs/2018MNRAS.478.2132C} {478, 2132}

\bibitem[\protect\citeauthoryear{{Ciliegi} et~al.,}{{Ciliegi}
  et~al.}{1999}]{Ciliegi1999}
{Ciliegi} P.,  et~al., 1999, \mn@doi [\mnras]
  {10.1046/j.1365-8711.1999.02103.x}, \href
  {https://ui.adsabs.harvard.edu/abs/1999MNRAS.302..222C} {302, 222}

\bibitem[\protect\citeauthoryear{{Condon}}{{Condon}}{1992}]{Condon1992}
{Condon} J.~J.,  1992, \mn@doi [\araa] {10.1146/annurev.aa.30.090192.003043},
  \href {https://ui.adsabs.harvard.edu/abs/1992ARA&A..30..575C} {30, 575}

\bibitem[\protect\citeauthoryear{{Condon}, {Cotton}, {Greisen}, {Yin},
  {Perley}, {Taylor}  \& {Broderick}}{{Condon} et~al.}{1998}]{Condon1998}
{Condon} J.~J.,  {Cotton} W.~D.,  {Greisen} E.~W.,  {Yin} Q.~F.,  {Perley}
  R.~A.,  {Taylor} G.~B.,   {Broderick} J.~J.,  1998, \mn@doi [\aj]
  {10.1086/300337}, \href
  {https://ui.adsabs.harvard.edu/abs/1998AJ....115.1693C} {115, 1693}

\bibitem[\protect\citeauthoryear{{Condon} et~al.,}{{Condon}
  et~al.}{2012}]{Condon2012}
{Condon} J.~J.,  et~al., 2012, \mn@doi [\apj] {10.1088/0004-637X/758/1/23},
  \href {https://ui.adsabs.harvard.edu/abs/2012ApJ...758...23C} {758, 23}

\bibitem[\protect\citeauthoryear{{Cooray} \& {Sheth}}{{Cooray} \&
  {Sheth}}{2002}]{Cooray2002}
{Cooray} A.,  {Sheth} R.,  2002, \mn@doi [\physrep]
  {10.1016/S0370-1573(02)00276-4}, \href
  {https://ui.adsabs.harvard.edu/abs/2002PhR...372....1C} {372, 1}

\bibitem[\protect\citeauthoryear{{Dav{\'e}}, {Angl{\'e}s-Alc{\'a}zar},
  {Narayanan}, {Li}, {Rafieferantsoa}  \& {Appleby}}{{Dav{\'e}}
  et~al.}{2019}]{Simba}
{Dav{\'e}} R.,  {Angl{\'e}s-Alc{\'a}zar} D.,  {Narayanan} D.,  {Li} Q.,
  {Rafieferantsoa} M.~H.,   {Appleby} S.,  2019, \mn@doi [\mnras]
  {10.1093/mnras/stz937}, \href
  {https://ui.adsabs.harvard.edu/abs/2019MNRAS.486.2827D} {486, 2827}

\bibitem[\protect\citeauthoryear{{Davies} et~al.,}{{Davies}
  et~al.}{2017}]{Davies2017}
{Davies} L.~J.~M.,  et~al., 2017, \mn@doi [\mnras] {10.1093/mnras/stw3080},
  \href {https://ui.adsabs.harvard.edu/abs/2017MNRAS.466.2312D} {466, 2312}

\bibitem[\protect\citeauthoryear{{Davies} et~al.,}{{Davies}
  et~al.}{2018}]{Davies2018}
{Davies} L.~J.~M.,  et~al., 2018, \mn@doi [\mnras] {10.1093/mnras/sty1553},
  \href {https://ui.adsabs.harvard.edu/abs/2018MNRAS.480..768D} {480, 768}

\bibitem[\protect\citeauthoryear{{Davies} et~al.,}{{Davies}
  et~al.}{2021}]{Davies2021}
{Davies} L.~J.~M.,  et~al., 2021, \mn@doi [\mnras] {10.1093/mnras/stab1601},
  \href {https://ui.adsabs.harvard.edu/abs/2021MNRAS.506..256D} {506, 256}

\bibitem[\protect\citeauthoryear{{Delhaize} et~al.,}{{Delhaize}
  et~al.}{2017}]{Delhaize2017}
{Delhaize} J.,  et~al., 2017, \mn@doi [\aap] {10.1051/0004-6361/201629430},
  \href {https://ui.adsabs.harvard.edu/abs/2017A&A...602A...4D} {602, A4}

\bibitem[\protect\citeauthoryear{{Delvecchio} et~al.,}{{Delvecchio}
  et~al.}{2021}]{Delvecchio2021}
{Delvecchio} I.,  et~al., 2021, \mn@doi [\aap] {10.1051/0004-6361/202039647},
  \href {https://ui.adsabs.harvard.edu/abs/2021A&A...647A.123D} {647, A123}

\bibitem[\protect\citeauthoryear{{Donley} et~al.,}{{Donley}
  et~al.}{2012}]{Donley2012}
{Donley} J.~L.,  et~al., 2012, \mn@doi [\apj] {10.1088/0004-637X/748/2/142},
  \href {https://ui.adsabs.harvard.edu/abs/2012ApJ...748..142D} {748, 142}

\bibitem[\protect\citeauthoryear{{Eddington}}{{Eddington}}{1913}]{Eddington1913}
{Eddington} A.~S.,  1913, \mn@doi [\mnras] {10.1093/mnras/73.5.359}, \href
  {https://ui.adsabs.harvard.edu/abs/1913MNRAS..73..359E} {73, 359}

\bibitem[\protect\citeauthoryear{{Fanaroff} \& {Riley}}{{Fanaroff} \&
  {Riley}}{1974}]{FanaroffRiley}
{Fanaroff} B.~L.,  {Riley} J.~M.,  1974, \mn@doi [\mnras]
  {10.1093/mnras/167.1.31P}, \href
  {https://ui.adsabs.harvard.edu/abs/1974MNRAS.167P..31F} {167, 31P}

\bibitem[\protect\citeauthoryear{{Fixsen} et~al.,}{{Fixsen}
  et~al.}{2011}]{Fixsen2011}
{Fixsen} D.~J.,  et~al., 2011, \mn@doi [\apj] {10.1088/0004-637X/734/1/5},
  \href {https://ui.adsabs.harvard.edu/abs/2011ApJ...734....5F} {734, 5}

\bibitem[\protect\citeauthoryear{{Fomalont}, {Kellermann}, {Cowie}, {Capak},
  {Barger}, {Partridge}, {Windhorst}  \& {Richards}}{{Fomalont}
  et~al.}{2006}]{Fomalont2006}
{Fomalont} E.~B.,  {Kellermann} K.~I.,  {Cowie} L.~L.,  {Capak} P.,  {Barger}
  A.~J.,  {Partridge} R.~B.,  {Windhorst} R.~A.,   {Richards} E.~A.,  2006,
  \mn@doi [\apjs] {10.1086/508169}, \href
  {https://ui.adsabs.harvard.edu/abs/2006ApJS..167..103F} {167, 103}

\bibitem[\protect\citeauthoryear{{Foreman-Mackey}, {Hogg}, {Lang}  \&
  {Goodman}}{{Foreman-Mackey} et~al.}{2013}]{emcee}
{Foreman-Mackey} D.,  {Hogg} D.~W.,  {Lang} D.,   {Goodman} J.,  2013, \mn@doi
  [\pasp] {10.1086/670067}, \href
  {https://ui.adsabs.harvard.edu/abs/2013PASP..125..306F} {125, 306}

\bibitem[\protect\citeauthoryear{{Galvin} et~al.,}{{Galvin}
  et~al.}{2018}]{Galvin2018}
{Galvin} T.~J.,  et~al., 2018, \mn@doi [\mnras] {10.1093/mnras/stx2613}, \href
  {https://ui.adsabs.harvard.edu/abs/2018MNRAS.474..779G} {474, 779}

\bibitem[\protect\citeauthoryear{{Garn}, {Green}, {Riley}  \&
  {Alexander}}{{Garn} et~al.}{2009}]{Garn2009}
{Garn} T.,  {Green} D.~A.,  {Riley} J.~M.,   {Alexander} P.,  2009, \mn@doi
  [\mnras] {10.1111/j.1365-2966.2009.15073.x}, \href
  {https://ui.adsabs.harvard.edu/abs/2009MNRAS.397.1101G} {397, 1101}

\bibitem[\protect\citeauthoryear{{Gehrels}}{{Gehrels}}{1986}]{Gehrels1986}
{Gehrels} N.,  1986, \mn@doi [\apj] {10.1086/164079}, \href
  {https://ui.adsabs.harvard.edu/abs/1986ApJ...303..336G} {303, 336}

\bibitem[\protect\citeauthoryear{{Gruppioni} et~al.,}{{Gruppioni}
  et~al.}{1999}]{Gruppioni1999}
{Gruppioni} C.,  et~al., 1999, \mn@doi [\mnras]
  {10.1046/j.1365-8711.1999.02415.x}, \href
  {https://ui.adsabs.harvard.edu/abs/1999MNRAS.305..297G} {305, 297}

\bibitem[\protect\citeauthoryear{{Gupta} et~al.,}{{Gupta} et~al.}{2017}]{ugmrt}
{Gupta} Y.,  et~al., 2017, \mn@doi [Current Science]
  {10.18520/cs/v113/i04/707-714}, \href
  {https://ui.adsabs.harvard.edu/abs/2017CSci..113..707G} {113, 707}

\bibitem[\protect\citeauthoryear{{G{\"u}rkan} et~al.,}{{G{\"u}rkan}
  et~al.}{2018}]{Gurkan2018}
{G{\"u}rkan} G.,  et~al., 2018, \mn@doi [\mnras] {10.1093/mnras/sty016}, \href
  {https://ui.adsabs.harvard.edu/abs/2018MNRAS.475.3010G} {475, 3010}

\bibitem[\protect\citeauthoryear{{Hale} et~al.,}{{Hale}
  et~al.}{2019}]{Hale2019}
{Hale} C.~L.,  et~al., 2019, \mn@doi [\aap] {10.1051/0004-6361/201833906},
  \href {https://ui.adsabs.harvard.edu/abs/2019A&A...622A...4H} {622, A4}

\bibitem[\protect\citeauthoryear{{Hale} et~al.,}{{Hale}
  et~al.}{2021}]{Hale2021}
{Hale} C.~L.,  et~al., 2021, \mn@doi [\pasa] {10.1017/pasa.2021.47}, \href
  {https://ui.adsabs.harvard.edu/abs/2021PASA...38...58H} {38, e058}

\bibitem[\protect\citeauthoryear{{Hardcastle} et~al.,}{{Hardcastle}
  et~al.}{2021}]{Hardcastle2021}
{Hardcastle} M.~J.,  et~al., 2021, \mn@doi [\aap]
  {10.1051/0004-6361/202038814}, \href
  {https://ui.adsabs.harvard.edu/abs/2021A&A...648A..10H} {648, A10}

\bibitem[\protect\citeauthoryear{{Harris} et~al.,}{{Harris}
  et~al.}{2020}]{numpy2}
{Harris} C.~R.,  et~al., 2020, \mn@doi [\nat] {10.1038/s41586-020-2649-2},
  \href {https://ui.adsabs.harvard.edu/abs/2020Natur.585..357H} {585, 357}

\bibitem[\protect\citeauthoryear{{Hasinger} et~al.,}{{Hasinger}
  et~al.}{2007}]{Hasinger2007}
{Hasinger} G.,  et~al., 2007, \mn@doi [\apjs] {10.1086/516576}, \href
  {https://ui.adsabs.harvard.edu/abs/2007ApJS..172...29H} {172, 29}

\bibitem[\protect\citeauthoryear{{Heckman} \& {Best}}{{Heckman} \&
  {Best}}{2014}]{Heckman2014}
{Heckman} T.~M.,  {Best} P.~N.,  2014, \mn@doi [\araa]
  {10.1146/annurev-astro-081913-035722}, \href
  {https://ui.adsabs.harvard.edu/abs/2014ARA&A..52..589H} {52, 589}

\bibitem[\protect\citeauthoryear{{Heywood}, {Jarvis}  \& {Condon}}{{Heywood}
  et~al.}{2013}]{Heywood2013}
{Heywood} I.,  {Jarvis} M.~J.,   {Condon} J.~J.,  2013, \mn@doi [\mnras]
  {10.1093/mnras/stt843}, \href
  {https://ui.adsabs.harvard.edu/abs/2013MNRAS.432.2625H} {432, 2625}

\bibitem[\protect\citeauthoryear{{Heywood}, {Hale}, {Jarvis}, {Makhathini},
  {Peters}, {Sebokolodi}  \& {Smirnov}}{{Heywood} et~al.}{2020}]{Heywood2020}
{Heywood} I.,  {Hale} C.~L.,  {Jarvis} M.~J.,  {Makhathini} S.,  {Peters}
  J.~A.,  {Sebokolodi} M.~L.~L.,   {Smirnov} O.~M.,  2020, \mn@doi [\mnras]
  {10.1093/mnras/staa1770}, \href
  {https://ui.adsabs.harvard.edu/abs/2020MNRAS.496.3469H} {496, 3469}

\bibitem[\protect\citeauthoryear{{Heywood} et~al.,}{{Heywood}
  et~al.}{2022}]{Heywood2022}
{Heywood} I.,  et~al., 2022, \mn@doi [\mnras] {10.1093/mnras/stab3021}, \href
  {https://ui.adsabs.harvard.edu/abs/2022MNRAS.509.2150H} {509, 2150}

\bibitem[\protect\citeauthoryear{{Hopkins}, {Afonso}, {Chan}, {Cram},
  {Georgakakis}  \& {Mobasher}}{{Hopkins} et~al.}{2003}]{Hopkins2003}
{Hopkins} A.~M.,  {Afonso} J.,  {Chan} B.,  {Cram} L.~E.,  {Georgakakis} A.,
  {Mobasher} B.,  2003, \mn@doi [\aj] {10.1086/345974}, \href
  {https://ui.adsabs.harvard.edu/abs/2003AJ....125..465H} {125, 465}

\bibitem[\protect\citeauthoryear{{Hotan} et~al.,}{{Hotan}
  et~al.}{2021}]{ASKAP3}
{Hotan} A.~W.,  et~al., 2021, \mn@doi [\pasa] {10.1017/pasa.2021.1}, \href
  {https://ui.adsabs.harvard.edu/abs/2021PASA...38....9H} {38, e009}

\bibitem[\protect\citeauthoryear{{Hunter}}{{Hunter}}{2007}]{matplotlib}
{Hunter} J.~D.,  2007, \mn@doi [Computing in Science and Engineering]
  {10.1109/MCSE.2007.55}, \href
  {https://ui.adsabs.harvard.edu/abs/2007CSE.....9...90H} {9, 90}

\bibitem[\protect\citeauthoryear{{Ibar}, {Ivison}, {Biggs}, {Lal}, {Best}  \&
  {Green}}{{Ibar} et~al.}{2009}]{Ibar2009}
{Ibar} E.,  {Ivison} R.~J.,  {Biggs} A.~D.,  {Lal} D.~V.,  {Best} P.~N.,
  {Green} D.~A.,  2009, \mn@doi [\mnras] {10.1111/j.1365-2966.2009.14866.x},
  \href {https://ui.adsabs.harvard.edu/abs/2009MNRAS.397..281I} {397, 281}

\bibitem[\protect\citeauthoryear{{Intema}, {Jagannathan}, {Mooley}  \&
  {Frail}}{{Intema} et~al.}{2017}]{Intema2017}
{Intema} H.~T.,  {Jagannathan} P.,  {Mooley} K.~P.,   {Frail} D.~A.,  2017,
  \mn@doi [\aap] {10.1051/0004-6361/201628536}, \href
  {https://ui.adsabs.harvard.edu/abs/2017A&A...598A..78I} {598, A78}

\bibitem[\protect\citeauthoryear{{Jarvis} \& {Rawlings}}{{Jarvis} \&
  {Rawlings}}{2004}]{Jarvis2004}
{Jarvis} M.~J.,  {Rawlings} S.,  2004, \mn@doi [\nar]
  {10.1016/j.newar.2004.09.006}, \href
  {https://ui.adsabs.harvard.edu/abs/2004NewAR..48.1173J} {48, 1173}

\bibitem[\protect\citeauthoryear{{Jarvis} et~al.,}{{Jarvis}
  et~al.}{2010}]{Jarvis2010}
{Jarvis} M.~J.,  et~al., 2010, \mn@doi [\mnras]
  {10.1111/j.1365-2966.2010.17772.x}, \href
  {https://ui.adsabs.harvard.edu/abs/2010MNRAS.409...92J} {409, 92}

\bibitem[\protect\citeauthoryear{{Jarvis} et~al.,}{{Jarvis}
  et~al.}{2013}]{Jarvis2013}
{Jarvis} M.~J.,  et~al., 2013, \mn@doi [\mnras] {10.1093/mnras/sts118}, \href
  {https://ui.adsabs.harvard.edu/abs/2013MNRAS.428.1281J} {428, 1281}

\bibitem[\protect\citeauthoryear{{Jarvis} et~al.,}{{Jarvis}
  et~al.}{2016}]{Jarvis2016}
{Jarvis} M.,  et~al., 2016, in MeerKAT Science: On the Pathway to the SKA. p.~6
  (\mn@eprint {arXiv} {1709.01901})

\bibitem[\protect\citeauthoryear{{Johnston} et~al.,}{{Johnston}
  et~al.}{2007}]{ASKAP1}
{Johnston} S.,  et~al., 2007, \mn@doi [\pasa] {10.1071/AS07033}, \href
  {https://ui.adsabs.harvard.edu/abs/2007PASA...24..174J} {24, 174}

\bibitem[\protect\citeauthoryear{{Johnston} et~al.,}{{Johnston}
  et~al.}{2008}]{ASKAP2}
{Johnston} S.,  et~al., 2008, \mn@doi [Experimental Astronomy]
  {10.1007/s10686-008-9124-7}, \href
  {https://ui.adsabs.harvard.edu/abs/2008ExA....22..151J} {22, 151}

\bibitem[\protect\citeauthoryear{{Jonas}}{{Jonas}}{2009}]{MeerKAT1}
{Jonas} J.~L.,  2009, \mn@doi [IEEE Proceedings] {10.1109/JPROC.2009.2020713},
  \href {https://ui.adsabs.harvard.edu/abs/2009IEEEP..97.1522J} {97, 1522}

\bibitem[\protect\citeauthoryear{{Joye} \& {Mandel}}{{Joye} \&
  {Mandel}}{2003}]{ds9_2}
{Joye} W.~A.,  {Mandel} E.,  2003, in {Payne} H.~E.,  {Jedrzejewski} R.~I.,
  {Hook} R.~N.,  eds,  Astronomical Society of the Pacific Conference Series
  Vol. 295, Astronomical Data Analysis Software and Systems XII. p.~489

\bibitem[\protect\citeauthoryear{{Kellermann}, {Fomalont}, {Mainieri},
  {Padovani}, {Rosati}, {Shaver}, {Tozzi}  \& {Miller}}{{Kellermann}
  et~al.}{2008}]{Kellermann2008}
{Kellermann} K.~I.,  {Fomalont} E.~B.,  {Mainieri} V.,  {Padovani} P.,
  {Rosati} P.,  {Shaver} P.,  {Tozzi} P.,   {Miller} N.,  2008, \mn@doi [\apjs]
  {10.1086/591055}, \href
  {https://ui.adsabs.harvard.edu/abs/2008ApJS..179...71K} {179, 71}

\bibitem[\protect\citeauthoryear{{Laigle} et~al.,}{{Laigle}
  et~al.}{2016}]{Laigle2016}
{Laigle} C.,  et~al., 2016, \mn@doi [\apjs] {10.3847/0067-0049/224/2/24}, \href
  {https://ui.adsabs.harvard.edu/abs/2016ApJS..224...24L} {224, 24}

\bibitem[\protect\citeauthoryear{{Lonsdale} et~al.,}{{Lonsdale}
  et~al.}{2003}]{Lonsdale2003}
{Lonsdale} C.~J.,  et~al., 2003, \mn@doi [\pasp] {10.1086/376850}, \href
  {https://ui.adsabs.harvard.edu/abs/2003PASP..115..897L} {115, 897}

\bibitem[\protect\citeauthoryear{{Lovell}, {Geach}, {Dav{\'e}}, {Narayanan}  \&
  {Li}}{{Lovell} et~al.}{2021}]{Lovell2021}
{Lovell} C.~C.,  {Geach} J.~E.,  {Dav{\'e}} R.,  {Narayanan} D.,   {Li} Q.,
  2021, \mn@doi [\mnras] {10.1093/mnras/staa4043}, \href
  {https://ui.adsabs.harvard.edu/abs/2021MNRAS.502..772L} {502, 772}

\bibitem[\protect\citeauthoryear{{Mandal} et~al.,}{{Mandal}
  et~al.}{2021}]{Mandal2021}
{Mandal} S.,  et~al., 2021, \mn@doi [\aap] {10.1051/0004-6361/202039998}, \href
  {https://ui.adsabs.harvard.edu/abs/2021A&A...648A...5M} {648, A5}

\bibitem[\protect\citeauthoryear{{Matthews}, {Condon}, {Cotton}  \&
  {Mauch}}{{Matthews} et~al.}{2021a}]{Matthews2021}
{Matthews} A.~M.,  {Condon} J.~J.,  {Cotton} W.~D.,   {Mauch} T.,  2021a,
  \mn@doi [\apj] {10.3847/1538-4357/abdd37}, \href
  {https://ui.adsabs.harvard.edu/abs/2021ApJ...909..193M} {909, 193}

\bibitem[\protect\citeauthoryear{{Matthews}, {Condon}, {Cotton}  \&
  {Mauch}}{{Matthews} et~al.}{2021b}]{Matthews2021b}
{Matthews} A.~M.,  {Condon} J.~J.,  {Cotton} W.~D.,   {Mauch} T.,  2021b,
  \mn@doi [\apj] {10.3847/1538-4357/abfaf6}, \href
  {https://ui.adsabs.harvard.edu/abs/2021ApJ...914..126M} {914, 126}

\bibitem[\protect\citeauthoryear{{Mauch} \& {Sadler}}{{Mauch} \&
  {Sadler}}{2007}]{Mauch2007}
{Mauch} T.,  {Sadler} E.~M.,  2007, \mn@doi [\mnras]
  {10.1111/j.1365-2966.2006.11353.x}, \href
  {https://ui.adsabs.harvard.edu/abs/2007MNRAS.375..931M} {375, 931}

\bibitem[\protect\citeauthoryear{{Mauch} et~al.,}{{Mauch}
  et~al.}{2020}]{Mauch2020}
{Mauch} T.,  et~al., 2020, \mn@doi [\apj] {10.3847/1538-4357/ab5d2d}, \href
  {https://ui.adsabs.harvard.edu/abs/2020ApJ...888...61M} {888, 61}

\bibitem[\protect\citeauthoryear{{Mauduit} et~al.,}{{Mauduit}
  et~al.}{2012}]{Mauduit2012}
{Mauduit} J.~C.,  et~al., 2012, \mn@doi [\pasp] {10.1086/666945}, \href
  {https://ui.adsabs.harvard.edu/abs/2012PASP..124..714M} {124, 714}

\bibitem[\protect\citeauthoryear{{McConnell} et~al.,}{{McConnell}
  et~al.}{2020}]{racs}
{McConnell} D.,  et~al., 2020, \mn@doi [\pasa] {10.1017/pasa.2020.41}, \href
  {https://ui.adsabs.harvard.edu/abs/2020PASA...37...48M} {37, e048}

\bibitem[\protect\citeauthoryear{{McCracken} et~al.,}{{McCracken}
  et~al.}{2012}]{McCracken2012}
{McCracken} H.~J.,  et~al., 2012, \mn@doi [\aap] {10.1051/0004-6361/201219507},
  \href {https://ui.adsabs.harvard.edu/abs/2012A&A...544A.156M} {544, A156}

\bibitem[\protect\citeauthoryear{{McMullin}, {Waters}, {Schiebel}, {Young}  \&
  {Golap}}{{McMullin} et~al.}{2007}]{CASA}
{McMullin} J.~P.,  {Waters} B.,  {Schiebel} D.,  {Young} W.,   {Golap} K.,
  2007, in {Shaw} R.~A.,  {Hill} F.,   {Bell} D.~J.,  eds,  Astronomical
  Society of the Pacific Conference Series Vol. 376, Astronomical Data Analysis
  Software and Systems XVI. p.~127

\bibitem[\protect\citeauthoryear{{Merson} et~al.,}{{Merson}
  et~al.}{2013}]{Merson2013}
{Merson} A.~I.,  et~al., 2013, \mn@doi [\mnras] {10.1093/mnras/sts355}, \href
  {https://ui.adsabs.harvard.edu/abs/2013MNRAS.429..556M} {429, 556}

\bibitem[\protect\citeauthoryear{{Mohan} \& {Rafferty}}{{Mohan} \&
  {Rafferty}}{2015}]{PyBDSF}
{Mohan} N.,  {Rafferty} D.,  2015, {PyBDSF: Python Blob Detection and Source
  Finder} (\mn@eprint {ascl} {1502.007})

\bibitem[\protect\citeauthoryear{{Murphy} \& {Chary}}{{Murphy} \&
  {Chary}}{2018}]{Murphy2018}
{Murphy} E.~J.,  {Chary} R.-R.,  2018, \mn@doi [\apj]
  {10.3847/1538-4357/aac2b6}, \href
  {https://ui.adsabs.harvard.edu/abs/2018ApJ...861...27M} {861, 27}

\bibitem[\protect\citeauthoryear{{Ni} et~al.,}{{Ni} et~al.}{2021}]{Ni2021}
{Ni} Q.,  et~al., 2021, \mn@doi [\apjs] {10.3847/1538-4365/ac0dc6}, \href
  {https://ui.adsabs.harvard.edu/abs/2021ApJS..256...21N} {256, 21}

\bibitem[\protect\citeauthoryear{{Norris} et~al.,}{{Norris}
  et~al.}{2021}]{Norris2021}
{Norris} R.~P.,  et~al., 2021, \mn@doi [\pasa] {10.1017/pasa.2021.42}, \href
  {https://ui.adsabs.harvard.edu/abs/2021PASA...38...46N} {38, e046}

\bibitem[\protect\citeauthoryear{{Ocran}, {Taylor}, {Vaccari},
  {Ishwara-Chandra}  \& {Prandoni}}{{Ocran} et~al.}{2020}]{Ocran2020}
{Ocran} E.~F.,  {Taylor} A.~R.,  {Vaccari} M.,  {Ishwara-Chandra} C.~H.,
  {Prandoni} I.,  2020, \mn@doi [\mnras] {10.1093/mnras/stz2954}, \href
  {https://ui.adsabs.harvard.edu/abs/2020MNRAS.491.1127O} {491, 1127}

\bibitem[\protect\citeauthoryear{{Offringa} et~al.,}{{Offringa}
  et~al.}{2014}]{Offringa2014}
{Offringa} A.~R.,  et~al., 2014, \mn@doi [\mnras] {10.1093/mnras/stu1368},
  \href {https://ui.adsabs.harvard.edu/abs/2014MNRAS.444..606O} {444, 606}

\bibitem[\protect\citeauthoryear{{Oliver} et~al.,}{{Oliver}
  et~al.}{2012}]{Oliver2012}
{Oliver} S.~J.,  et~al., 2012, \mn@doi [\mnras]
  {10.1111/j.1365-2966.2012.20912.x}, \href
  {https://ui.adsabs.harvard.edu/abs/2012MNRAS.424.1614O} {424, 1614}

\bibitem[\protect\citeauthoryear{{Owen} \& {Morrison}}{{Owen} \&
  {Morrison}}{2008}]{Owen2008}
{Owen} F.~N.,  {Morrison} G.~E.,  2008, \mn@doi [\aj]
  {10.1088/0004-6256/136/5/1889}, \href
  {https://ui.adsabs.harvard.edu/abs/2008AJ....136.1889O} {136, 1889}

\bibitem[\protect\citeauthoryear{{Padovani}}{{Padovani}}{2016}]{Padovani2016}
{Padovani} P.,  2016, \mn@doi [\aapr] {10.1007/s00159-016-0098-6}, \href
  {https://ui.adsabs.harvard.edu/abs/2016A&ARv..24...13P} {24, 13}

\bibitem[\protect\citeauthoryear{{Padovani}, {Bonzini}, {Kellermann}, {Miller},
  {Mainieri}  \& {Tozzi}}{{Padovani} et~al.}{2015}]{Padovani2015}
{Padovani} P.,  {Bonzini} M.,  {Kellermann} K.~I.,  {Miller} N.,  {Mainieri}
  V.,   {Tozzi} P.,  2015, \mn@doi [\mnras] {10.1093/mnras/stv1375}, \href
  {https://ui.adsabs.harvard.edu/abs/2015MNRAS.452.1263P} {452, 1263}

\bibitem[\protect\citeauthoryear{{Planck Collaboration} et~al.,}{{Planck
  Collaboration} et~al.}{2016}]{Planck2016}
{Planck Collaboration} et~al., 2016, \mn@doi [\aap]
  {10.1051/0004-6361/201525830}, \href
  {https://ui.adsabs.harvard.edu/abs/2016A&A...594A..13P} {594, A13}

\bibitem[\protect\citeauthoryear{{Prandoni}, {Guglielmino}, {Morganti},
  {Vaccari}, {Maini}, {R{\"o}ttgering}, {Jarvis}  \& {Garrett}}{{Prandoni}
  et~al.}{2018}]{Prandoni2018}
{Prandoni} I.,  {Guglielmino} G.,  {Morganti} R.,  {Vaccari} M.,  {Maini} A.,
  {R{\"o}ttgering} H.~J.~A.,  {Jarvis} M.~J.,   {Garrett} M.~A.,  2018, \mn@doi
  [\mnras] {10.1093/mnras/sty2521}, \href
  {https://ui.adsabs.harvard.edu/abs/2018MNRAS.481.4548P} {481, 4548}

\bibitem[\protect\citeauthoryear{{Prescott} et~al.,}{{Prescott}
  et~al.}{2018}]{Prescott2018}
{Prescott} M.,  et~al., 2018, \mn@doi [\mnras] {10.1093/mnras/sty1789}, \href
  {https://ui.adsabs.harvard.edu/abs/2018MNRAS.480..707P} {480, 707}

\bibitem[\protect\citeauthoryear{{Reich} \& {Reich}}{{Reich} \&
  {Reich}}{1986}]{Reich1986}
{Reich} P.,  {Reich} W.,  1986, \aaps, \href
  {https://ui.adsabs.harvard.edu/abs/1986A&AS...63..205R} {63, 205}

\bibitem[\protect\citeauthoryear{{Richards}}{{Richards}}{2000}]{Richards2000}
{Richards} E.~A.,  2000, \mn@doi [\apj] {10.1086/308684}, \href
  {https://ui.adsabs.harvard.edu/abs/2000ApJ...533..611R} {533, 611}

\bibitem[\protect\citeauthoryear{{Robitaille}}{{Robitaille}}{2019}]{aplpy1}
{Robitaille} T.,  2019, {APLpy v2.0: The Astronomical Plotting Library in
  Python}, \mn@doi{10.5281/zenodo.2567476}

\bibitem[\protect\citeauthoryear{{Robitaille} \& {Bressert}}{{Robitaille} \&
  {Bressert}}{2012}]{aplpy2}
{Robitaille} T.,  {Bressert} E.,  2012, {APLpy: Astronomical Plotting Library
  in Python} (\mn@eprint {ascl} {1208.017})

\bibitem[\protect\citeauthoryear{{Sabater} et~al.,}{{Sabater}
  et~al.}{2021}]{Sabater2021}
{Sabater} J.,  et~al., 2021, \mn@doi [\aap] {10.1051/0004-6361/202038828},
  \href {https://ui.adsabs.harvard.edu/abs/2021A&A...648A...2S} {648, A2}

\bibitem[\protect\citeauthoryear{{Seymour} et~al.,}{{Seymour}
  et~al.}{2008}]{Seymour2008}
{Seymour} N.,  et~al., 2008, \mn@doi [\mnras]
  {10.1111/j.1365-2966.2008.13166.x}, \href
  {https://ui.adsabs.harvard.edu/abs/2008MNRAS.386.1695S} {386, 1695}

\bibitem[\protect\citeauthoryear{{Shimwell} et~al.,}{{Shimwell}
  et~al.}{2019}]{Shimwell2019}
{Shimwell} T.~W.,  et~al., 2019, \mn@doi [\aap] {10.1051/0004-6361/201833559},
  \href {https://ui.adsabs.harvard.edu/abs/2019A&A...622A...1S} {622, A1}

\bibitem[\protect\citeauthoryear{{Shimwell} et~al.,}{{Shimwell}
  et~al.}{2022}]{Shimwell2022}
{Shimwell} T.~W.,  et~al., 2022, arXiv e-prints, \href
  {https://ui.adsabs.harvard.edu/abs/2022arXiv220211733S} {p. arXiv:2202.11733}

\bibitem[\protect\citeauthoryear{{Smirnov} \& {Tasse}}{{Smirnov} \&
  {Tasse}}{2015}]{Smirnov2015}
{Smirnov} O.~M.,  {Tasse} C.,  2015, \mn@doi [\mnras] {10.1093/mnras/stv418},
  \href {https://ui.adsabs.harvard.edu/abs/2015MNRAS.449.2668S} {449, 2668}

\bibitem[\protect\citeauthoryear{{Smith} et~al.,}{{Smith}
  et~al.}{2021}]{Smith2021}
{Smith} D.~J.~B.,  et~al., 2021, \mn@doi [\aap] {10.1051/0004-6361/202039343},
  \href {https://ui.adsabs.harvard.edu/abs/2021A&A...648A...6S} {648, A6}

\bibitem[\protect\citeauthoryear{{Smithsonian Astrophysical
  Observatory}}{{Smithsonian Astrophysical Observatory}}{2000}]{ds9_1}
{Smithsonian Astrophysical Observatory} 2000, {SAOImage DS9: A utility for
  displaying astronomical images in the X11 window environment} (\mn@eprint
  {ascl} {0003.002})

\bibitem[\protect\citeauthoryear{{Smol{\v{c}}i{\'c}}
  et~al.,}{{Smol{\v{c}}i{\'c}} et~al.}{2017a}]{Smolcic2017}
{Smol{\v{c}}i{\'c}} V.,  et~al., 2017a, \mn@doi [\aap]
  {10.1051/0004-6361/201628704}, \href
  {https://ui.adsabs.harvard.edu/abs/2017A&A...602A...1S} {602, A1}

\bibitem[\protect\citeauthoryear{{Smol{\v{c}}i{\'c}}
  et~al.,}{{Smol{\v{c}}i{\'c}} et~al.}{2017b}]{Smolcic2017b}
{Smol{\v{c}}i{\'c}} V.,  et~al., 2017b, \mn@doi [\aap]
  {10.1051/0004-6361/201630223}, \href
  {https://ui.adsabs.harvard.edu/abs/2017A&A...602A...2S} {602, A2}

\bibitem[\protect\citeauthoryear{{Tabatabaei} et~al.,}{{Tabatabaei}
  et~al.}{2017}]{Tabatabaei2017}
{Tabatabaei} F.~S.,  et~al., 2017, \mn@doi [\apj]
  {10.3847/1538-4357/836/2/185}, \href
  {https://ui.adsabs.harvard.edu/abs/2017ApJ...836..185T} {836, 185}

\bibitem[\protect\citeauthoryear{{Tasse}, {R{\"o}ttgering}, {Best}, {Cohen},
  {Pierre}  \& {Wilman}}{{Tasse} et~al.}{2007}]{Tasse2007}
{Tasse} C.,  {R{\"o}ttgering} H.~J.~A.,  {Best} P.~N.,  {Cohen} A.~S.,
  {Pierre} M.,   {Wilman} R.,  2007, \mn@doi [\aap]
  {10.1051/0004-6361:20066986}, \href
  {https://ui.adsabs.harvard.edu/abs/2007A&A...471.1105T} {471, 1105}

\bibitem[\protect\citeauthoryear{{Tasse} et~al.,}{{Tasse}
  et~al.}{2018}]{Tasse2018}
{Tasse} C.,  et~al., 2018, \mn@doi [\aap] {10.1051/0004-6361/201731474}, \href
  {https://ui.adsabs.harvard.edu/abs/2018A&A...611A..87T} {611, A87}

\bibitem[\protect\citeauthoryear{{Tasse} et~al.,}{{Tasse}
  et~al.}{2021}]{Tasse2021}
{Tasse} C.,  et~al., 2021, \mn@doi [\aap] {10.1051/0004-6361/202038804}, \href
  {https://ui.adsabs.harvard.edu/abs/2021A&A...648A...1T} {648, A1}

\bibitem[\protect\citeauthoryear{{Taylor}}{{Taylor}}{2005}]{topcat1}
{Taylor} M.~B.,  2005, in {Shopbell} P.,  {Britton} M.,   {Ebert} R.,  eds,
  Astronomical Society of the Pacific Conference Series Vol. 347, Astronomical
  Data Analysis Software and Systems XIV. p.~29

\bibitem[\protect\citeauthoryear{{Taylor}}{{Taylor}}{2011}]{topcat2}
{Taylor} M.,  2011, {TOPCAT: Tool for OPerations on Catalogues And Tables}
  (\mn@eprint {ascl} {1101.010})

\bibitem[\protect\citeauthoryear{{Thomas}, {Dav{\'e}}, {Angl{\'e}s-Alc{\'a}zar}
   \& {Jarvis}}{{Thomas} et~al.}{2019}]{Thomas2019}
{Thomas} N.,  {Dav{\'e}} R.,  {Angl{\'e}s-Alc{\'a}zar} D.,   {Jarvis} M.,
  2019, \mn@doi [\mnras] {10.1093/mnras/stz1703}, \href
  {https://ui.adsabs.harvard.edu/abs/2019MNRAS.487.5764T} {487, 5764}

\bibitem[\protect\citeauthoryear{{Thomas}, {Dav{\'e}}, {Jarvis}  \&
  {Angl{\'e}s-Alc{\'a}zar}}{{Thomas} et~al.}{2021}]{Thomas2021}
{Thomas} N.,  {Dav{\'e}} R.,  {Jarvis} M.~J.,   {Angl{\'e}s-Alc{\'a}zar} D.,
  2021, \mn@doi [\mnras] {10.1093/mnras/stab654}, \href
  {https://ui.adsabs.harvard.edu/abs/2021MNRAS.503.3492T} {503, 3492}

\bibitem[\protect\citeauthoryear{{Thompson}, {Clark}, {Wade}  \&
  {Napier}}{{Thompson} et~al.}{1980}]{VLA}
{Thompson} A.~R.,  {Clark} B.~G.,  {Wade} C.~M.,   {Napier} P.~J.,  1980,
  \mn@doi [\apjs] {10.1086/190688}, \href
  {https://ui.adsabs.harvard.edu/abs/1980ApJS...44..151T} {44, 151}

\bibitem[\protect\citeauthoryear{{Vernstrom}, {Scott}  \& {Wall}}{{Vernstrom}
  et~al.}{2011}]{Vernstrom2011}
{Vernstrom} T.,  {Scott} D.,   {Wall} J.~V.,  2011, \mn@doi [\mnras]
  {10.1111/j.1365-2966.2011.18990.x}, \href
  {https://ui.adsabs.harvard.edu/abs/2011MNRAS.415.3641V} {415, 3641}

\bibitem[\protect\citeauthoryear{{Vernstrom}, {Scott}, {Wall}, {Condon},
  {Cotton}, {Kellermann}  \& {Perley}}{{Vernstrom}
  et~al.}{2016}]{Vernstrom2016}
{Vernstrom} T.,  {Scott} D.,  {Wall} J.~V.,  {Condon} J.~J.,  {Cotton} W.~D.,
  {Kellermann} K.~I.,   {Perley} R.~A.,  2016, \mn@doi [\mnras]
  {10.1093/mnras/stw1836}, \href
  {https://ui.adsabs.harvard.edu/abs/2016MNRAS.462.2934V} {462, 2934}

\bibitem[\protect\citeauthoryear{{Virtanen} et~al.,}{{Virtanen}
  et~al.}{2020}]{scipy}
{Virtanen} P.,  et~al., 2020, \mn@doi [Nature Methods]
  {10.1038/s41592-019-0686-2}, \href
  {https://ui.adsabs.harvard.edu/abs/2020NatMe..17..261V} {17, 261}

\bibitem[\protect\citeauthoryear{{White}, {Becker}, {Helfand}  \&
  {Gregg}}{{White} et~al.}{1997}]{White1997}
{White} R.~L.,  {Becker} R.~H.,  {Helfand} D.~J.,   {Gregg} M.~D.,  1997,
  \mn@doi [\apj] {10.1086/303564}, \href
  {https://ui.adsabs.harvard.edu/abs/1997ApJ...475..479W} {475, 479}

\bibitem[\protect\citeauthoryear{{White}, {Jarvis}, {H{\"a}u{\ss}ler}  \&
  {Maddox}}{{White} et~al.}{2015}]{White2015}
{White} S.~V.,  {Jarvis} M.~J.,  {H{\"a}u{\ss}ler} B.,   {Maddox} N.,  2015,
  \mn@doi [\mnras] {10.1093/mnras/stv134}, \href
  {https://ui.adsabs.harvard.edu/abs/2015MNRAS.448.2665W} {448, 2665}

\bibitem[\protect\citeauthoryear{{White}, {Jarvis}, {Kalfountzou},
  {Hardcastle}, {Verma}, {Cao Orjales}  \& {Stevens}}{{White}
  et~al.}{2017}]{White2017}
{White} S.~V.,  {Jarvis} M.~J.,  {Kalfountzou} E.,  {Hardcastle} M.~J.,
  {Verma} A.,  {Cao Orjales} J.~M.,   {Stevens} J.,  2017, \mn@doi [\mnras]
  {10.1093/mnras/stx284}, \href
  {https://ui.adsabs.harvard.edu/abs/2017MNRAS.468..217W} {468, 217}

\bibitem[\protect\citeauthoryear{{Whittam}, {Prescott}, {McAlpine}, {Jarvis}
  \& {Heywood}}{{Whittam} et~al.}{2018}]{Whittam2018}
{Whittam} I.~H.,  {Prescott} M.,  {McAlpine} K.,  {Jarvis} M.~J.,   {Heywood}
  I.,  2018, \mn@doi [\mnras] {10.1093/mnras/sty1787}, \href
  {https://ui.adsabs.harvard.edu/abs/2018MNRAS.480..358W} {480, 358}

\bibitem[\protect\citeauthoryear{{Whittam} et~al.,}{{Whittam}
  et~al.}{2022}]{Whittam2022}
{Whittam} I.~H.,  et~al., 2022, \mn@doi [\mnras] {10.1093/mnras/stac2140},
  \href {https://ui.adsabs.harvard.edu/abs/2022MNRAS.516..245W} {516, 245}

\bibitem[\protect\citeauthoryear{{Williams} et~al.,}{{Williams}
  et~al.}{2018}]{Williams2018}
{Williams} W.~L.,  et~al., 2018, \mn@doi [\mnras] {10.1093/mnras/sty026}, \href
  {https://ui.adsabs.harvard.edu/abs/2018MNRAS.475.3429W} {475, 3429}

\bibitem[\protect\citeauthoryear{{Williams} et~al.,}{{Williams}
  et~al.}{2021}]{Williams2021}
{Williams} W.~L.,  et~al., 2021, \mn@doi [\aap] {10.1051/0004-6361/202141745},
  \href {https://ui.adsabs.harvard.edu/abs/2021A&A...655A..40W} {655, A40}

\bibitem[\protect\citeauthoryear{{Wilman} et~al.,}{{Wilman}
  et~al.}{2008}]{Wilman2008}
{Wilman} R.~J.,  et~al., 2008, \mn@doi [\mnras]
  {10.1111/j.1365-2966.2008.13486.x}, \href
  {https://ui.adsabs.harvard.edu/abs/2008MNRAS.388.1335W} {388, 1335}

\bibitem[\protect\citeauthoryear{{Wilman}, {Jarvis}, {Mauch}, {Rawlings}  \&
  {Hickey}}{{Wilman} et~al.}{2010}]{Wilman2010}
{Wilman} R.~J.,  {Jarvis} M.~J.,  {Mauch} T.,  {Rawlings} S.,   {Hickey} S.,
  2010, \mn@doi [\mnras] {10.1111/j.1365-2966.2010.16453.x}, \href
  {https://ui.adsabs.harvard.edu/abs/2010MNRAS.405..447W} {405, 447}

\bibitem[\protect\citeauthoryear{{Zehavi} et~al.,}{{Zehavi}
  et~al.}{2004}]{Zehavi2004}
{Zehavi} I.,  et~al., 2004, \mn@doi [\apj] {10.1086/386535}, \href
  {https://ui.adsabs.harvard.edu/abs/2004ApJ...608...16Z} {608, 16}

\bibitem[\protect\citeauthoryear{{Zwart}, {Santos}  \& {Jarvis}}{{Zwart}
  et~al.}{2015}]{Zwart2015}
{Zwart} J. T.~L.,  {Santos} M.,   {Jarvis} M.~J.,  2015, \mn@doi [\mnras]
  {10.1093/mnras/stv1716}, \href
  {https://ui.adsabs.harvard.edu/abs/2015MNRAS.453.1740Z} {453, 1740}

\bibitem[\protect\citeauthoryear{da Costa-Luis et~al.,}{da~Costa-Luis
  et~al.}{2021}]{tqdm}
da Costa-Luis C.,  et~al., 2021, {tqdm: A fast, Extensible Progress Bar for
  Python and CLI}, \mn@doi{10.5281/zenodo.5109730}, \url
  {https://doi.org/10.5281/zenodo.5109730}

\bibitem[\protect\citeauthoryear{{de Gasperin}, {Intema}  \& {Frail}}{{de
  Gasperin} et~al.}{2018}]{deGasperin2018}
{de Gasperin} F.,  {Intema} H.~T.,   {Frail} D.~A.,  2018, \mn@doi [\mnras]
  {10.1093/mnras/stx3125}, \href
  {https://ui.adsabs.harvard.edu/abs/2018MNRAS.474.5008D} {474, 5008}

\bibitem[\protect\citeauthoryear{{de Zotti}, {Massardi}, {Negrello}  \&
  {Wall}}{{de Zotti} et~al.}{2010}]{deZotti2010}
{de Zotti} G.,  {Massardi} M.,  {Negrello} M.,   {Wall} J.,  2010, \mn@doi
  [\aapr] {10.1007/s00159-009-0026-0}, \href
  {https://ui.adsabs.harvard.edu/abs/2010A&ARv..18....1D} {18, 1}

\bibitem[\protect\citeauthoryear{{van Haarlem} et~al.,}{{van Haarlem}
  et~al.}{2013}]{LOFAR}
{van Haarlem} M.~P.,  et~al., 2013, \mn@doi [\aap]
  {10.1051/0004-6361/201220873}, \href
  {https://ui.adsabs.harvard.edu/abs/2013A&A...556A...2V} {556, A2}

\bibitem[\protect\citeauthoryear{{van der Vlugt} et~al.,}{{van der Vlugt}
  et~al.}{2021}]{vandervlugt2021}
{van der Vlugt} D.,  et~al., 2021, \mn@doi [\apj] {10.3847/1538-4357/abcaa3},
  \href {https://ui.adsabs.harvard.edu/abs/2021ApJ...907....5V} {907, 5}

\bibitem[\protect\citeauthoryear{{van der Walt}, {Colbert}  \&
  {Varoquaux}}{{van der Walt} et~al.}{2011}]{numpy1}
{van der Walt} S.,  {Colbert} S.~C.,   {Varoquaux} G.,  2011, \mn@doi
  [Computing in Science and Engineering] {10.1109/MCSE.2011.37}, \href
  {https://ui.adsabs.harvard.edu/abs/2011CSE....13b..22V} {13, 22}

\makeatother
\end{thebibliography}



\appendix
\section{Comparison with VLA 3GHz Sources}
\label{sec:appendix1}
\begin{figure*}
    \centering
    \begin{subfigure}{.24\textwidth}
    \includegraphics[width=0.97\textwidth]{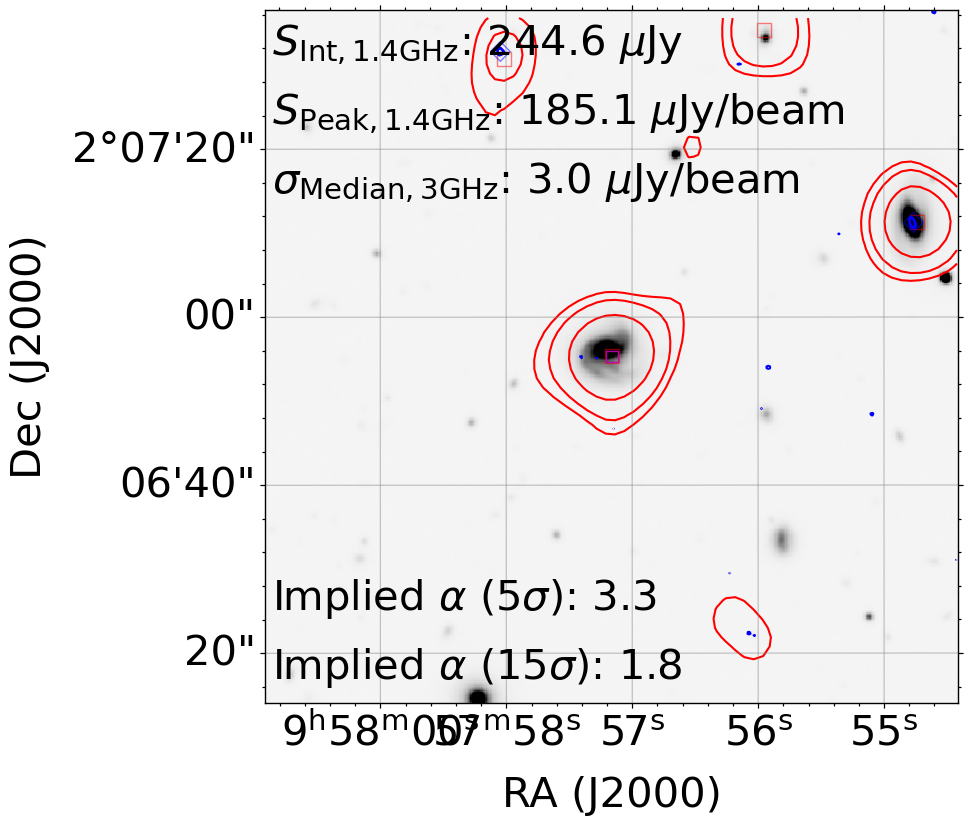}
    \end{subfigure}%
    \begin{subfigure}{.24\textwidth}
    \includegraphics[width=0.97\textwidth]{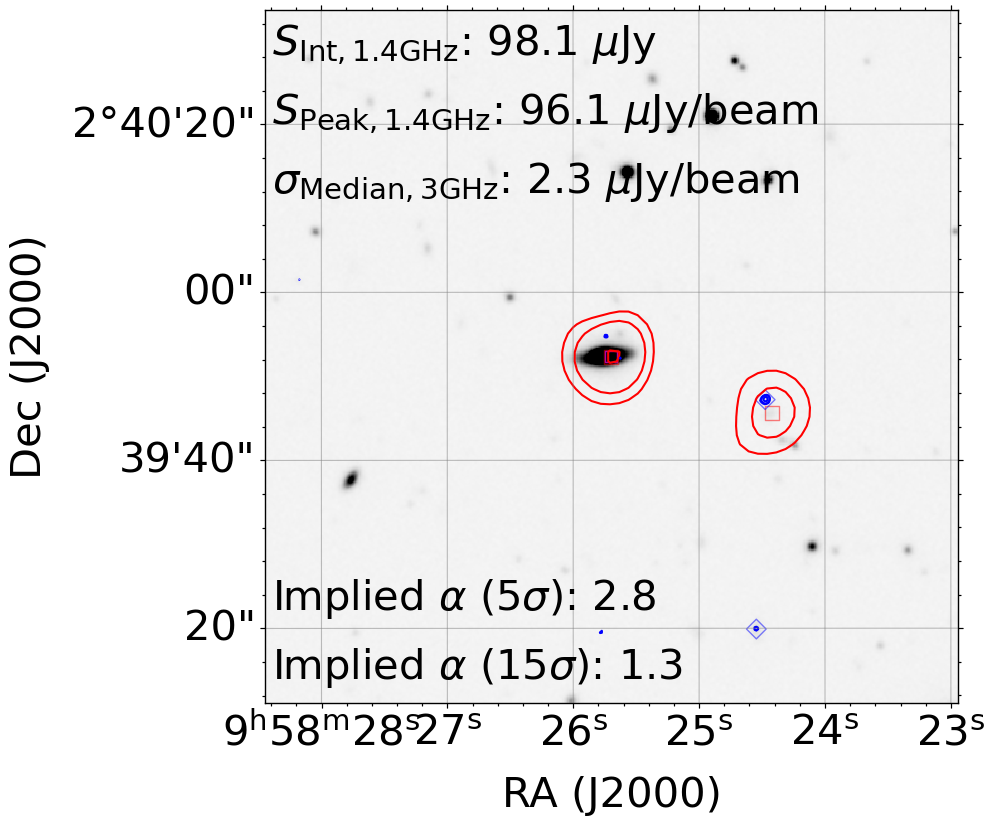}
    \end{subfigure}%
        \begin{subfigure}{.24\textwidth}
    \includegraphics[width=0.97\textwidth]{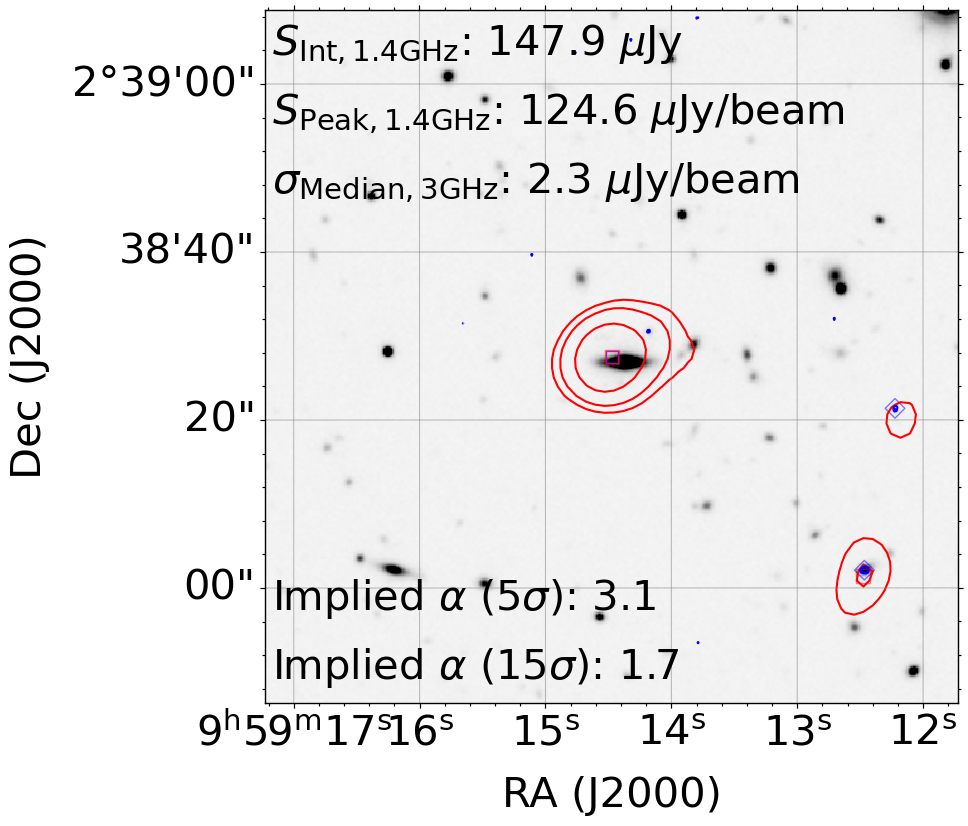}
    \end{subfigure}%
        \begin{subfigure}{.24\textwidth}
    \includegraphics[width=0.97\textwidth]{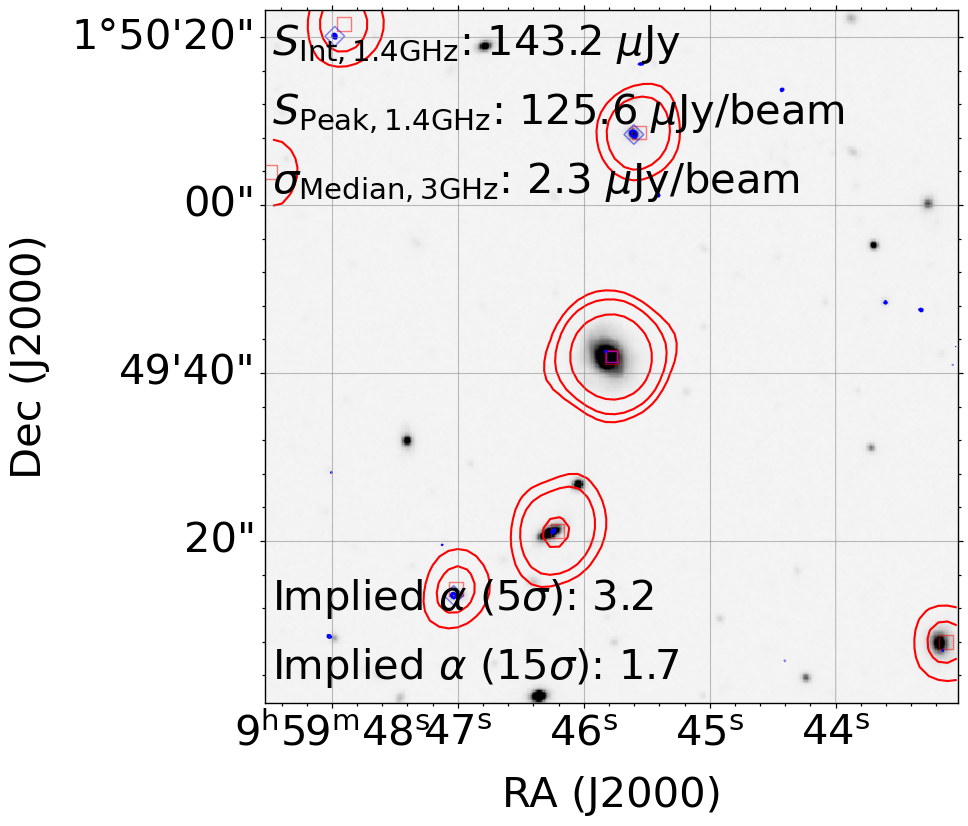}
\end{subfigure}%
            \linebreak
    \begin{subfigure}{.24\textwidth}
    \includegraphics[width=0.97\textwidth]{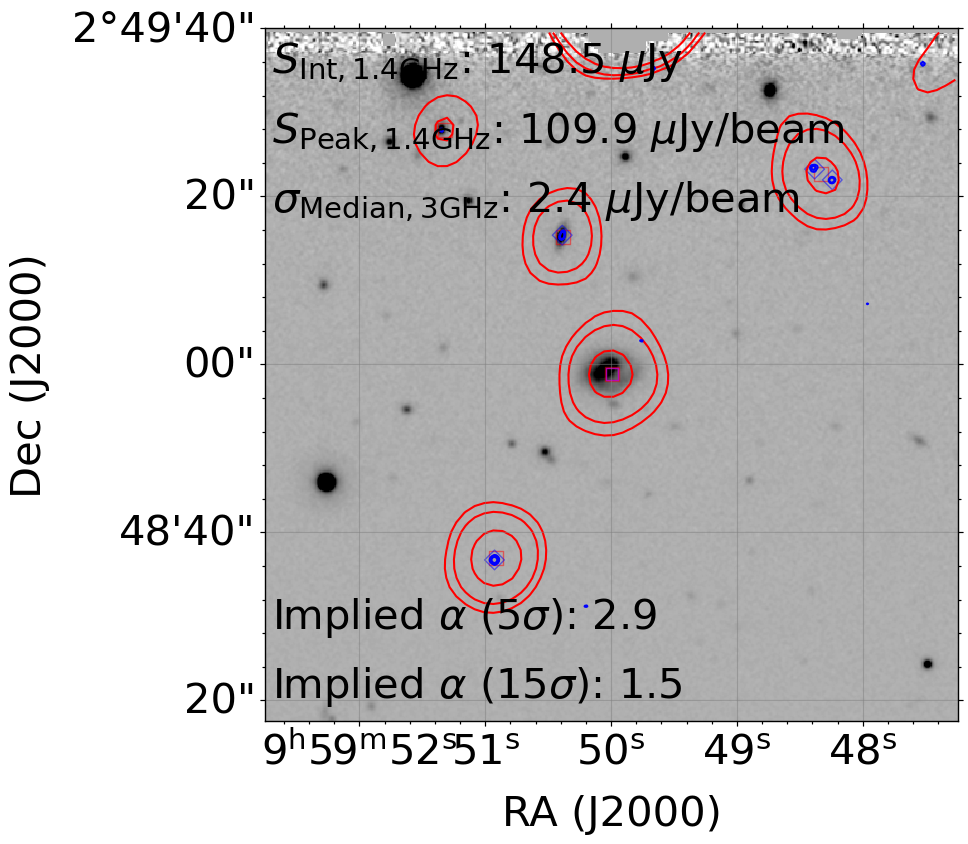}
    \end{subfigure}%
    \begin{subfigure}{.24\textwidth}
    \includegraphics[width=0.97\textwidth]{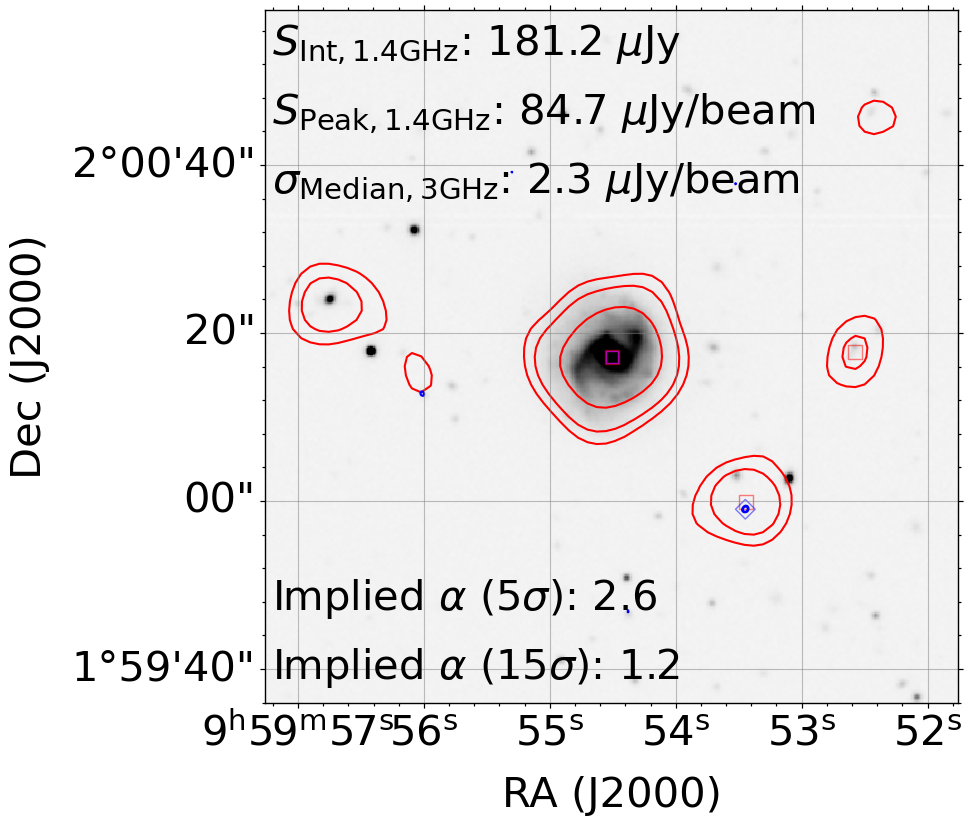}
    \end{subfigure}%
        \begin{subfigure}{.24\textwidth}
    \includegraphics[width=0.97\textwidth]{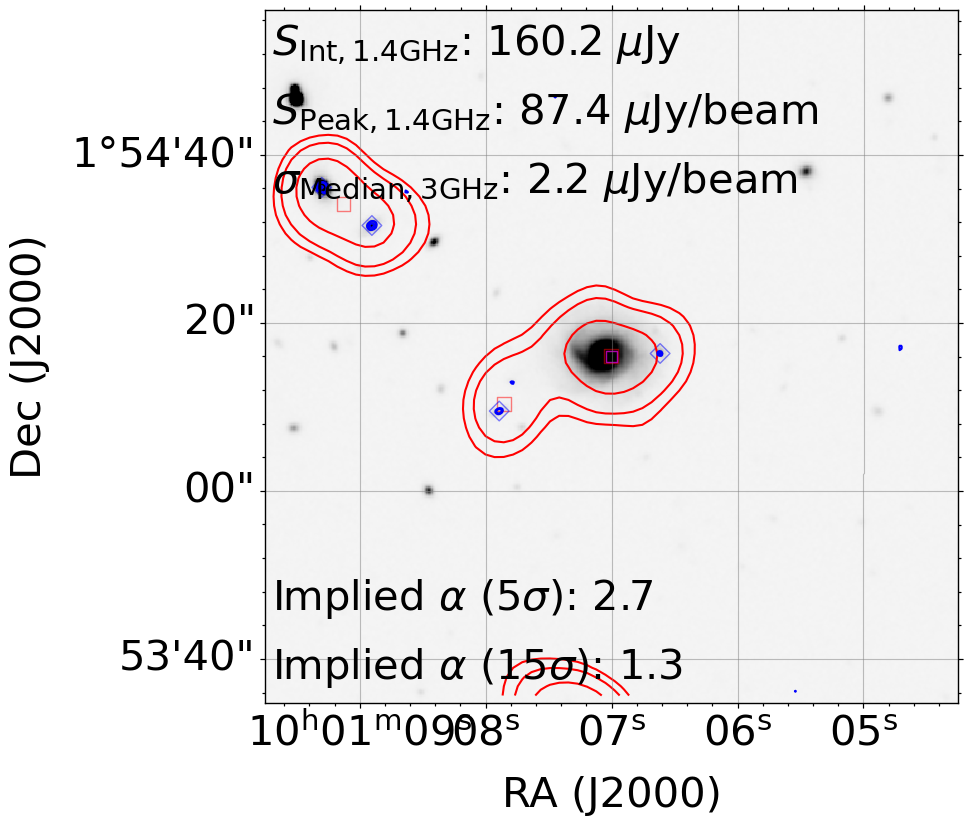}
    \end{subfigure}%
        \begin{subfigure}{.24\textwidth}
    \includegraphics[width=0.97\textwidth]{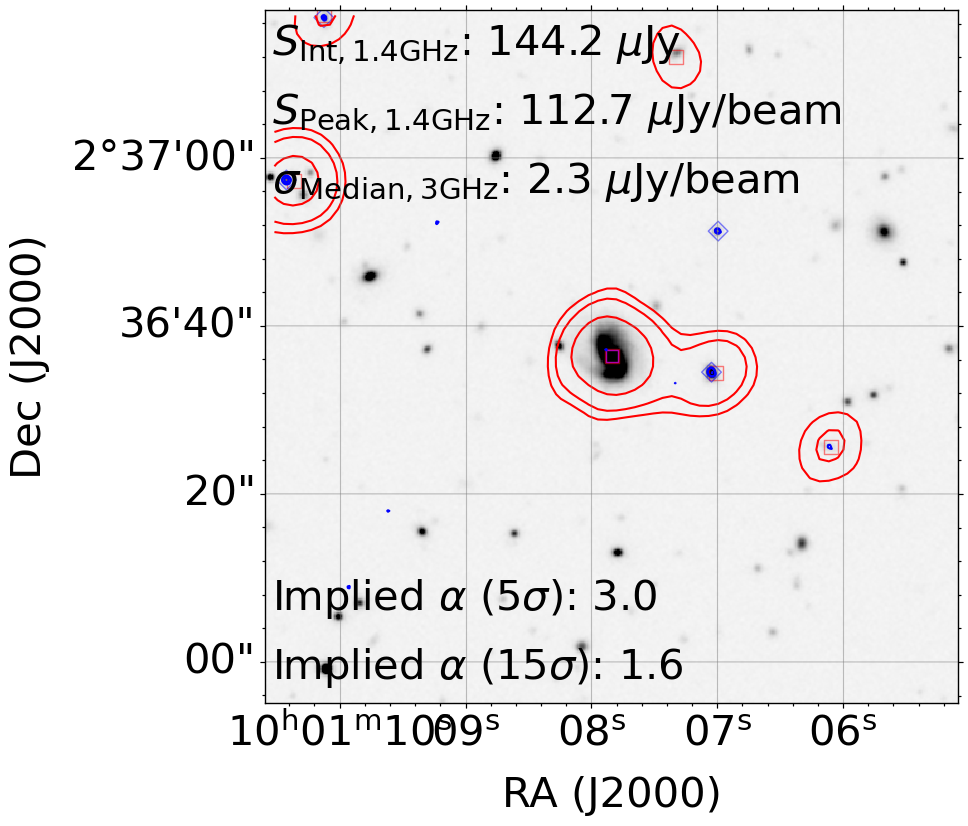}
    \end{subfigure}%
                \linebreak
    \begin{subfigure}{.24\textwidth}
    \includegraphics[width=0.97\textwidth]{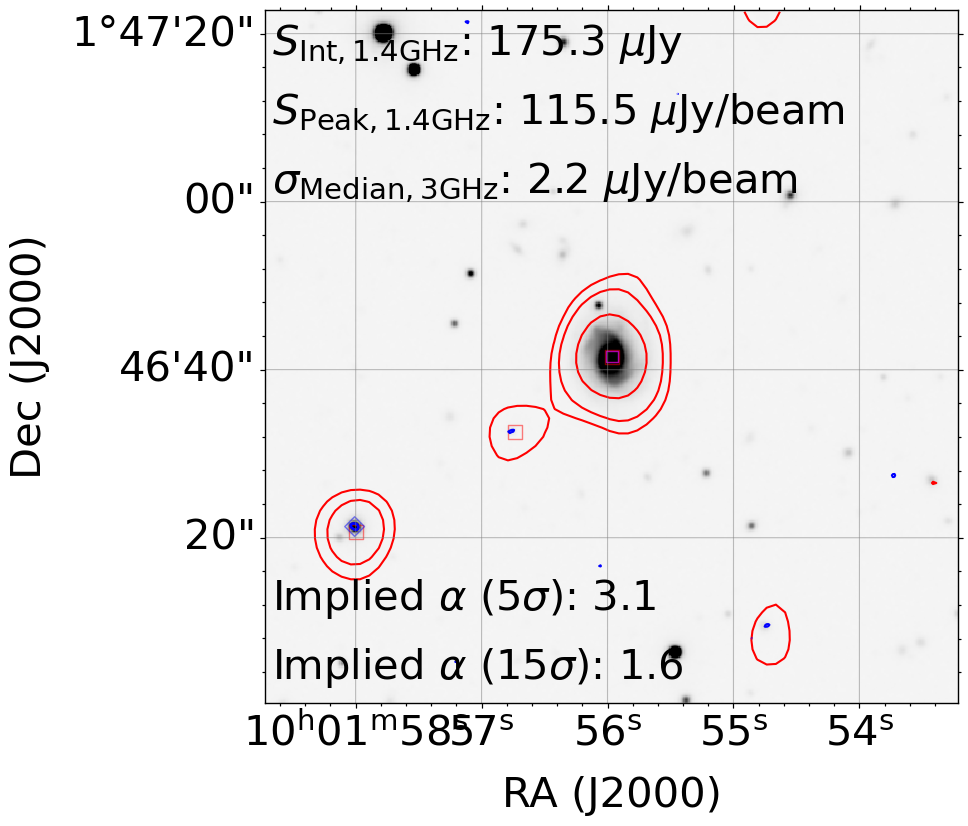}
    \end{subfigure}%
    \begin{subfigure}{.24\textwidth}
    \includegraphics[width=0.97\textwidth]{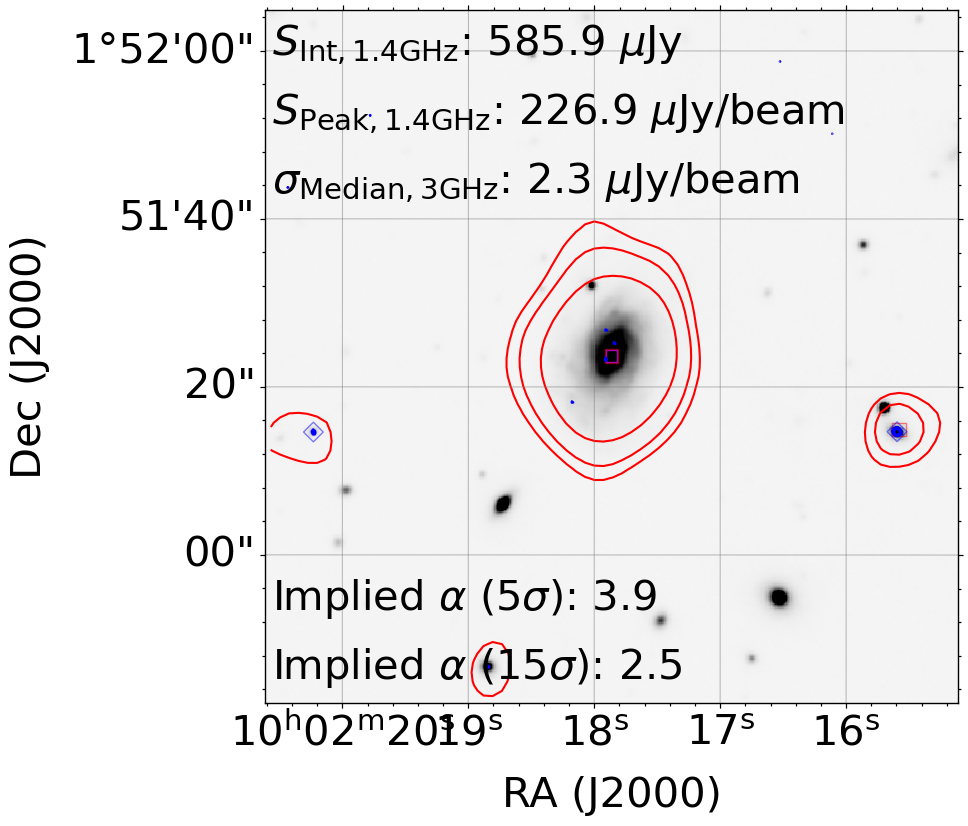}
    \end{subfigure}%
        \begin{subfigure}{.24\textwidth}
    \includegraphics[width=0.97\textwidth]{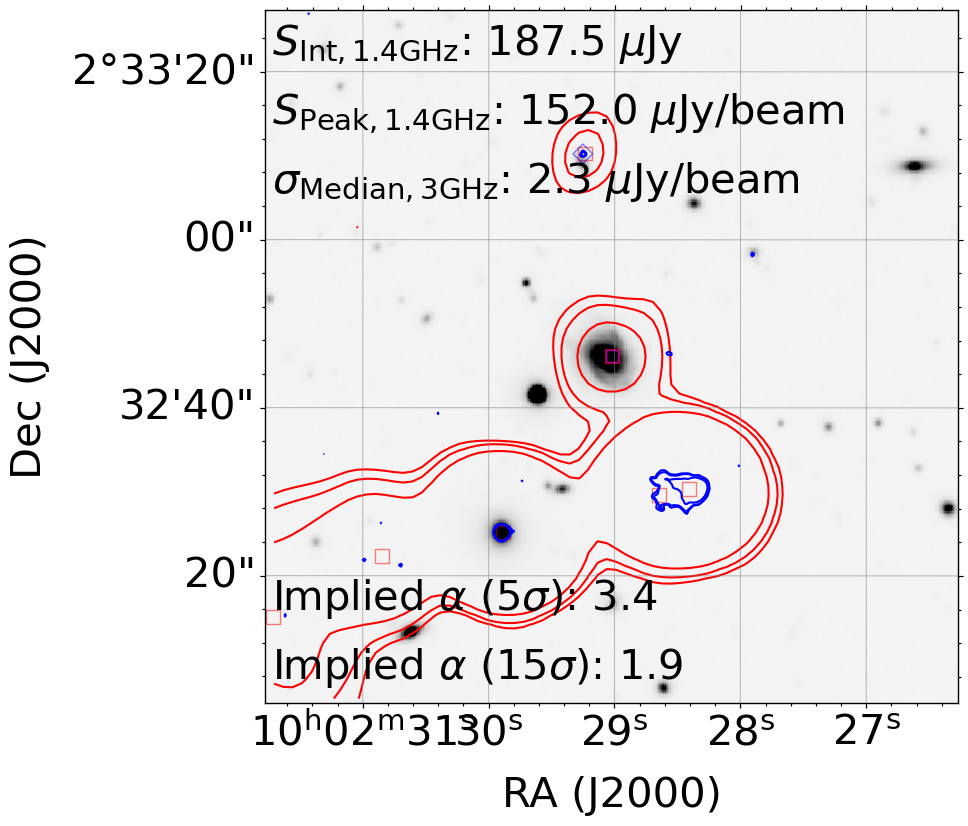}
    \end{subfigure}%
        \begin{subfigure}{.24\textwidth}
    \includegraphics[width=0.97\textwidth]{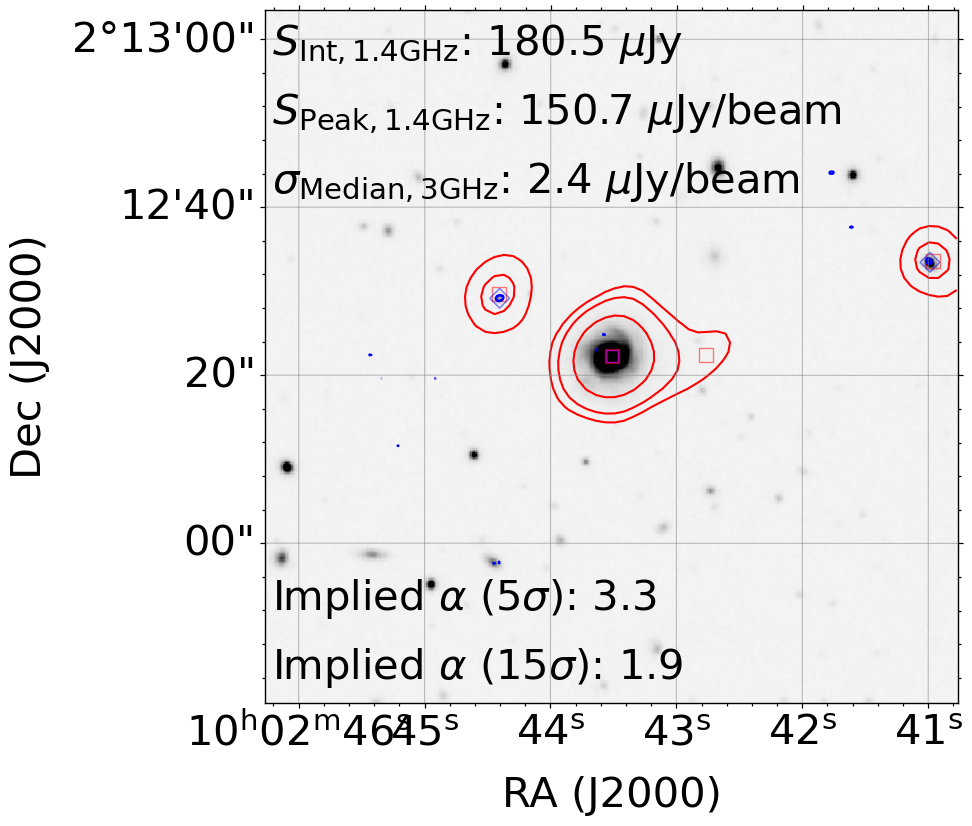}
    \end{subfigure}%
    \caption{{Example overlays for 12 sources whose host galaxies appear to have extended morphologies and are visible in the MIGHTEE images (red contours at 3, 5 and 10$\sigma$ and sources shown as red squares) but are not detected within the VLA 3GHz COSMOS image (blue contours at 4, 5 and 10$\sigma$ and sources shown as blue diamonds). These radio contours are overlaid on {{$K_S$}} band images from UltraVISTA \protect \citep{McCracken2012} DR4. The source which being investigated is in the centre of the image. Included in the figure {are} the 1.4 GHz integrated {($S_{\textrm{Int, 1.4 GHz}}$)} and peak {($S_{\textrm{Peak, 1.4 GHz}}$)} flux densities from MIGHTEE, the median 3 GHz rms {($\sigma_{\textrm{Median, 3 GHz}}$)} within the cutout from \protect \cite{Smolcic2017} and the implied lower limit on $\alpha$ assuming a $5\sigma$ and $15\sigma$ detection using the peak flux densities.}  }
    \label{fig:missing}
\end{figure*}

{As discussed in Section \ref{sec:discuss_sc}, one potential reason {for larger source counts at} faint flux densities compared to \cite{Smolcic2017} could relate to emission being resolved out by the VLA observations, resulting in missing sources or a reduction in the flux density observed from these sources. If not accounted for sufficiently, this could affect source count measurements. {Given that the MIGHTEE {{Early Science}} data covers the COSMOS field, we made a brief investigation} of this. Specifically, we examined sources in the MIGHTEE catalogues with peak flux densities $S_{ \textrm{1.4GHz, MIGHTEE}} \geq$60 $\muup$Jy that do not have a VLA 3 GHz COSMOS \citep{Smolcic2017} counterpart source within a 5\arcsec \ match radius. {Whilst many MIGHTEE sources have a counterpart or are not expected to due to sensitivity limits, a small number of sources were found that had limited or no 3 GHz emission and had extended host source morphologies. We} show 12 example overlays of these in Figure \ref{fig:missing}. For each source we indicate both the MIGHTEE scaled 1.4 GHz integrated and peak flux densities, which are in the range of $\sim100-600$ $\muup$Jy (integrated) and $\sim80-230$ $\muup$Jy {beam$^{-1}$} (peak). We also measure the median rms within the same VLA 3 GHz cutout and use this to determine what the measured spectral index would be from the peak flux densities assuming that the maximum emission of the source in the VLA 3 GHz image was at a 5$\sigma$ and 15$\sigma$ detection level. As shown in Figure 16 of \cite{Smolcic2017}, completeness of their catalogue is $\sim$50\% at $\sim$5$\sigma$ and rises to $\sim$90\% completeness at 15$\sigma$\footnote{{These completeness levels at a SNR assumes the median rms of 2.3 $\muup$Jy beam$^{-1}$}, though this rms level varies across the field.}. }

{The examples shown are some of the most extreme cases which have an implied limit on of $\alpha>1$ even based on 15-sigma limits and peak flux densities. The spectral indices measured from integrated flux densities or at 5$\sigma$ would give even steeper measurements of $\alpha$. Although sources could potentially have steep spectral indices, it could also imply that there is missing emission due to the baselines configurations used in the observations of \cite{Smolcic2017}, which may be less sensitive to large angular scales. If extended emission is being resolved out in the images for these and other sources, this could lead to an underestimation in flux densities and could affect source count measurements. If these potential effects are under accounted for in \cite{Smolcic2017}, this may explain why the source counts from \cite{Smolcic2017} appear to be {{underestimated}} compared to other deep radio observations in this work and that of \cite{Mauch2020, Matthews2021} and \cite{vandervlugt2021}. } {However, while Fig \ref{fig:missing} provides some indicative examples, {as stated earlier the majority of sources have counterparts or may not {necessarily} be expected {to, given the relative sensitivity limits}. A} full investigation of this issue is beyond the scope of this paper, and other factors may play a role. \cite{Smolcic2017} calculate their completeness to be less than 100\% at {15$\sigma$ and sources like these} may already be accounted for in the completeness corrections used in \cite{Smolcic2017}{, which do include methods to account for resolution bias}. {Factors} such as source finder incompleteness, {source variability}, flux offsets in the data and intrinsic steep spectral indices may also play a role.}


\bsp	
\label{lastpage}
\end{document}